\documentclass[useAMS,usenatbib,a4paper]{mn2e}
\newcommand {\aplt} {{\raise-.5ex\hbox{$\buildrel<\over\sim$}}} 
\usepackage{amsmath}
\usepackage{deluxetable}
\usepackage{graphicx}
\usepackage{url}

 \title[Open cluster wide binaries]{Wide binaries are rare in open clusters}

   \author[N.R.\ Deacon et al.]{N.R. Deacon$^{1}$, A.L. Kraus$^{2}$ \\
   $^1$Max Planck Institute for Astronomy, Konigstuhl 17, Heidelberg, 69117, Germany\\
   $^2$ Department of Astronomy, The University of Texas at Austin, Austin, TX 78712, USA\\}
 
\begin{document}
 \date{}
 \pagerange{\pageref{firstpage}--\pageref{lastpage}} \pubyear{2015}
 \maketitle
 \label{firstpage}
  \begin{abstract}
The population statistics of binary stars are an important output of star formation models. However populations of wide binaries evolve over time due to interactions within a system's birth environment and the unfolding of wide, hierarchical triple systems. Hence the wide binary populations observed in star forming regions or OB associations may not accurately reflect the wide binary populations that will eventually reach the field. We use Gaia DR2 data to select members of three open clusters, Alpha~Per, the Pleiades and Praesepe and to flag cluster members that are likely unresolved binaries due to overluminosity or elevated astrometric noise. We then identify the resolved wide binary population in each cluster, separating it from coincident pairings of unrelated cluster members. We find that these clusters have an average wide binary fraction in the 300-3000\,AU projected separation range of 2.1$\pm^{0.4}_{0.2}$\% increasing to 3.0$\pm^{0.8}_{0.7}$\% for primaries with masses in the 0.5--1.5\,$M_{\odot}$ range. This is significantly below the observed field wide binary fraction, but shows some wide binaries survive in these dynamically highly processed environments. We compare our results with another open cluster (the Hyades) and two populations of young stars that likely originated in looser associations (Young Moving Groups and the Pisces-Eridanus stream). We find that the Hyades also has a deficit of wide binaries while the products of looser associations have wide binary fractions at or above field level.
  \end{abstract}
%

\section{Introduction}
Binary systems are common, as almost half of solar-type stars (44$\pm$2\%; \citealt{Raghavan2010}, see also more recent work by \citealt{Moe2017}) have one or more companions. Multiple systems have been shown to be more common around higher-mass stars with a trend to wider systems around more massive primaries (see reviews by \citealt{Duchene2013}). Around a quarter of solar-type stars have a companion wider than 100\,AU \citep{Raghavan2010} with 4.4\% of stars similar to the Sun having companions wider than 2000\,AU \citep{Tokovinin2012}. 

\cite{Brandner1998} suggested that binary frequency and the separation distribution of binaries could vary significantly with star formation environment. As the field population is the combination of the outputs of countless star formation events, the field binary population is a superposition of the binary populations of all of these events (see discussion in \citealt{Patience2002}). Star formation is broadly categorised into clustered enviroments that are the likely progenitors of the open clusters we see in the solar neighbourhood and lower density distributed environments such as Sco-Cen or Taurus-Auriga. This latter star formation mode may be the progenitor of solar neighbourhood young moving groups. 

In lower-density star forming regions such as Taurus-Auriga \citep{Kraus2011} and Ophiuchus \citep{Cheetham2015}, wide systems are seen to exist at similar or higher frequencies to the field. \cite{Scally1999} suggested that the higher-mass Orion Nebular Cluster (ONC) had virtually no binaries wider than 1000\,AU. \cite{Reipurth2007} found that the ONC had a binary fraction of 8.8$\pm$1.1\% between 67.5 and 675 AU. More recently \cite{Jerabkova2019} found that the binary frequency per unit log separation was approximately 5\% for binaries in the 1000-3000\,AU range. However these studies of wide binarity in young populations may not translate directly to the binaries these populations will eventually contribute to the field.

 \cite{Reipurth2012} postulate that wide binaries are often triple systems which have evolved to one tight pair and one wide companion. This is supported by the apparently high frequency of higher-order multiplicity seen in wide systems \citep{Law2010,Allen2012}.
As noted by \cite{Elliott2015}, the \cite{Reipurth2012} model suggests that many young wide binaries will be unstable and will not survive to field age. Thus while the population of wide binaries in star forming regions is an vital test for the output of star formation simulations, these systems may have undergone significant evolution or disruption by the time they reach field age. This means that population of wide binaries in young $<100$\,Myr populations may not match the population of wide binaries from that star formation event that will eventually reach the field. Hence if one wishes to test the contribution a population of stars makes to the field wide binary population it is better to target intermediate-aged (100\,Myr--1\,Gyr) populations as these will suffer less future evolution due to either internal angular momentum evolution of the binary/hierarchical triple or due to external influences such as encounters with stars in the association/cluster or in the field.

We chose to study the wide binary populations of three well-studied open clusters, Alpha~Per, the Pleiades and Praesepe. All three have had their memberships extensively studied using a variety of techniques. Alpha~Per is a young cluster (85\,Myr; \citealt{Navascues2004}) lying at low Galactic latitude making it the most challenging of our three clusters to study. That said there have been multiple studies of its membership \citep{Jones1991,Rebolo1992,D.BarradoyNavascues2002,Deacon2004,Lodieu2012a,Lodieu2019a}. The Pleiades is the best-studied of our three clusters \citep{Hambly1991,Deacon2004,Stauffer2007,Lodieu2012,Bouy2015,Rebull2016,Olivares2018,Lodieu2019a}. It's closeness and young age (125\,Myr; \citealt{Stauffer1998}) have made it an ideal target to search for low-mass objects culminating in the discovery of some of the first brown dwarfs \citep{Rebolo1995,Basri1996} and even probing down to the planet-brown~dwarf boundary \citep{ZapateroOsorio2018}. Praesepe is the most distant and oldest of our clusters (790\,Myr; \citealt{Brandt2015}) and as such it has proved an ideal target to study a stellar population at close to field age \citep{Hambly1995,Hodgkin1999,Kraus2007,Baker2010,Boudreault2012,Rebull2017,Gao2019,Lodieu2019a}.

Binary stars in clusters can be identified by their overluminosity \citep{Stauffer1984,Rubenstein1997,Khalaj2013,Sheikhi2016}, lunar occultation \citep{Richichi2012}, radial velocity techniques \citep{NeillReid2000} or by direct imaging \citep{Bouvier1997a,Patience2002,Garcia2015,Hillenbrand2018}. However these studies are typically limited to relatively close separations with only the direct imaging surveys going out to separations of a few hundred~AU.

In this work we first identify cluster members using Gaia~DR2 data \citep{Prusti2016,Brown2018}. We then characterise each cluster, measuring the system mass function (the mass function without a correction for unresolved binarity) and flagging objects that may be unresolved binaries. After doing this we identify pairs of objects and disentangle the population of wide binaries in each cluster from the population of coincident pairings between unrelated cluster members. We then estimate the wide binary fractions of each cluster. These wide binary fractions are then compared to those in other clusters and associations. We use this comparison to draw conclusions on the origin of the field wide binary population.
\section{Cluster membership}

\subsection{Selecting cluster members}
We implemented a probabilistic cluster member search. This method built on that undertaken by \cite{Deacon2004} which is based on \cite{Hambly1995} and \cite{Sanders1971}. Our method takes into account the 3D nature of the cluster while also including a statistical treatment of background contamination. Full details of the likelihood calculation are given in Appendix~\ref{memb_like}.

We selected a sample of potential members from each cluster using the Gaia Archive\footnote{\url{https://gea.esac.esa.int/archive/}} \citep{Prusti2016,Brown2018}. Gaia provides exquisitely accurate astrometry with typical proper motion uncertainties of 0.1 milliarcseconds per year for $G<15$ rising to 1\,milliarcsecond per year at $G=20$ as well as high-quality parallaxes. For each cluster we used the cluster centre quoted in the Simbad database\footnote{http://simbad.u-strasbg.fr}. We then selected all stars in Gaia DR2 out to a radius of five degrees from the cluster centres. We list these cluster centres along with the approximate proper motions of each cluster in Table~\ref{cluster_details}. 

\begin{table*}
\caption{\label{cluster_details} The three open clusters we studied. The listed proper motions are the approximate Gaia proper motions of the cluster centres.}
\begin{center}
\begin{tabular}{|ccrrrrr}
\hline
Cluster&Cluster Centre&Search radius&$\mu_{\alpha}\cos{\delta}$&$\mu_{\delta}$&$\theta$&Number of\\
&&&&&&members\\
&&($^{\circ}$)&(mas/yr)&(mas/yr)&($^{\circ}$)\\
\hline
Alpha Per&03 26 42.0 +48 48 00$^a$&5&23&$-$26&48.0&801\\
Pleiades&03 47 00.0 +24 07 00$^b$&5&20&$-$46&66.5&1526\\
Praesepe&08 40 24.0 +19 40 00$^b$&5&$-$36&$-$13&160.0&1230\\
\hline
\multicolumn{6}{l}{$^a$ \protect\cite{Kharchenko2013}}\\
\multicolumn{6}{l}{$^b$ \protect\cite{Wu2009}}\\
\end{tabular}
\end{center}
\end{table*}%

 For each cluster we divided the data by observed Gaia $G$-band magnitude. Objects brighter than $G=10$\,mag were put into the brightest bin. We then used bins that were two magnitudes wide from $G=10$\,mag to $G=18$\,mag. The two faintest bins were set to be only one magnitude wide ($18<G<19$ and $19<G<20$) because astrometric errors increase faster at fainter magnitudes. We did not cut on Renormalised Unit Weighted Error (RUWE; \citealt{Lindegren2018a}) as we expect some binaries in the cluster to have elevated astrometric noise (see Section~\ref{astro_noise}). As wide binaries often have components that are themselves binaries \citep{Law2010,Allen2012}, cutting on RUWE would bias our search against some wide binaries.

\begin{figure*}
 \setlength{\unitlength}{1mm}
 \begin{tabular}{ccc}
 \includegraphics[scale=0.25]{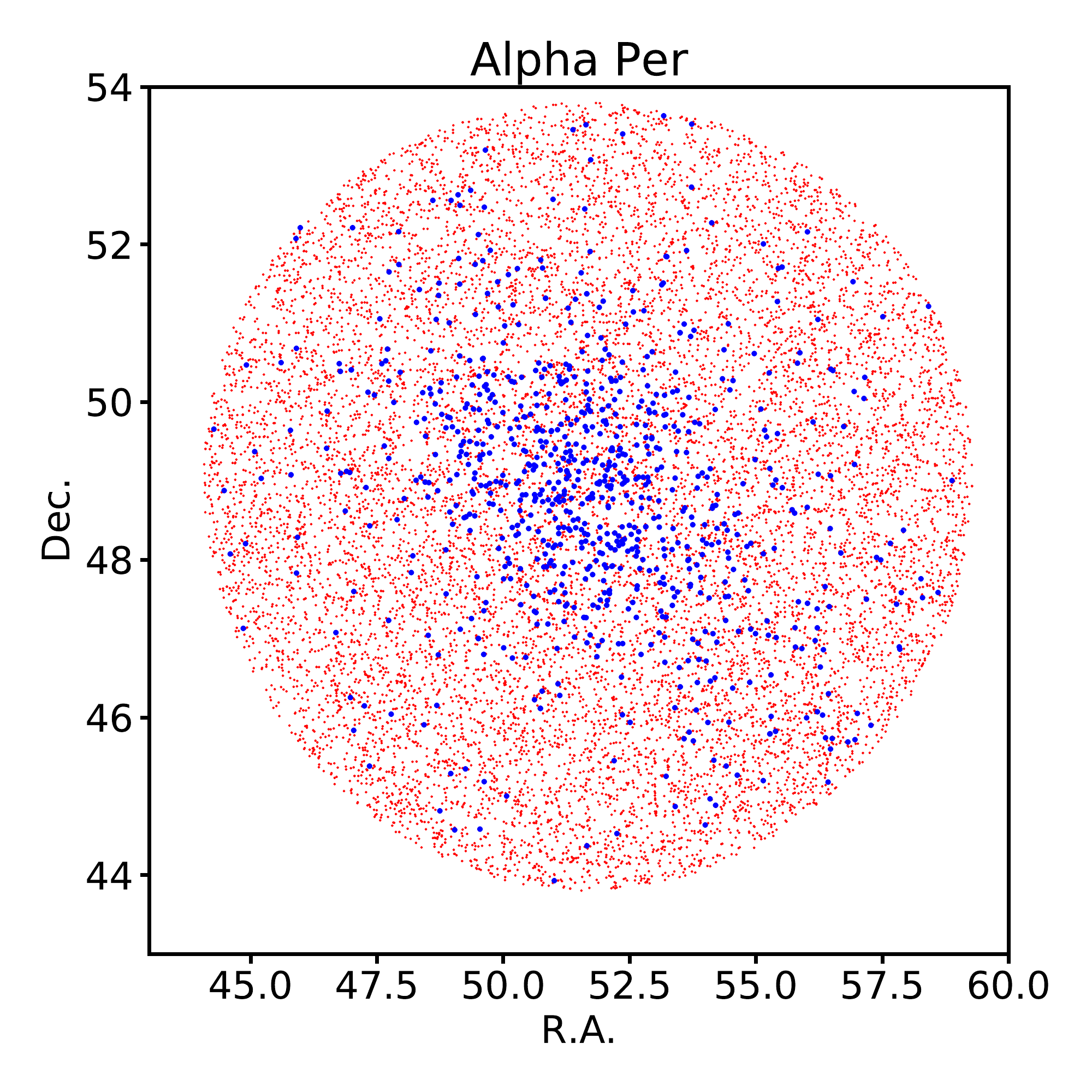}&
 \includegraphics[scale=0.25]{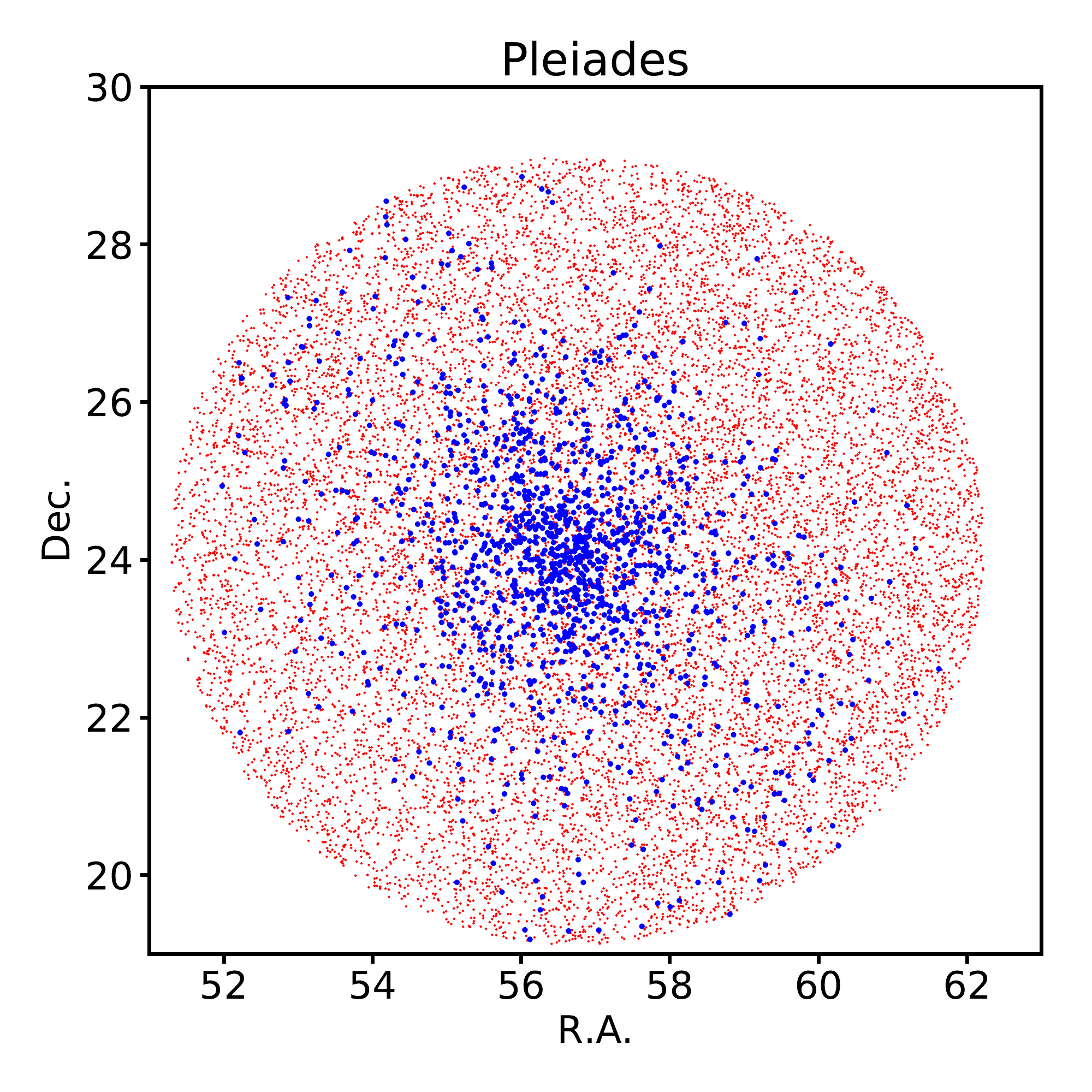}&
 \includegraphics[scale=0.25]{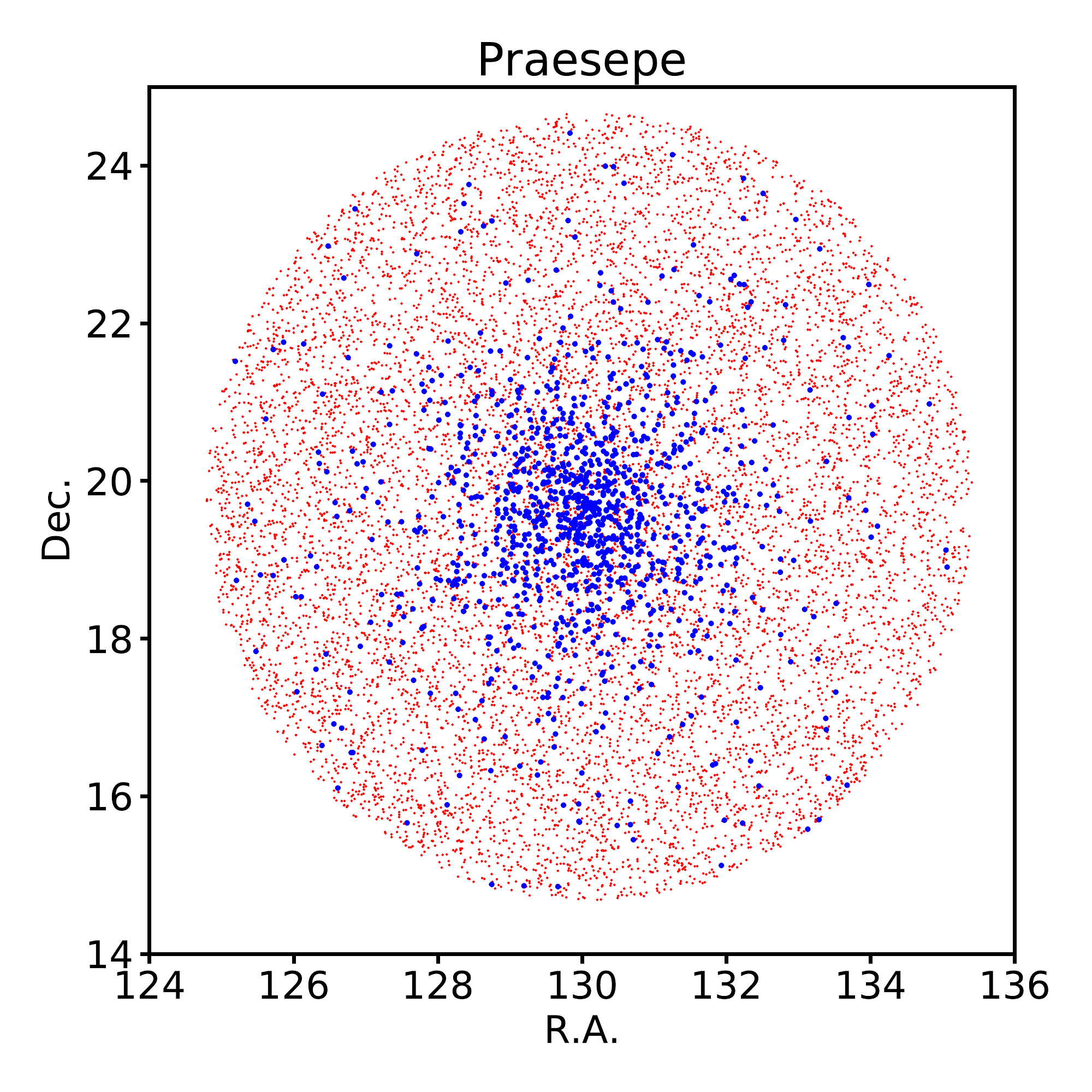}\\
 \includegraphics[scale=0.25]{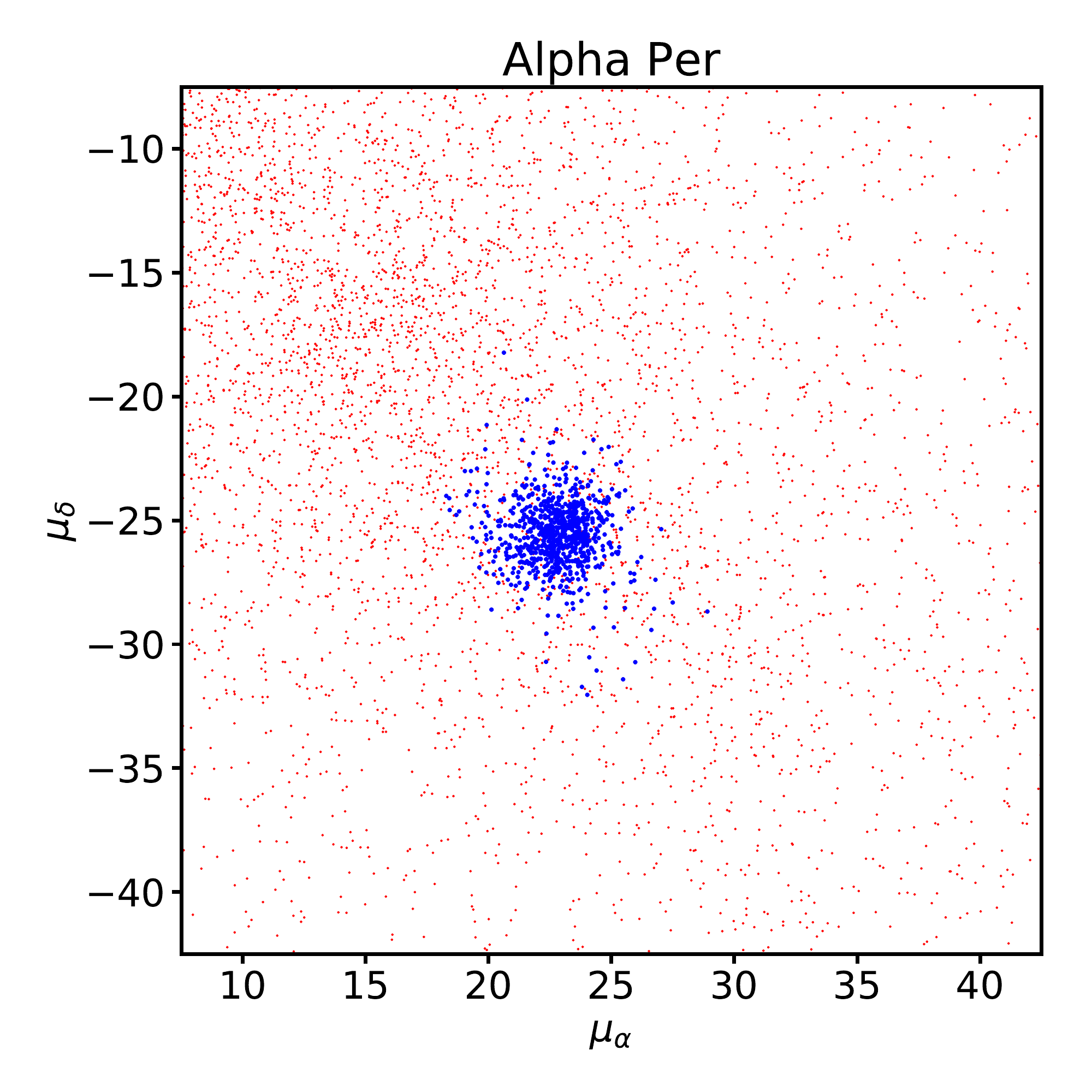}&
 \includegraphics[scale=0.25]{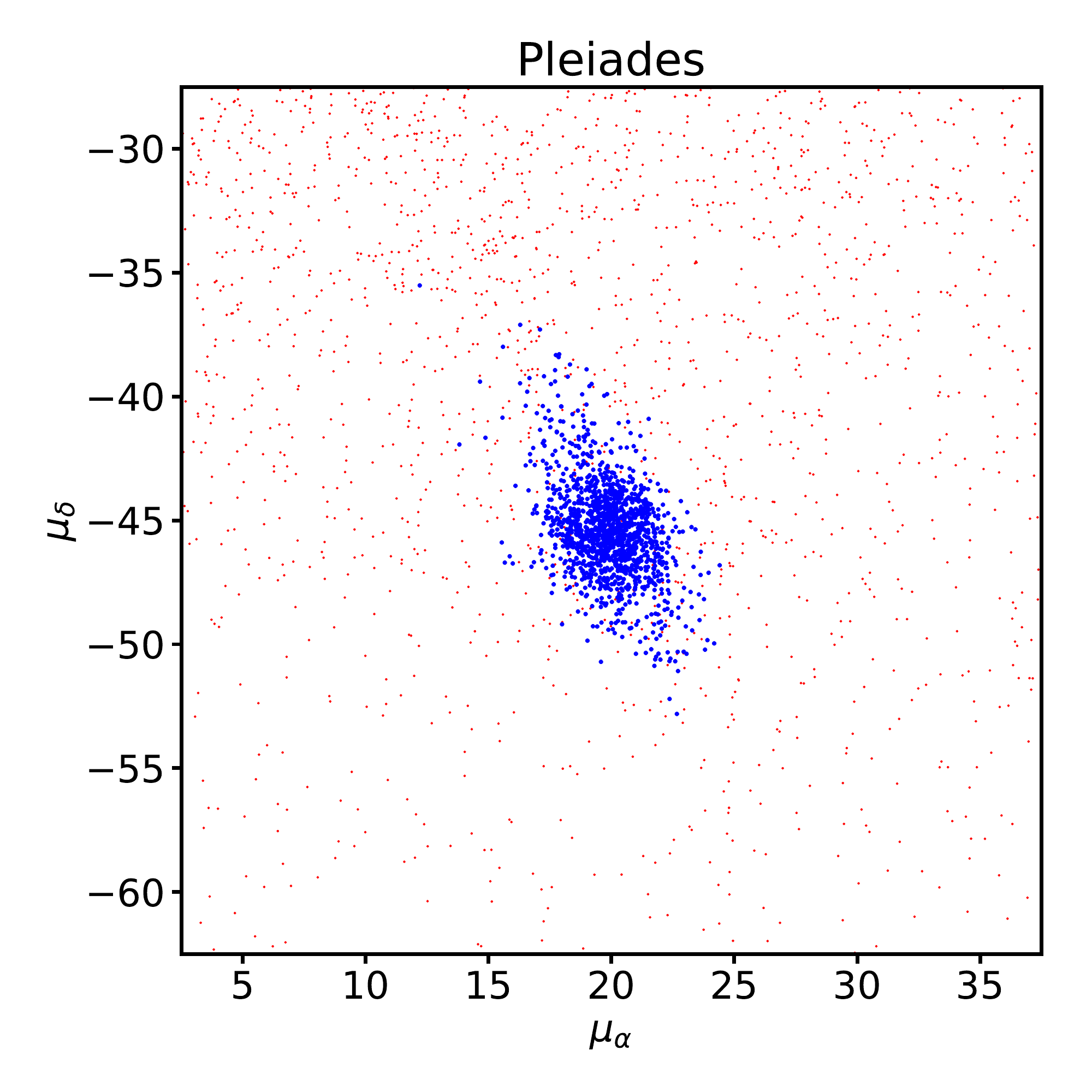}&
 \includegraphics[scale=0.25]{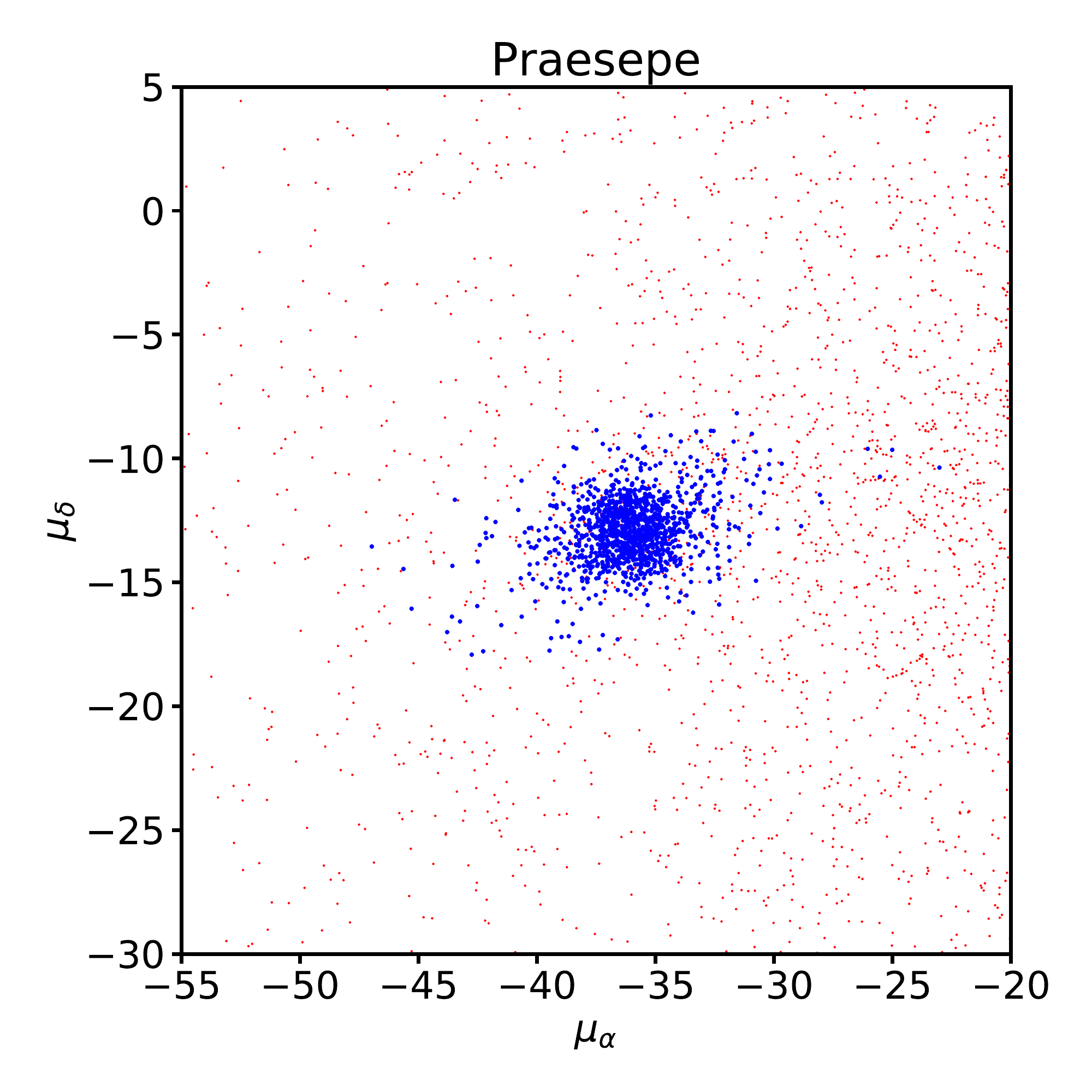}\\
 \includegraphics[scale=0.21]{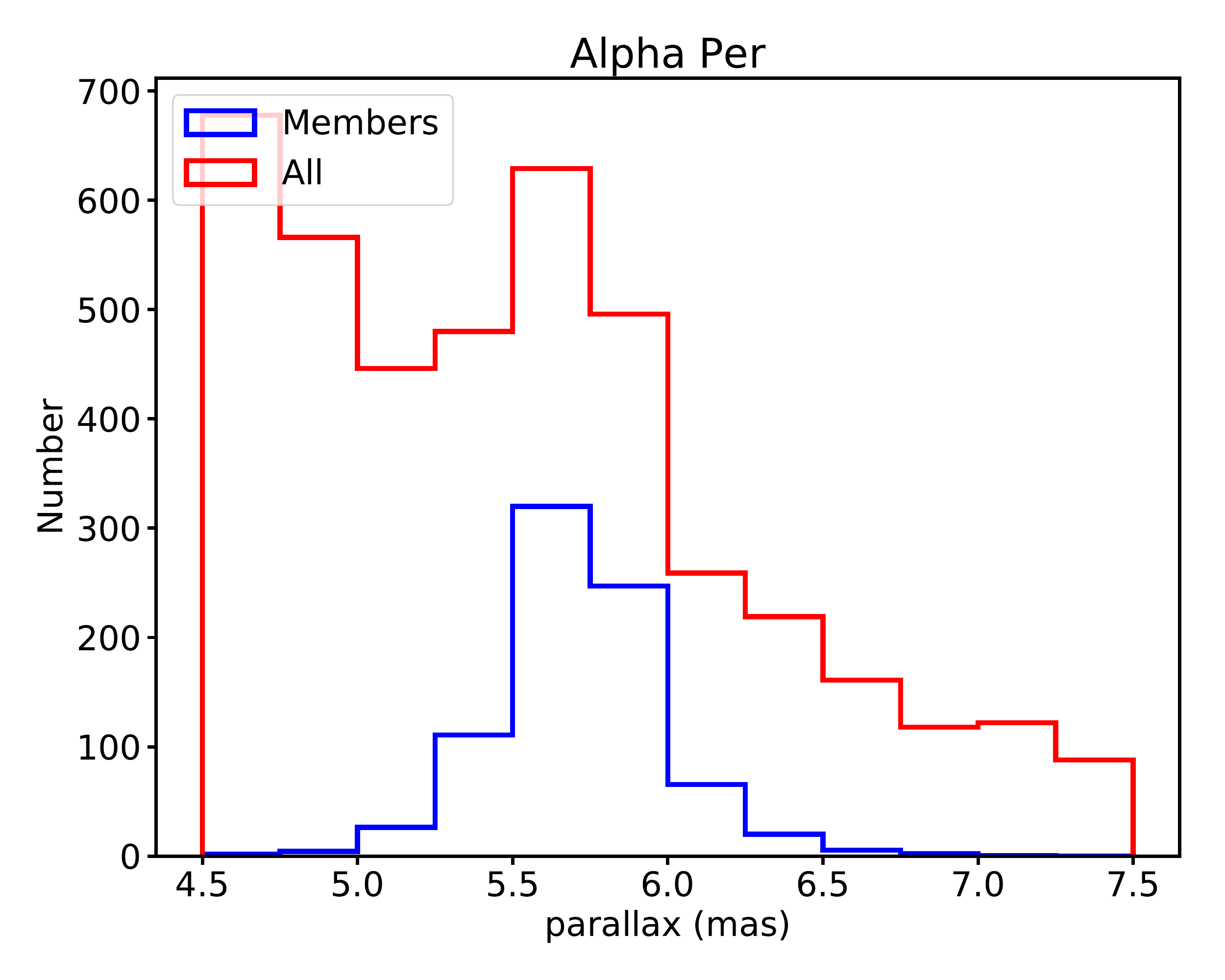}&
 \includegraphics[scale=0.21]{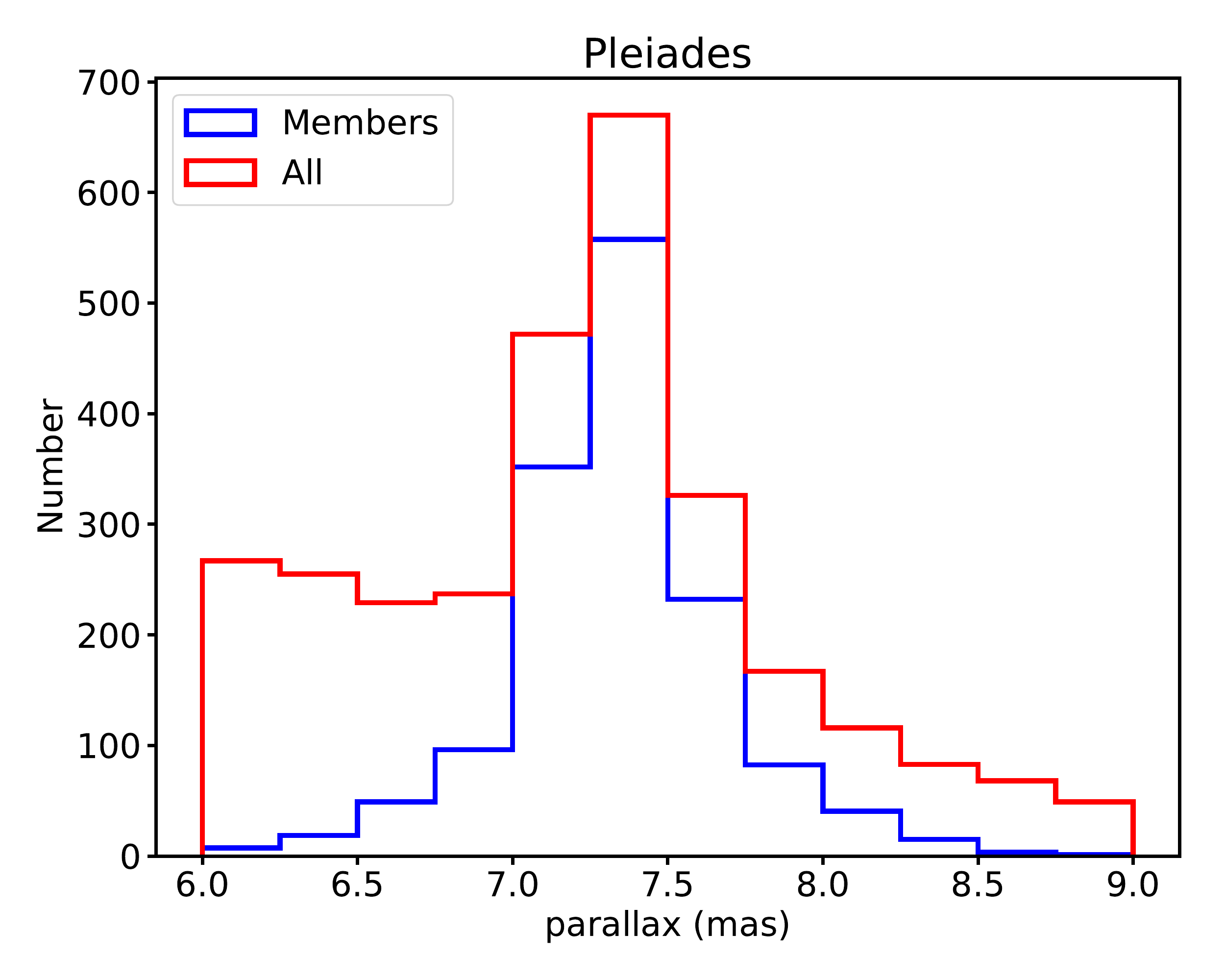}&
 \includegraphics[scale=0.21]{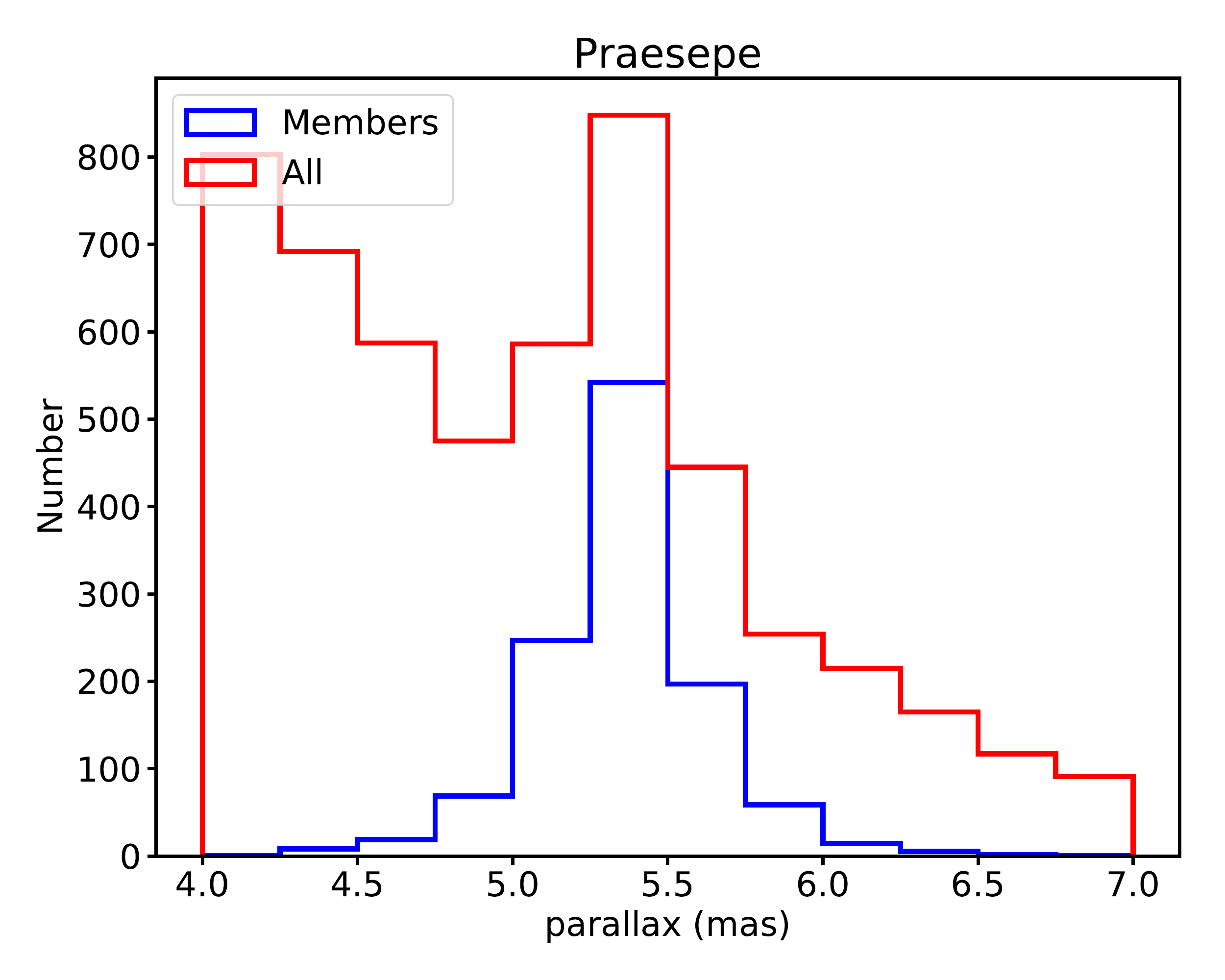}\\
 \includegraphics[scale=0.21]{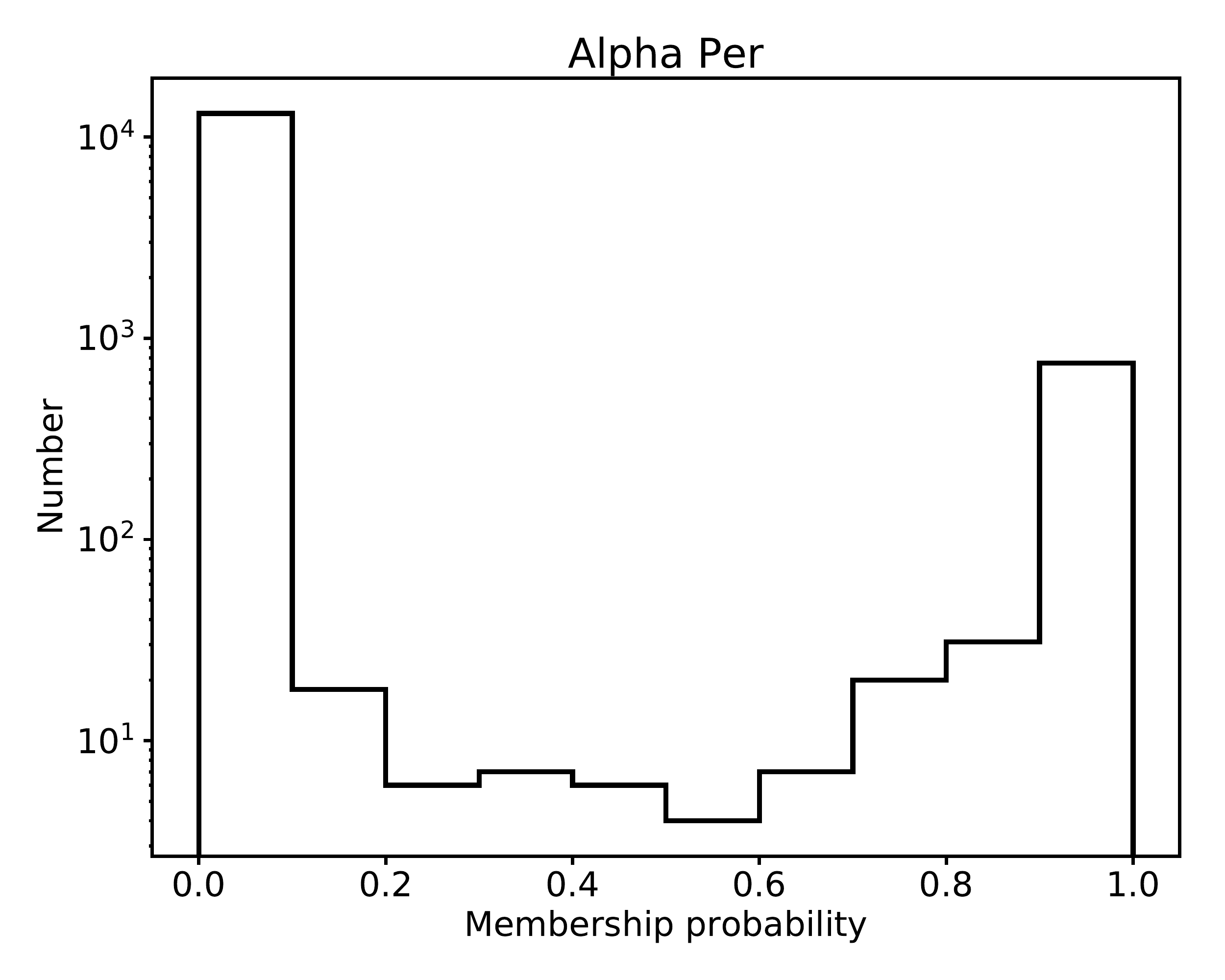}&
 \includegraphics[scale=0.21]{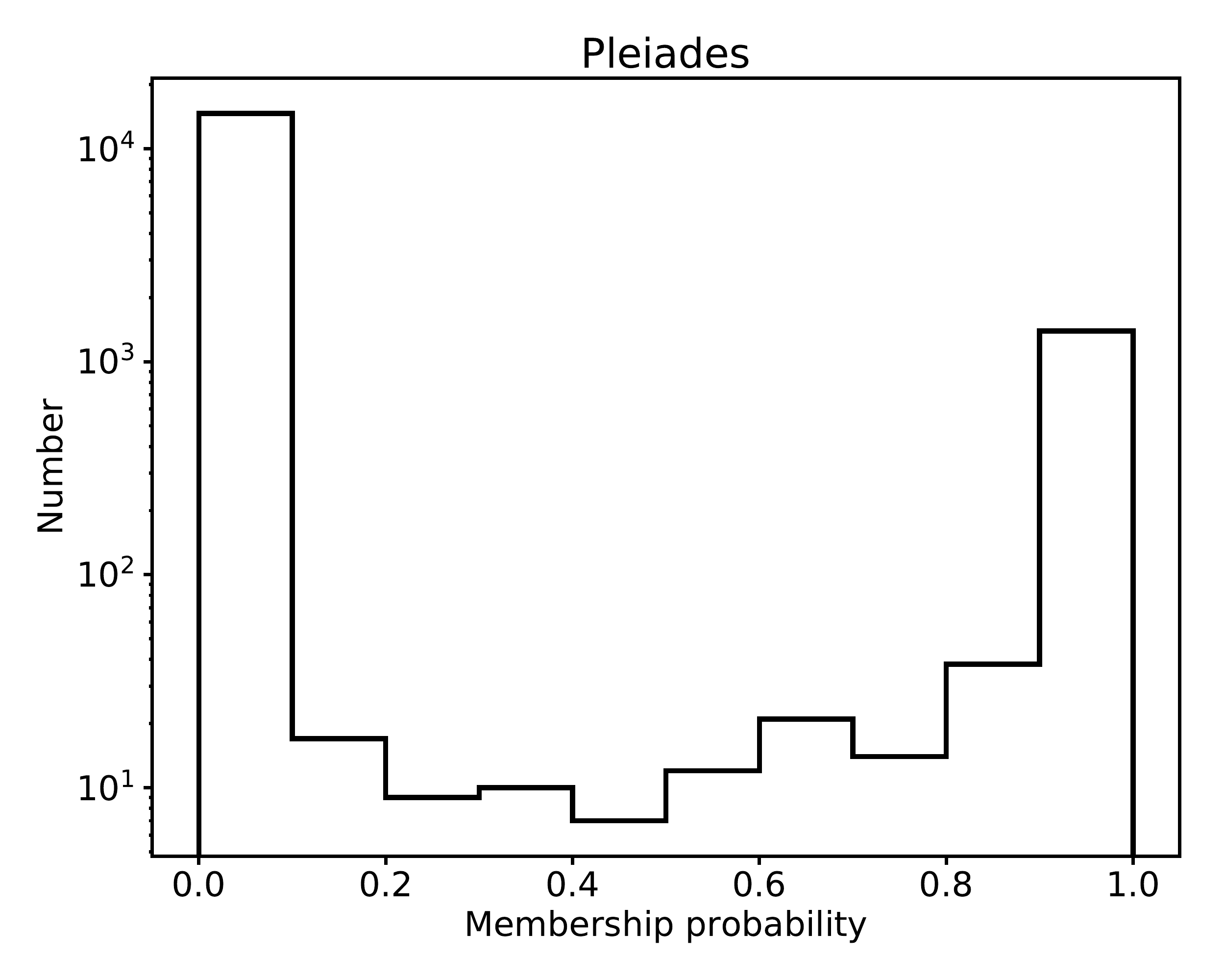}&
 \includegraphics[scale=0.21]{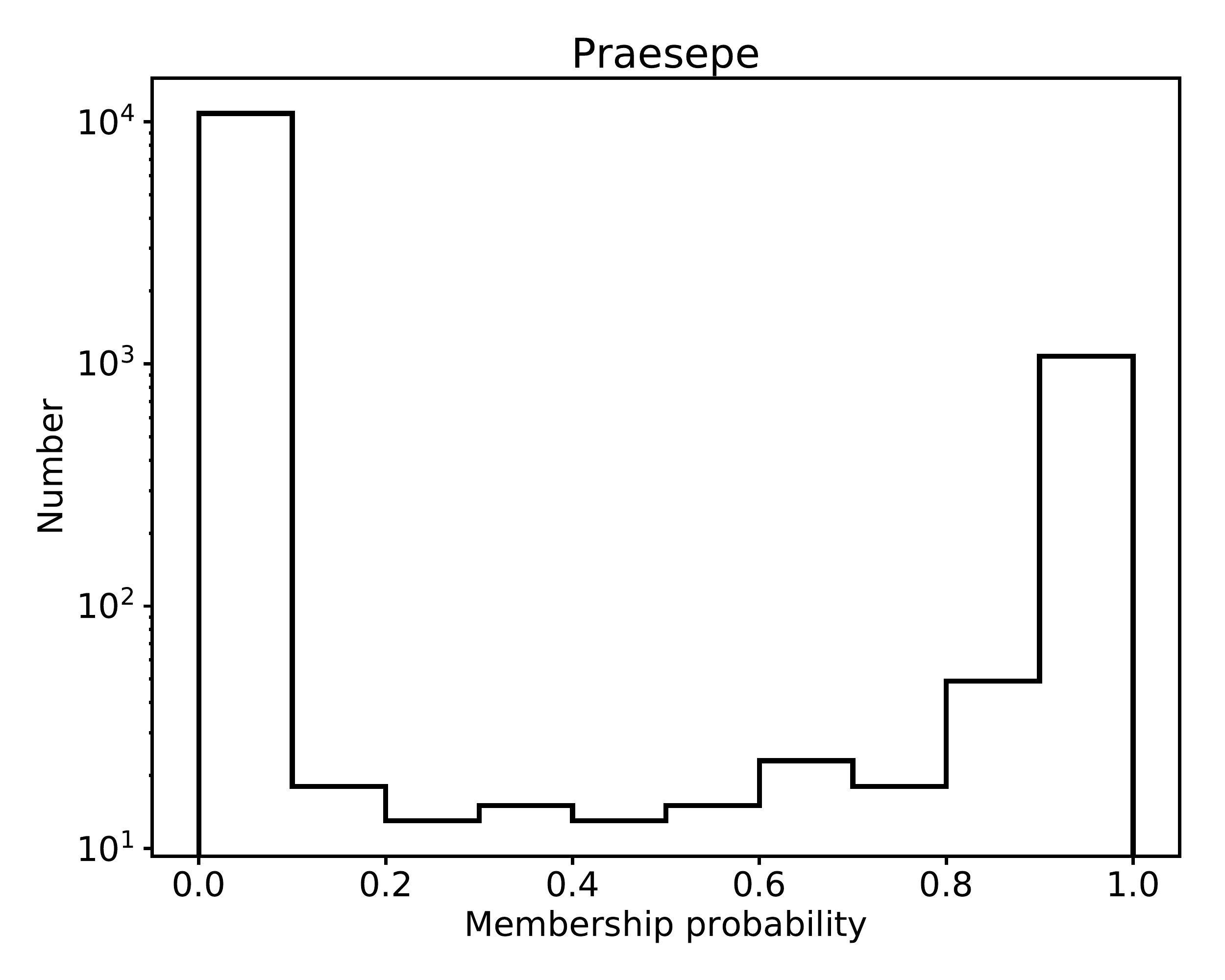}
 \end{tabular}
 \caption[]{Sky distribution, proper motion vector point diagrams, parallax histograms and membership probability histograms for the our three clusters. The blue points shown are objects with membership probabilities above 0.5.} 
  \label{vpds}
 \end{figure*}

We fitted likelihoods to each magnitude bin of each cluster, allowing us to calculate membership probabilities for each star in the area around our clusters. Our fitting code did not converge for the faintest bin of Alpha~Per, likely due to high background contamination caused by its low Galactic latitude. Thus our faint limit for Alpha~Per is $G=19$\, mag. We selected a star as a cluster member if it had a membership probability greater than 0.5. This left us with 815 Alpha~Per members, 1477 Pleiades members and 1181 Praesepe members. Figure~\ref{vpds} shows the sky distribution plots, proper motion vector point diagrams and parallax histograms for each cluster. Our method clearly selects well-defined cluster populations. Table~\ref{memb_tab} lists all objects for which we calculate membership probabilities in each cluster. 
\subsection{Completeness and contamination}
\label{memb_comp}
We study the completeness of our survey by using previous studies of our clusters to test how well we recover previously known objects. We expect to lose objects in a number of different ways. There will be objects which do not have a counterpart in Gaia. We will also lose objects where Gaia has only calculated a two parameter solution, making it impossible to calculate membership probabilities. There will also be objects which have proper motions and parallaxes that are so discrepant from the cluster values that they lie outside the normalisation limits for our likelihood. Additionally there are objects where the better-quality Gaia proper motions and the addition of parallax measurements reduce the object's membership probability so that it is below 0.5.

\cite{Arenou2018} {estimate that the Gaia survey completeness by comparing to the OGLE and HST data at different stellar densities. For the stellar densities found over most of they sky they recovered $>$99\% of OGLE stars to fainter than $G=20$\,mag}. This is likely true of our clusters as they do not lie in areas of extremely high stellar density. We tested our completeness by examining the recovery fraction between our work and studies using the UKIDSS Galactic Plane Survey (\citealt{Lodieu2012} in the Pleiades, \citealt{Lodieu2012a} in Alpha~Per and \citealt{Boudreault2012} in Praesepe). The Pleiades cluster has also been a target for the DANCe programme which uses a compilation of all previous observations with different surveys and telescopes to accurately measure the astrometric properties of candidate cluster members (see \citealt{Bouy2015} and \citealt{Olivares2018}). Finally we added the Praesepe membership survey of \cite{Kraus2007}. We note that none of these surveys will themselves be 100\% complete and that the recovery fractions listed here are only indicative of our true completeness.

Before beginning our completeness test we excluded any object in our comparison sample which fell outside the sky areas of our cluster search. We also excluded any object that would likely be faint enough to fall below our $G=20$\,mag limit. We therefore cut on $Z<18$\,mag (UKIDSS studies) and $J<16$\,mag (DANCe) for Praesepe and the Pleiades and cutting on $Z<17$\,mag for comparison with \cite{Lodieu2012a}'s studies of Alpha~Per.

Table~\ref{comp_tab} lists the number of objects we recover from each cluster along with the number of objects lost at each particular stage. We are losing 0.0--0.4\% of objects due to them lacking Gaia data and 1.1--3.9\% of objects due to them having only two parameter solutions in Gaia. We found that most of the objects we miss due to low (or lack of) membership probabilities have parallax measurements that are inconsistent with cluster membership with other objects excluded due to discrepant proper motions. However, we note the \cite{Olivares2018} candidate cluster members form a distribution on a proper motion vector point diagram that is elongated along the direction of cluster motion. This is to be expected as if the cluster members have the same space velocity, then the more distant members will have slightly lower proper motions and the closer members will have higher proper motions but all will move in roughly the same direction. Indeed our work finds a higher dispersion in proper motions in the direction of cluster motion ($\sigma_y$).

We also compared to the recent study of \cite{Lodieu2019a}. This study uses a more detailed 3D model for the cluster than our method. It also covers a much larger area and includes candidate members out to three times the tidal radii of each cluster. Our study by contrast is constrained to a smaller area around the cluster centre. Restricting ourselves to objects that fell within \cite{Lodieu2019a}'s tidal radius for each cluster and to objects bright enough to appear in our sample and which fall in our survey area, we found we recover 453/471  of \cite{Lodieu2019a}'s Alpha~Per members, 1195/1225 of their Pleiades members and 696/707 of their Praesepe members. This is a 97\% recovery rate across the three clusters. We also find that 814/815 of our Alpha~Per members, 1473/1477 of our Pleiades members and 1147/1181 of our Praesepe members appear in \cite{Lodieu2019a}'s member list. Hence, despite our differing membership selection techniques, our membership lists for the core of each of our three clusters are almost identical.

We estimate our contamination by applying our cluster fits to control fields at the same Galactic latitude as our clusters but offset by ten degrees in Galactic longitude. Each field was three degrees in radius and the stars had the same transformation and filtering steps applied to them before having their membership probabilities analysed. These control fields should contain no true cluster members. In the Alpha Per offset field we found 32 out of 4596 stars had $p_{bin}>0.5$, 18 in the Pleiades offset field out of 5074 and 11 out of 3810 in the Praesepe offset field. Adjusting for the higher number of stars in our cluster fields (13979, 16167 and 12071 for Alpha~Per, the Pleiades and Praesepe respectively) and we estimate that our cluster samples contain 97, 57 and 35 field interlopers for Alpha~Per, the Pleiades and Praesepe respectively.

\begin{table*}
\caption{\label{comp_tab} Details of comparisons with previous membership studies of each cluster. We list the total number of objects which were bright enough and fell in the right sky area for us to detect ($N_{objects}$) plus the number recovered ($N_{recovered}$), how many had low membership probabilities ($N_{p_{memb}<0.5}$), how many fell outside the proper motion and parallax bounds of our membership probability calculations ($N_{no p_{memb}}$), how many had only two parameter solutions in Gaia ($N_{2 par Gaia}$)and how many had no Gaia counterpart ($N_{no Gaia}$).}
\begin{tabular}{lcrrrrrrr}
\hline
Study&Cluster&$N_{objects}$&$N_{recovered}$&Recovery&$N_{p_{memb}<0.5}$&$N_{no p_{memb}}$&$N_{2 par Gaia}$&$N_{no Gaia}$\\
&&&&percentage\\
\hline
\cite{Lodieu2012a}&Alpha Per&726&474&65\%&148&78&23&3\\
\cite{Lodieu2019a}&Alpha Per&471&453&96\%&18&0&0&0\\

\cite{Lodieu2012}&Pleiades&1618&1346&83\%&156&71&45&0\\
\cite{Bouy2015}&Pleiades&1947&1393&72\%&436&62&52&4\\
\cite{Olivares2018}&Pleiades&2426&1356&56\%&896&86&81&4\\
\cite{Lodieu2019a}&Pleiades&1225&1195&98\%&29&1&0&0\\

\cite{Boudreault2012}&Praesepe&677&513&76\%&103&37&21&3\\
\cite{Kraus2007}&Praesepe&1134&914&81\%&161&18&41&0\\
\cite{Lodieu2019a}&Praesepe&707&681&96\%&26&0&0&0\\
\end{tabular}
\end{table*}

\subsection{Properties of cluster members}
\begin{figure*}
 \setlength{\unitlength}{1mm}
 \begin{tabular}{ccc}
 \includegraphics[scale=0.21]{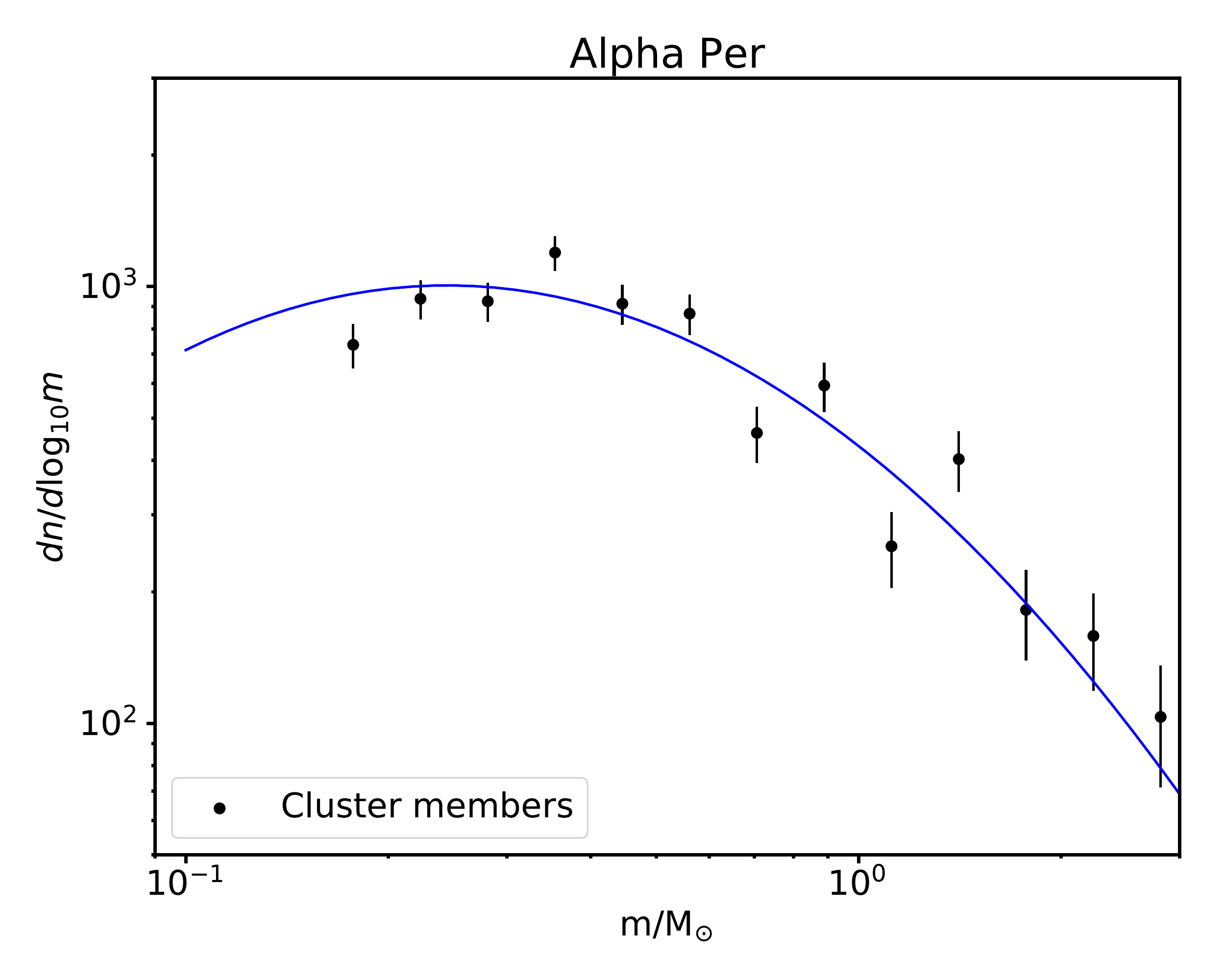}&
 \includegraphics[scale=0.21]{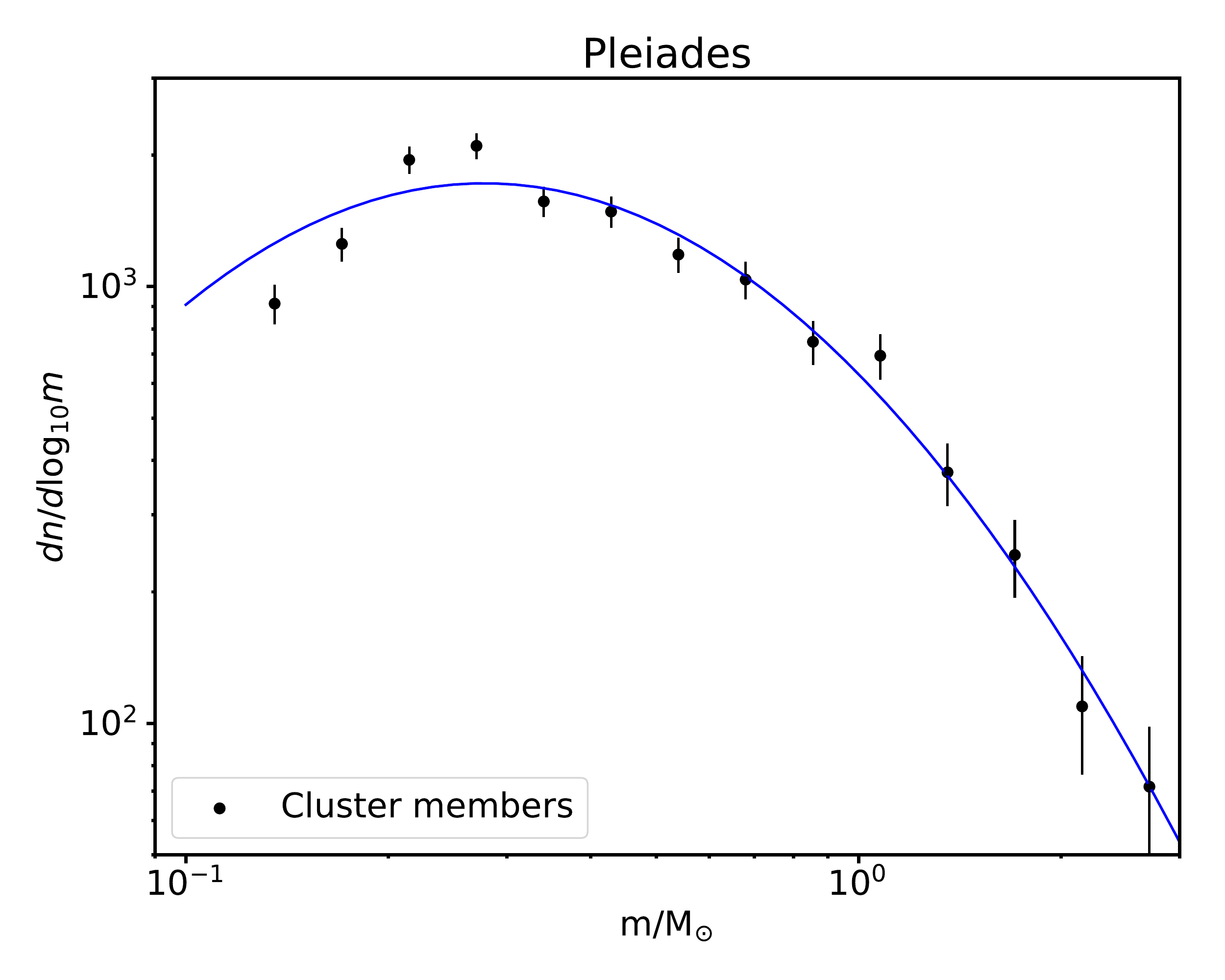}&
 \includegraphics[scale=0.21]{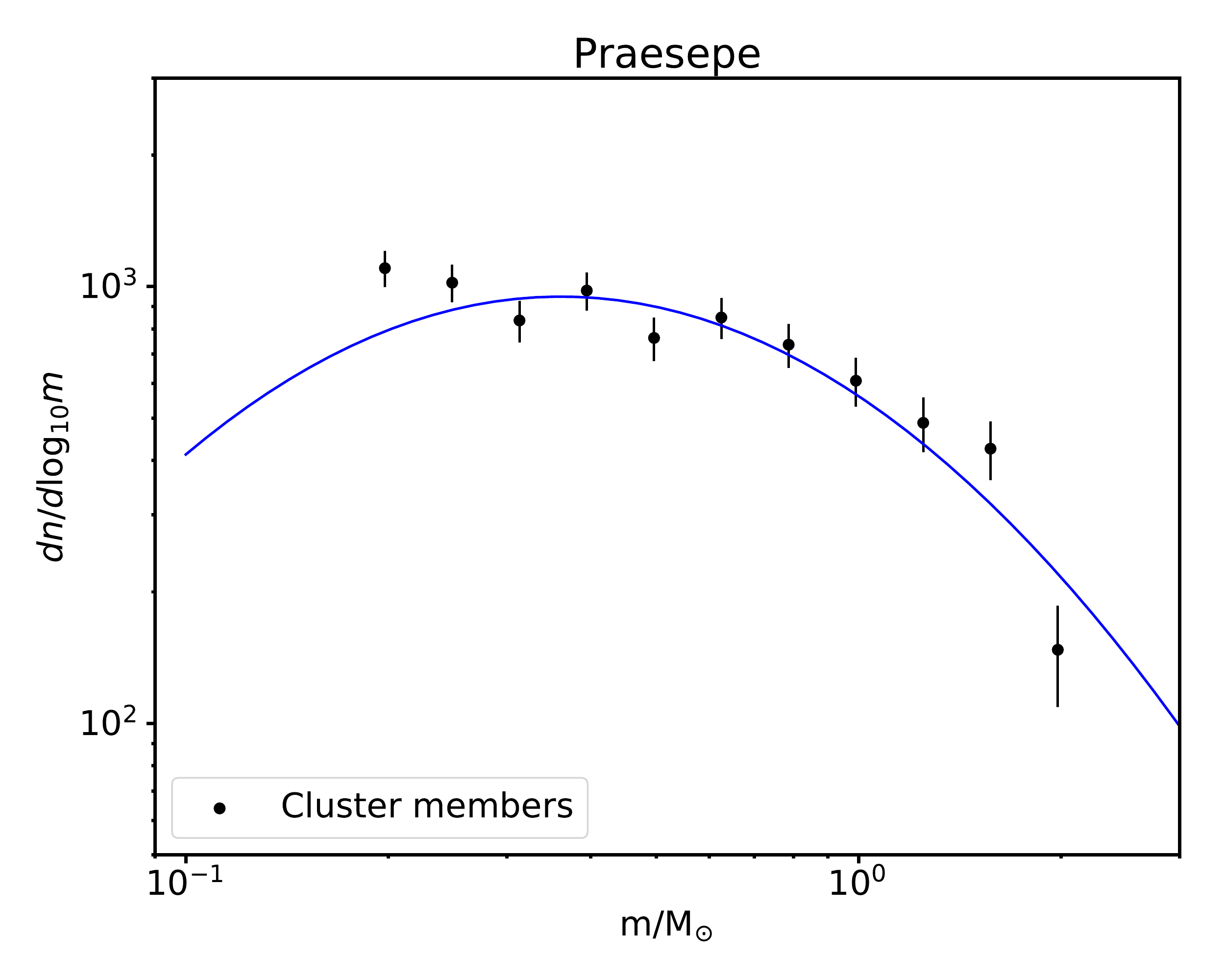}
 \end{tabular}
 \caption[]{System mass functions for the our three clusters. The blue lines show log-normal fits to each mass function. Points covered by red crosses were excluded from the system mass function fits.} 
  \label{mfs}
 \end{figure*}

We estimated the masses for each of our potential cluster members using absolute Gaia $G$ magnitudes calculated using each star's measured Gaia parallax. These absolute magnitudes were then converted into masses using isochrones from PARSEC stellar evolution models \citep{Marigo2017} with the appropriate age and metallicity for each cluster. For the Pleiades we used an age of 125\,Myr \citep{Stauffer1998} and a metallicity of $[Fe/H]=-0.01$ \citep{Netopil2016}; for Praesepe 790\,Myr \citep{Brandt2015} and $[Fe/H]=0.16$ \citep{Netopil2016}; and for Alpha Per 85\,Myr \citep{Navascues2004} and $[Fe/H]=0.14$ \citep{Netopil2016}. To produce a system mass function we then summed the membership probabilities of all stars in a particular mass bin weighting by the inverse of our incompleteness estimates from the previous section. We then used the {\scshape scipy curvefit} package to fit a log-normal function of the form 
\begin{equation}
\label{mf_form}
\xi(m)=\frac{dn}{d\log_{10}m}\propto e^{-\frac{(\log_{10}m-log_{10}m_c)^2}{2\sigma^2}}
\end{equation}
to each system mass function. We started our system mass function at the mass which was equivalent to $G=19$\,mag for the Pleiades and Praesepe and $G=18$\,mag for Alpha~Per. This was because we found that the Gaia data became incomplete at fainter magnitudes. We found parameter values of $\log_{10}m_c=-0.611\pm0.125$ ($m_c=0.24\,M_{\odot}$) \& $\sigma=0.470\pm0.065$ for Alpha~Per, $\log_{10}m_c=-0.556\pm0.038$ ($m_c=0.28\,M_{\odot}$) \& $\sigma=0.392\pm0.025$ for the Pleiades and $\log_{10}m_c=-0.442\pm0.083$ ($m_c=0.36\,M_{\odot}$) \& $\sigma=0.432\pm0.063$ for Praesepe. These values are broadly similar to the previously derived values of $\log_{10}m_c=-0.46\pm0.05$ \& $\sigma=0.45\pm0.02$ for the Alpha~Per \citep{Lodieu2012a}, $\log_{10}m_c=-0.62\pm0.02$ \& $\sigma=0.44\pm0.01$ for the Pleiades \citep{Lodieu2012} and  $\log_{10}m_c=-0.60$ \& $\sigma=0.55$ for the field system mass function of \cite{Chabrier2005}. Figure~\ref{mfs} shows our mass function for each cluster and Table~\ref{mf_tab} gives the values for the individual system mass function bins for each cluster.
 
\begin{table*}
\caption{\label{mf_tab} The individual bin values for our calculated system mass functions for each cluster.}
\begin{center}
\begin{tabular}{rrrrrr}
\hline
\multicolumn{2}{c}{Alpha Per}&\multicolumn{2}{c}{Pleiades}&\multicolumn{2}{c}{Praesepe}\\
mass range&$\frac{dn}{d\log_{10}m}$&mass range&$\frac{dn}{d\log_{10}m}$&mass range&$\frac{dn}{d\log_{10}m}$\\
(M$_{\odot}$)&&(M$_{\odot}$)&&(M$_{\odot}$)\\
 \hline
\multicolumn{2}{c}{\ldots}&0.120--0.151&914&\multicolumn{2}{c}{\ldots}\\
0.157--0.198&735&0.151--0.190&1252&\multicolumn{2}{c}{\ldots}\\
0.198--0.249&937&0.190--0.239&1949&0.175--0.220&1101\\
0.249--0.313&925&0.239--0.301&2098&0.220--0.277&1020\\
0.313--0.394&1196&0.301--0.379&1565&0.277--0.349&836\\
0.394--0.496&913&0.379--0.478&1484&0.349--0.440&979\\
0.496--0.625&867&0.478--0.601&1183&0.440--0.553&762\\
0.625--0.787&462&0.601--0.757&1036&0.553--0.697&849\\
0.787--0.991&593&0.757--0.953&747&0.697--0.877&735\\
0.991--1.247&254&0.953--1.200&694&0.877--1.104&608\\
1.247--1.570&403&1.200--1.511&376&1.104--1.390&487\\
1.570--1.977&182&1.511--1.902&243&1.390--1.750&425\\
1.977--2.488&159&1.902--2.394&109&1.750--2.050&147\\
2.488--3.133&104&2.394--3.014&72&\multicolumn{2}{c}{\ldots}\\
3.133--3.944&30&3.014--3.794&48&\multicolumn{2}{c}{\ldots}\\
3.944--4.965&58&\multicolumn{2}{c}{\ldots}&\multicolumn{2}{c}{\ldots}\\

\end{tabular}
\end{center}

\end{table*}
\begin{table*}
\caption{\label{memb_tab}Membership probabilities for objects with $p_{memb}>0.5$ for each cluster. The positions, proper motions and parallaxes are taken from Gaia DR2.}
\begin{center}
\begin{tabular}{|ccrrrrrrr}
\hline
Gaia source ID&R.A.&Dec.&$\mu_{\alpha}\cos{\delta}$&$\mu_{\delta}$&$\varpi$&$G$&mass&$p_{memb}$\\
&\multicolumn{2}{c}{(J2000 Ep=2015.5)}&(mas/yr)&(mas/yr)&(mas)&(mag)&(M$_{\odot}$)\\
\hline
\multicolumn{9}{c}{Alpha Per}\\
\hline
439440029866690944&02 57 05.23&+49 39 29.3&25.7$\pm$0.3&-24.6$\pm$0.2&5.6$\pm$0.1&17.3&0.27&0.97\\
439191437159059968&02 57 54.15&+48 52 43.1&24.9$\pm$0.1&-22.0$\pm$0.1&5.5$\pm$0.1&8.7&1.73&0.90\\
434488516689743232&02 59 24.83&+47 07 49.8&23.9$\pm$0.3&-22.3$\pm$0.2&5.7$\pm$0.1&17.2&0.28&0.89\\
437601096669764736&02 59 35.66&+48 12 19.0&24.6$\pm$0.2&-22.1$\pm$0.2&5.5$\pm$0.1&16.4&0.4&0.70\\
439161986569729280&03 00 50.16&+49 01 59.4&25.2$\pm$0.6&-22.7$\pm$0.5&5.4$\pm$0.3&18.7&0.17&0.54\\
\hline
\end{tabular}
\end{center}

\end{table*}%

\section{Binarity}
\subsection{Flagging potential unresolved pairs}
\label{un_res}
The components of wide binary systems can themselves be close binaries. Such higher order multiples are common and the components of wide binaries might even be more likely to be close pairs than isolated field stars \citep{Allen2012,Law2010}. The dynamical evolution of higher order multiplies has been suggested as a formation mechanism for wide binaries \citep{Reipurth2012}.
We identify possible higher-order multiples via two features: excess astrometric noise in Gaia, and overluminosity. 
\subsubsection{Stars with noisy astrometric solutions}
\label{astro_noise}
Binary companions introduce additional astrometric noise via several effects. When unresolved they can add shifts in the binary photocentre due to astrometric motion or due to photometric variability of one or both components. When resolved they can add additional astrometric datapoints that can confuse astrometric solution calculations. Finally, when they are partially resolved, PSF mismatch with the single-star PSF model results in higher scatter in the astrometric measurements. Gaia DR2 encapsulates the excess astrometric noise in the form of the Renormalised Weighted Error (RUWE; \citealt{Lindegren2018,Lindegren2018a}) which is similar to the the square root of the reduced $\chi^2$ statistic. RUWE measures the excess astrometric noise, accounting for the increase in excess noise in fainter objects and objects with extreme colours. To test how this excess noise statistic is elevated by close binarity we followed a similar approach to \cite{Rizzuto2018}. We took close, directly imaged binaries (with separations within 1 arcsecond) from \cite{Kraus2016} and binaries found to be unresolved in Gaia from \cite{Ziegler2018} and \cite{Lamman2020}. This sample in this latter work is a compilation of directly imaged binaries from \cite{Law2014a}, \cite{Baranec2016} and \cite{Ziegler2017} and counts binaries as resolved in Gaia if the secondary has a Gaia DR2 entry of its own.
We follow \cite{Rizzuto2018} in using the ROBO-AO LP600 magnitude ratio quoted in \cite{Ziegler2018} as the Gaia magnitude difference (as the LP600 filter is similar to the Gaia $G$-band filter). We then use a PARSEC \citep{Marigo2017} 2\,Gyr isochrone to convert the \cite{Kraus2016} $K$-band magnitude difference to Gaia $G$-band magnitude difference.

\begin{figure*}
 \begin{center}
 \setlength{\unitlength}{1mm}
 \begin{tabular}{cc}
 \includegraphics[scale=0.5]{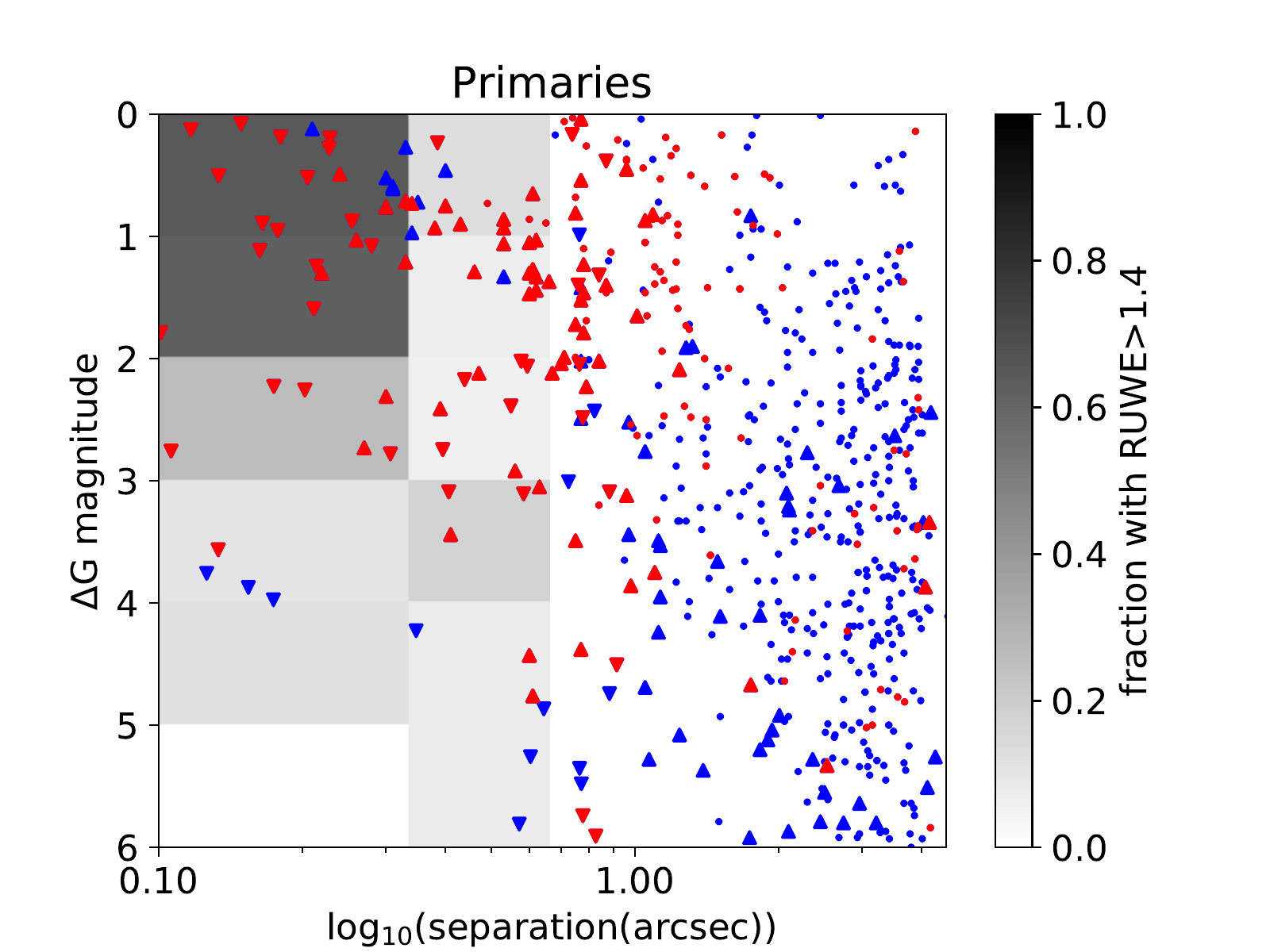}&
  \includegraphics[scale=0.5]{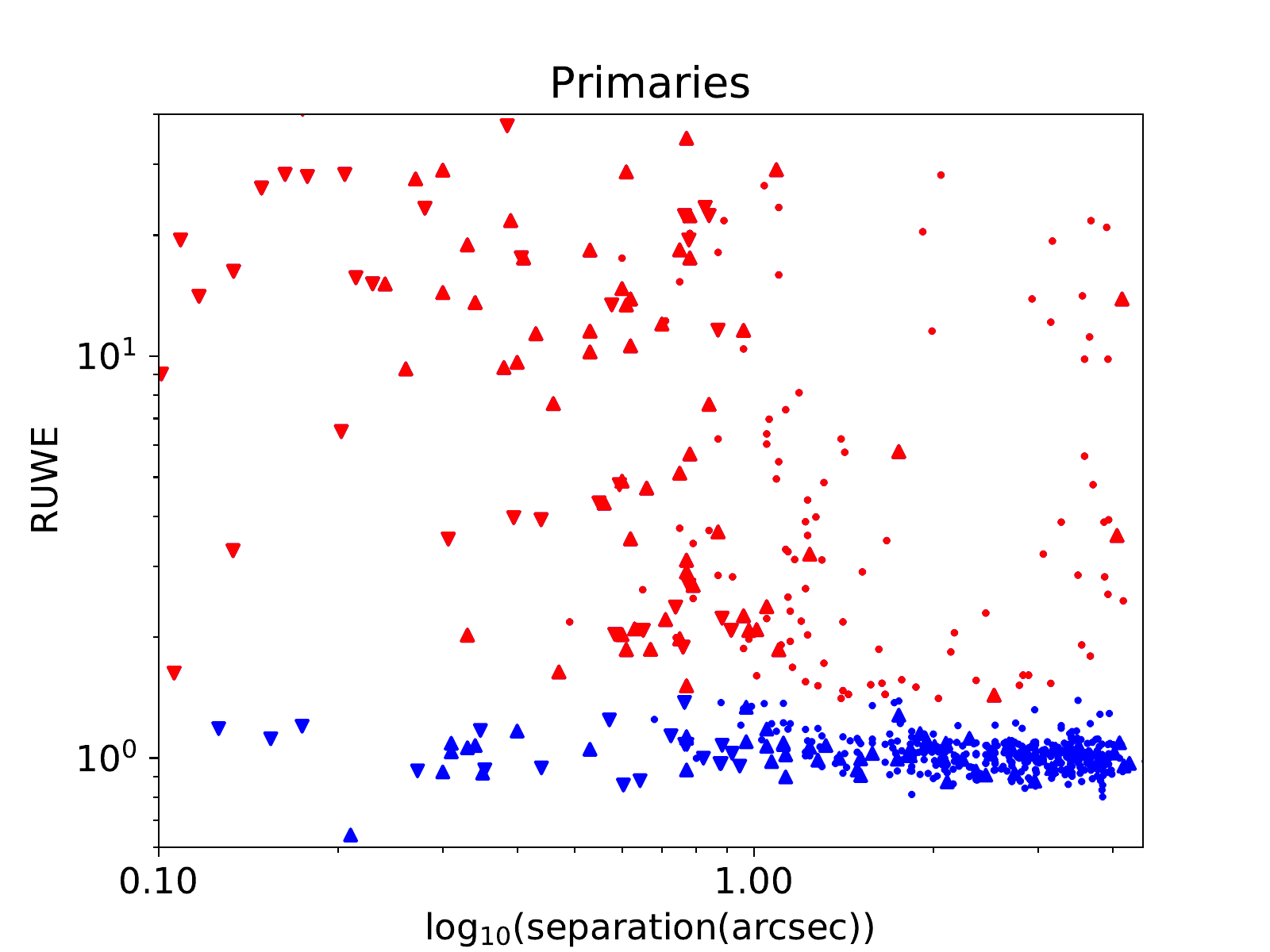}\\
 \includegraphics[scale=0.5]{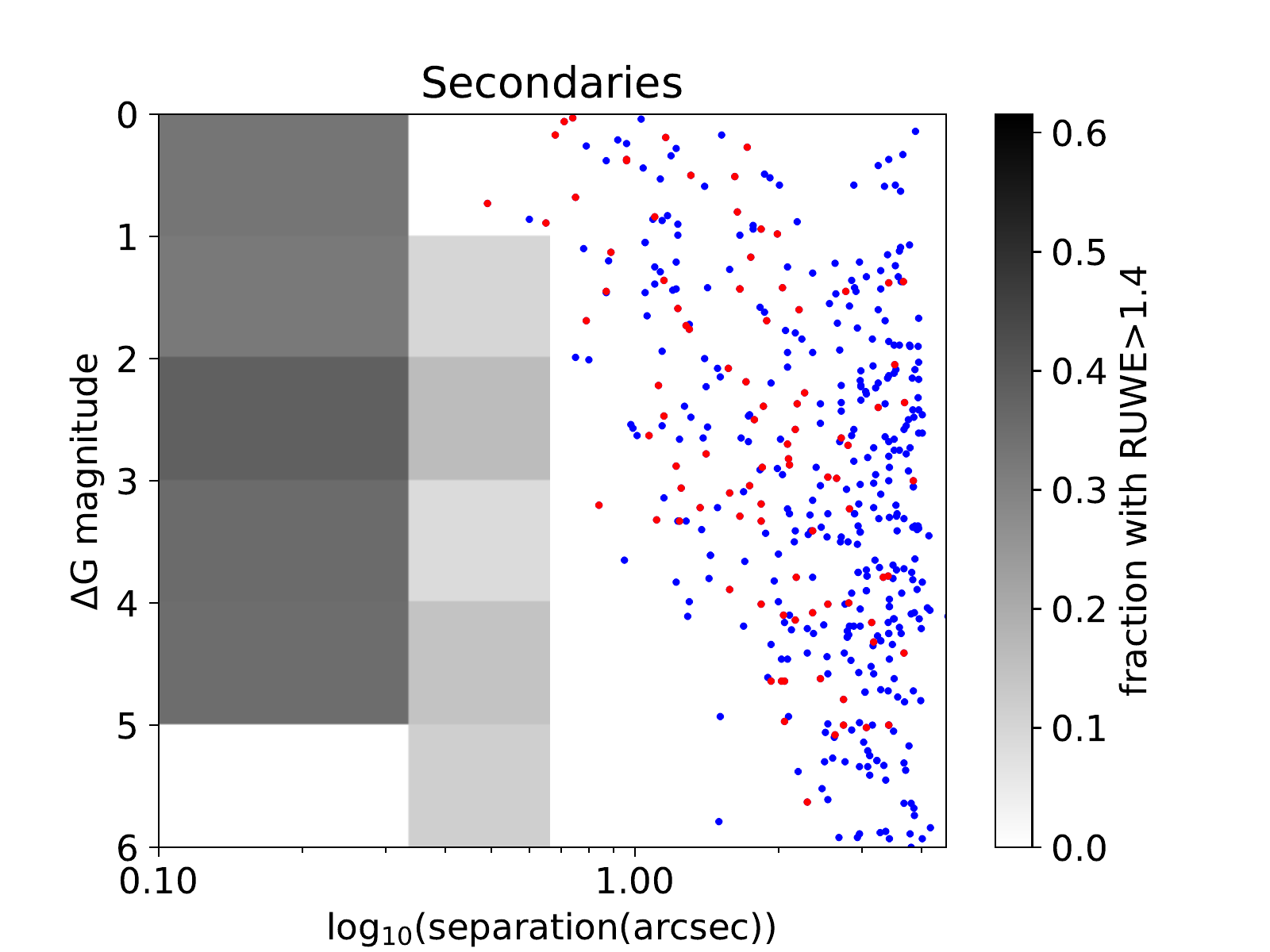}&
  \includegraphics[scale=0.5]{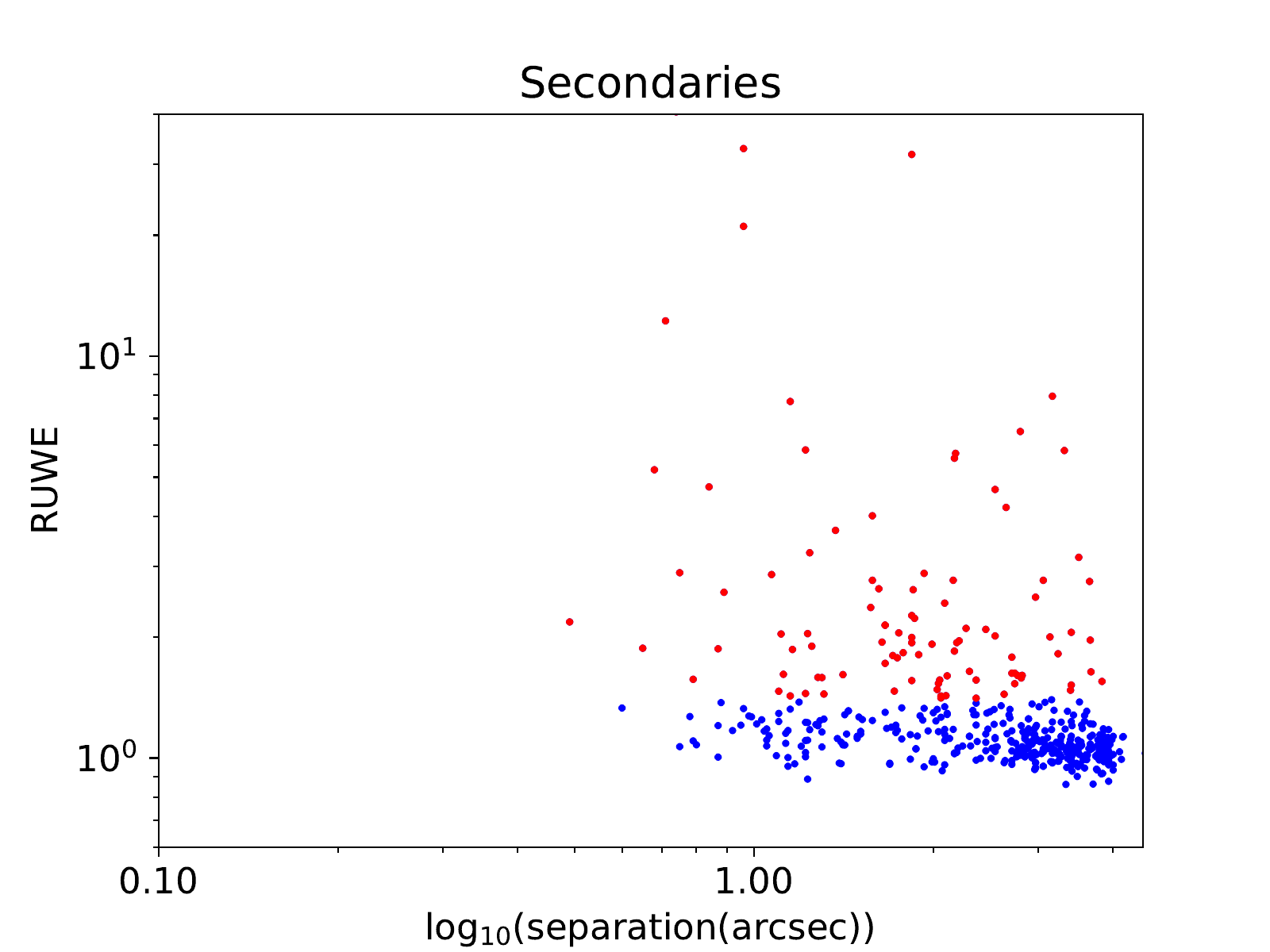}\\
 \end{tabular}
 \caption[]{A plot showing how the Renormalised Unit Weighted Error \protect\citep{Lindegren2018a} statistic in Gaia DR2 that measures excess astrometric noise as a function of binary separation and magnitude difference. The upward-pointing triangles are binaries from \protect\cite{Kraus2016}, the downward-pointing triangles are binaries from \protect\cite{Ziegler2018} which are not resolved in Gaia and the small circles are binaries from \protect\cite{Ziegler2018} resolved in Gaia. Red points are components or unresolved systems with RUWE$>$1.4.}  
  \label{ruwe_plot}
  \end{center}
 \end{figure*}

Figure~\ref{ruwe_plot} shows that binaries closer than one arcsecond with magnitude differences less than four magnitudes have elevated RUWE values. We chose 1.4 as the limit for elevated RUWE as it was used as the limit in \cite{Lindegren2018a}. Even at projected separations of $\rho \sim 0.1$\arcsec, we find that RUWE is elevated for binaries with magnitude differences less than three magnitudes. This complements the high recovery fraction in Gaia for binaries wider than one arcsecond  \citep{Ziegler2018}. We flag objects with RUWE$>$1.4 and with no detected binary companion within one arcsecond as possible unresolved binaries or unresolved components of wide binaries.

\subsubsection{Overluminous stars}
\label{overlum}
To identify stars which may be overluminous due to binarity we used a simple isochrone fitting process. We use the previously mentioned PARSEC stellar evolution models \citep{Marigo2017} to determine a star's absolute $G$ magnitude based on its Gaia $BP-RP$ colour. However we found that the isochrones did not match the data well, especially for lower-mass cluster members. To remedy this we estimated the median offset in absolute $G$-band magnitude between estimates from $BP-RP$ colours \& the PARSEC models and the values derived from Gaia parallaxes and apparent $G$ magnitude. We restricted ourselves to the brighter stars in each cluster where the isochrone still matched the observed colours and magnitudes reasonably well ($1<G_{abs}<12$\,mag. for Alpha~Per and the Pleiades, $1<G_{abs}<9.5$\,mag. for Praesepe) but with small offsets. We found that using colours plus the PARSEC model made stars in Alpha~Per and the Pleiades too faint (by 0.119 and 0.061 magnitudes respectively), while stars in Praesepe were marginally too bright (0.038\,mag.). These offsets were then used to correct our estimated magnitudes based on $BP-RP$ colour. We then estimated the approximate scatter of the main sequence by calculating the median absolute deviation between the calculated Gaia $G$-band absolute magnitudes and those estimated from colours (after removing the previously mentioned offset) for objects that lay below the cluster isochrone. This approach allows us to estimate the scatter on the cluster star's CMD distribution without being affected by overluminous binaries. We used 1.48 times these median absolute deviations as a robust estimate of the one sigma scatter \citep{Maronna2006}. These values were 0.151\,mag. for Alpha~Per, 0.162\,mag. for the Pleiades and 0.092\,mag. for Praesepe. Any star that had a calculated Gaia $G$-band absolute magnitude which was brighter than the estimated absolute $G$-band from colour by more than three times the scatter for the appropriate cluster was flagged as a possible overluminous binary. These are shown in Figure~\ref{unresolved_cmd}. When calculating the offsets and scatters we limited ourselves to $1<G_{abs}<9.5$\,mag. in all clusters. However we found that using these calculated values we were able to reliably flag overluminous objects in Alpha~Per and the Pleiades with $1<G_{abs}<12$\,mag. while in Praesepe we were still limited to $1<G_{abs}<9.5$\,mag. It is clear that we select stars that appear overluminous but that we miss a small number of clearly overluminous stars, especially in the $G_{abs}$=5-8\,mag. range. This is likely due to to the fact that the offset between the absolute magnitudes estimated from the $BP-RP$ colours and the absolute magnitudes calculated from observed magnitude and parallax varies over the CMD. 
\subsection{Identifying comoving pairs}
We searched the Gaia database for comoving  companions to objects in our sample with separations up to 300 arcseconds. In works such as \cite{Deacon2014}, wide binaries are identified by their common proper motion. Studying wide binaries in clusters using Gaia changes this strategy in two ways. Firstly all members of an open cluster are moving through space together. This means that there is a large population of other cluster members which are not binary companions to any particular star in that cluster but which will have common proper motion when paired with it. Secondly Gaia's proper motions are so accurate that orbital motion outside the measurement errors of the proper motions must be taken into account. A $\sim$1000AU companion to a solar mass star will likely have an orbital velocity of $\sim$1\,km/s. At the distance of our clusters this equates to a tangential motion of $\sim$1\,mas/yr, significantly larger than the proper motion measurement errors. We perform a conservative cut on our data by adapting the astrometric difference defined by \cite{Deacon2017a}. This is the quadrature sum of the number of standard deviations that each pair of measurements for the proposed binary differ by. Including our orbital velocity term it is defined so that,
 \begin{equation}
 n_{\sigma}^2 = \frac{(\mu_{\alpha 1} - \mu_{\alpha 2})^2}{(\sigma_{\mu \alpha 1}^2 + \sigma_{\mu \alpha 2}^2 +\mu_{orb,\alpha}^2)} + \frac{(\mu_{\delta 1} - \mu_{\delta 2})^2}{(\sigma_{\mu \delta 1}^2 + \sigma_{\mu \delta 2}^2+\mu_{orb,\delta}^2))} + \frac{(\pi_{1} - \pi_{2})^2}{(\sigma_{\pi 1}^2 + \sigma_{\pi 2}^2)}
 \label{dast_eq}
 \end{equation}
Such that $\mu_{orb,\alpha}$ and $\mu_{orb,\delta}$ are set to be the one dimensional mean projections of the orbital velocity difference of each pair of stars separated by 1000\,AU and situated at the mean distance of each cluster. We ignored covariance terms between proper motion and parallax.
  \begin{equation*}
  \begin{aligned}
  \mu_{orb,\alpha}=&\mu_{orb,\delta}&=29.8 km\,s^{-1}\times \left(\frac{1000.0}{4.74 d_{cluster}(pc)}\right)\\
  &&\left(\frac{1}{\sqrt{3}}\right)\left(\sqrt{\frac{(M_1+ M_2)}{\rho(AU)}}\right)
   \end{aligned}
   \end{equation*}
 Where these proper motions are expressed in milliarcseconds per year. The first factor in the above equation converts from m/s to mas/yr at the distance of the cluster, the second factor accounts for the motion being split between three dimensions and the third factor is the total combined orbital velocity of the two stars. For both components of any potential binary, we use masses estimated from cluster isochrones. We select only pairs with $n_{\sigma}<5$.
\subsection{Defining the wide binary population}
Our sample of pairs will consist of a mix of true wide binaries and coincident pairings between unrelated cluster members, ignoring the presence of non-members in our sample because they are a negligible fraction of the total (see Section~\ref{memb_comp}). We considered two populations of objects, field stars and cluster members. We then fitted likelihood distributions for both the cluster and field populations as a function of proper motion and parallax. The probability of cluster membership was then calculated for each star using these likelihood distributions.  We take a similar approach where our pairs could be drawn from two different populations, true binaries and coincident pairings between cluster members. As we make use of the cluster density profile our logic is similar to that set out in \cite{Hambly1995}. The full outline of the mathematical model for our likelihood is given in Appendix~\ref{likemath}. We use a likelihood that takes the form,
\begin{equation}
\begin{aligned}
\phi=&f_{comp}\phi_{comp}+(1-f_{comp})\phi_{coincident}\\
\phi_{comp}=&\frac{e^{-r/r_0}}{2\pi r_0^2(1-e^{(-r_{max}/r_0)}(1+r_{max}/r_0))}\frac{1}{x\log(\frac{x_{max}}{x_{min}})}\\
\phi_{coincident}=&\frac{e^{-2r/r_0}c_{coincident,m_2}\xi(m_2)}{2\pi (2r_0)^2(1-e^{(-2r/r_0)}(1+2r_{max}/r_0))}\\
&\frac{2x}{(x_{max}^2-x_{min}^2)}
\normalsize
\end{aligned}
\end{equation}
Where $r$ is the distance of the binary from the cluster centre, $m_1$ and $m_2$ are the masses of the primary and secondary component and $x$ is the separation of the binary on the sky in arcseconds. For the system mass function $\xi(m_2)$ of each cluster we use the log-normal system mass functions we previously fitted for each cluster and use the perviously estimated masses for each star.

For the characteristic radii of the clusters $r_0$ we derive values using the distribution of stars in our membership selection and cluster centres taken from the Simbad database. We fitted exponential distribution to histograms of $r$ the distance from the cluster centre for each of our three clusters. We derived values of $r_0$ of 53.03 arcminutes for the Pleiades, 43.27 arcminutes for Praesepe and 62.77 arcminutes for Alpha Per.

For each cluster we maximised the likelihood as a function of $f_{comp}$. We did this by differentiating the log-likelihood such that,
\begin{equation}
\sum_i \frac{\partial \log\phi_i}{\partial f_{comp}}=0
\end{equation}
Summing over all $i$ pairs in our sample. This gives us,
\begin{equation}
\sum_i \frac{1}{\phi_i}(\phi_{comp,i}-\phi_{coincident,i})=0
\end{equation}
We solved the above equation using a simple bisection algorithm yielding values of $f_{comp}$ of 0.076 for the Pleiades, 0.036 for Praesepe and 0.075 for Alpha Per. Note that $f_{comp}$ the fraction of pairs that we believe to be real binaries and is not the same as $f_{bin}$ the actual binary fraction.

We were then able to estimate the probability that any particular pair is a binary using the equation,

\begin{equation}
p_i=\frac{\phi_{comp}(r_i,x_i,m_{1,i},m_{2,i})}{\phi(r_i,x_i,m_{1,i},m_{2,i})}
\end{equation}

\section{Results}
 The separation histograms for our dataset are shown in Figure~\ref{sep_hists}. In each we see a log-flat distribution of true binaries that is eventually swamped by coincident pairings at separations of around 3000\,AU. There may be a population of true binaries at even wider separations, but the number of coincident pairs makes it hard to draw conclusions beyond ~3000AU. We find 20 binaries with $p_{bin}>0.5$ in Alpha~Per, 47 in the Pleiades and 28 in Praesepe. These are listed in Table~\ref{pairs}. One of our probable pairings include a previously unknown companion to Asterope, one of the naked-eye members of the Pleiades.  We note that we have a handful of very wide binaries ($>$5000\,AU). These are found in the outskirts of each cluster where the density of cluster members, and hence the probability of chance alignment, is lower.

\subsection{Completeness}
\label{bin_comp}
To estimate our completeness we use the work of \cite{Hillenbrand2018} as a comparison survey. This work searched for companions to Pleiades and Praesepe members using adaptive optics. While most of their companions were at projected separations of less than an arcsecond and undetected by us, we recovered two of the four companions to stars that were in both our member selections and \cite{Hillenbrand2018}'s sample which had separations of 1.0--1.5 arcseconds, two of the three with 1.5--2.0 arcsecond separations and all three of the companions wider than that. Some of the \cite{Hillenbrand2018} targets did not appear in our sample as they fell outside our sky search area for each cluster, had no astrometric solution in Gaia or were not selected as members by our algorithm. For companions from \cite{Hillenbrand2018} that are wider than two arcseconds we do not recover four Pleiades systems. Both components in all of these systems are excluded as Pleiades members due to their parallaxes with three (s5035799, HII~659 and s5197248) having proper motions that lie outside our cluster distribution. Only DH~800 is listed as having a membership probability greater than 0.01 for either component. Only one of the components of these four binaries (the secondary of HII~659) is listed as having an elevated Renormalised Weighted Error (RUWE; \citealt{Lindegren2018,Lindegren2018a}) indicating that it is not poor-quality astrometric solutions driven by confusion that are causing these binaries to be ruled out. Indeed the primary of HII~659 also has a parallax that is discrepant with Pleiades membership ($\varpi=5.92\pm0.05$\,mas for the primary versus $\varpi=5.48\pm0.34$\,mas for the secondary). We also recover all five binaries wider than two arcseconds in \cite{Bouvier1997a}. 

To test if missing the four wider Pleiades systems from \cite{Hillenbrand2018} is significant we extracted the SuperCOSMOS \citep{Hambly2001} proper motions for these systems. SuperCOSMOS uses plate data with a much lower resolution and longer time baseline than Gaia. We therefore would expect binarity to affect the astrometric solutions differently from the way it affects Gaia data. We found that three of the objects (s5035799, HII~659 and s5197248) had proper motions that lay outside the proper motion distribution of Pleiades members found by \cite{Deacon2004} while obviously DH~800 was selected as a Pleiades member. This suggests that the absence of three wide Pleiades binaries from \cite{Hillenbrand2018} is likely due to those binaries being field interlopers or objects in the kinematic outskirts of the cluster which we exclude from our Pleiades member sample rather than any bias we have against selecting wide binary components as cluster members. The additional proper motion information of Gaia only serves to further refine the input member list that was available to \cite{Hillenbrand2018}, it does not call the existence of any of their binaries into question.

\begin{figure}
\begin{tabular}{c}
 \setlength{\unitlength}{1mm}
 \includegraphics[scale=0.5]{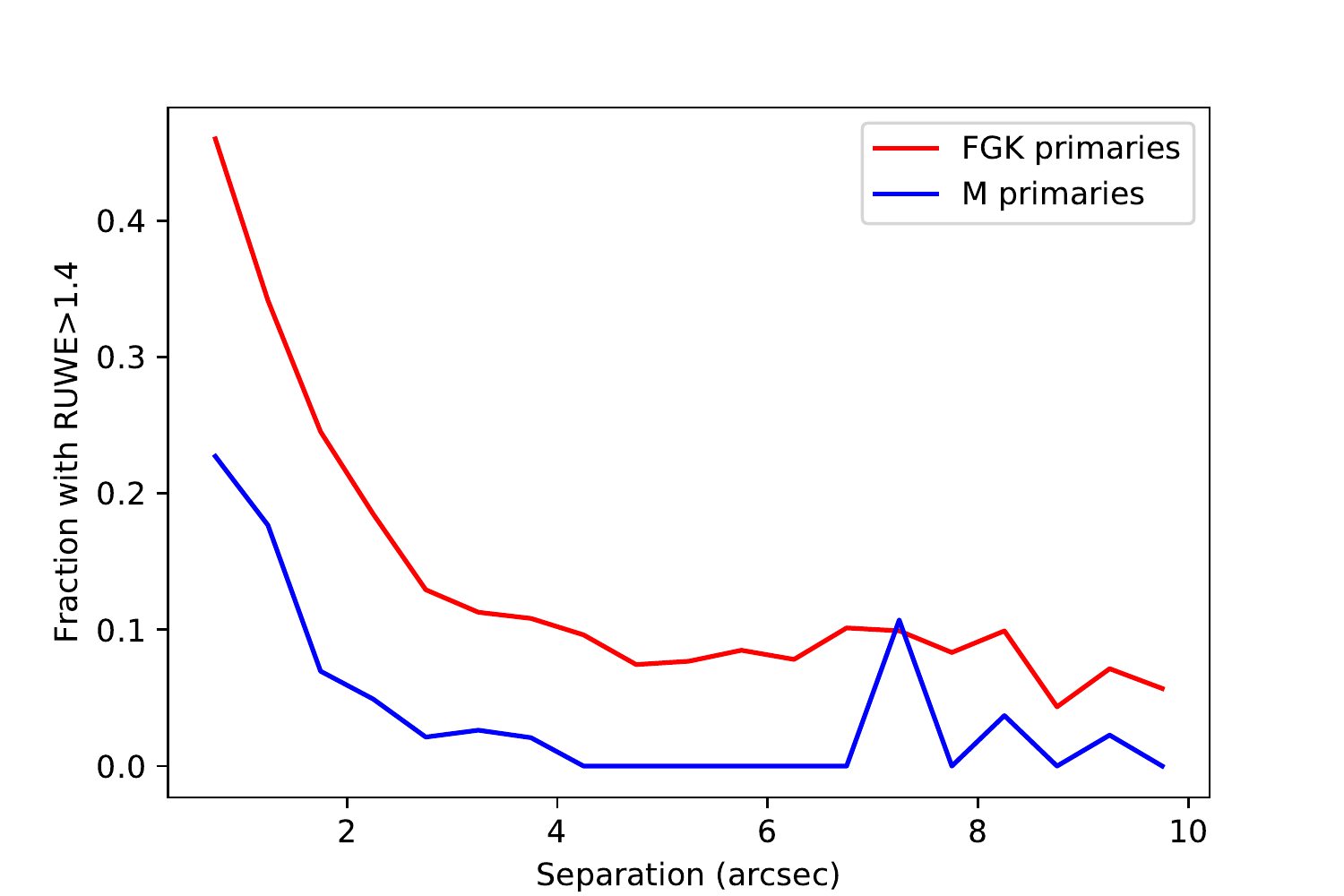}
 \end{tabular}
 \caption[]{Testing the incompleteness of our survey as a function of binary separation. Here we estimate the fraction of companions to objects in the vicinity of all three clusters that have noisy astrometric solutions. } 
  \label{ast_comp}
 \end{figure}

It is possible to further test if we are losing true binaries which have components that are missing from our cluster membership lists due to poor-quality astrometric solutions. To test this we examined all pairs of objects with separations up to ten arcseconds in the area around all three clusters. We chose to examine the secondary  (i.e. fainter) components as these will be more likely to be affected by scattered light from a bright companion. Firstly we estimated the fraction of pairs as a function of separation that have noisy ($RUWE>1.4$) astrometric solutions for their secondary components (see the left-hand panel of Figure~\ref{ast_comp}). This shows that for separations below two arcseconds there is a large fraction of objects with raised RUWE values. However beyond around three arcseconds this fraction reverts to the background level.

To further test our completeness we examine the work of \cite{Ziegler2018}. They identified Gaia detections for components of directly imaged binaries (taken from \citealt{Law2014a}, \citealt{Baranec2016} and \citealt{Ziegler2017}). Their results show that almost all binaries wider than $\sim2$ arcseconds and with magnitude differences less than 5 magnitudes have both components detected. Their sample contains few sources wider than 4 arcseconds. \cite{Arenou2018} use stars from the Washington Double Star catalogue \citep{Mason2001} to show that binaries wider than around 4 arcseconds are complete in Gaia (their Figure~8). Hence we assume that we are complete to our faint limits for binaries wider than four arcseconds. 

\begin{figure}
 \setlength{\unitlength}{1mm}
 \includegraphics[scale=0.41]{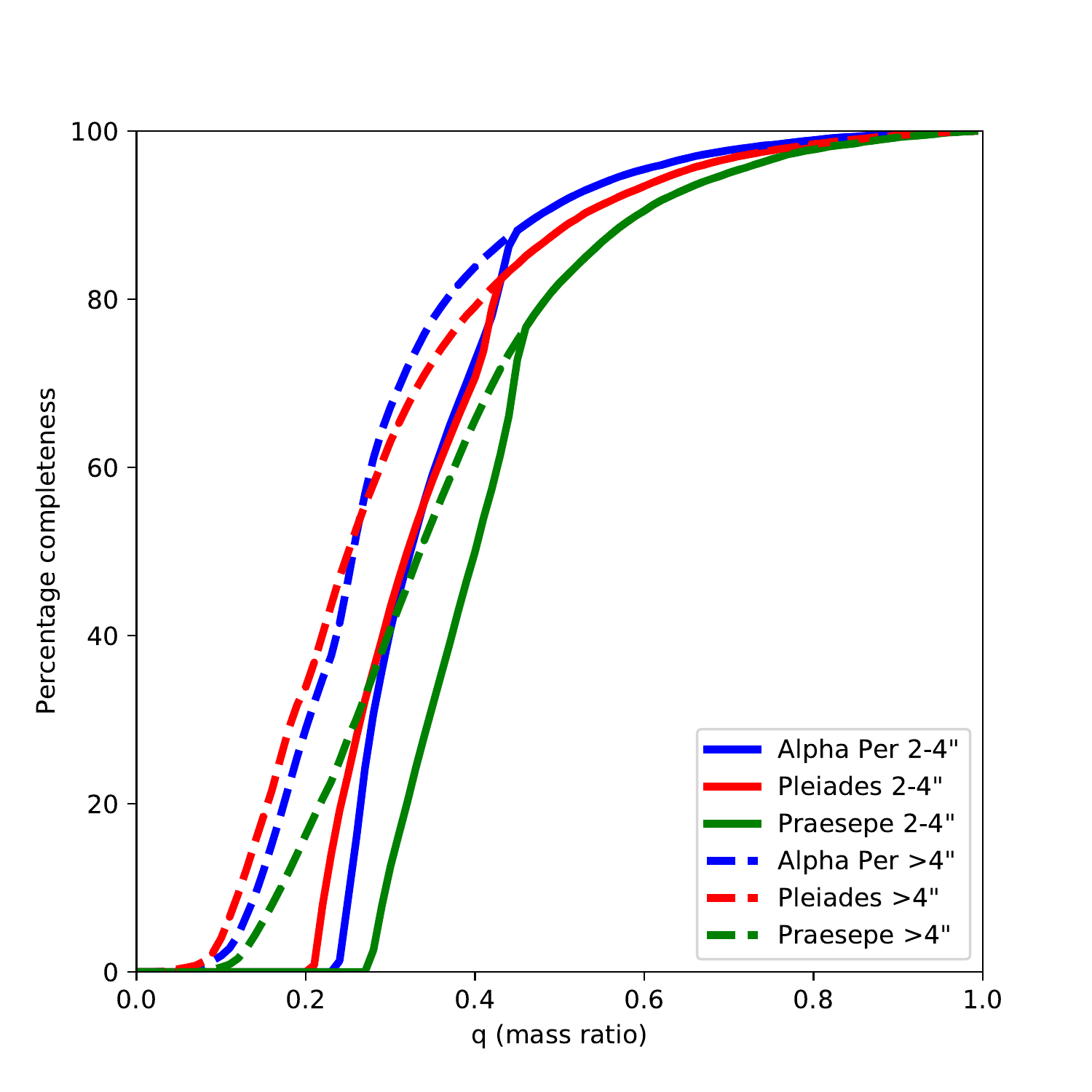}
 \caption[]{Completeness as a function of mass ratio for our three clusters both for companions in the 2--4 arcsecond range and in the $>$4 arcsecond range.} 
  \label{mass_comp}
 \end{figure}

\begin{figure*}
 \setlength{\unitlength}{1mm}
 \begin{tabular}{ccc}
 \includegraphics[scale=0.21]{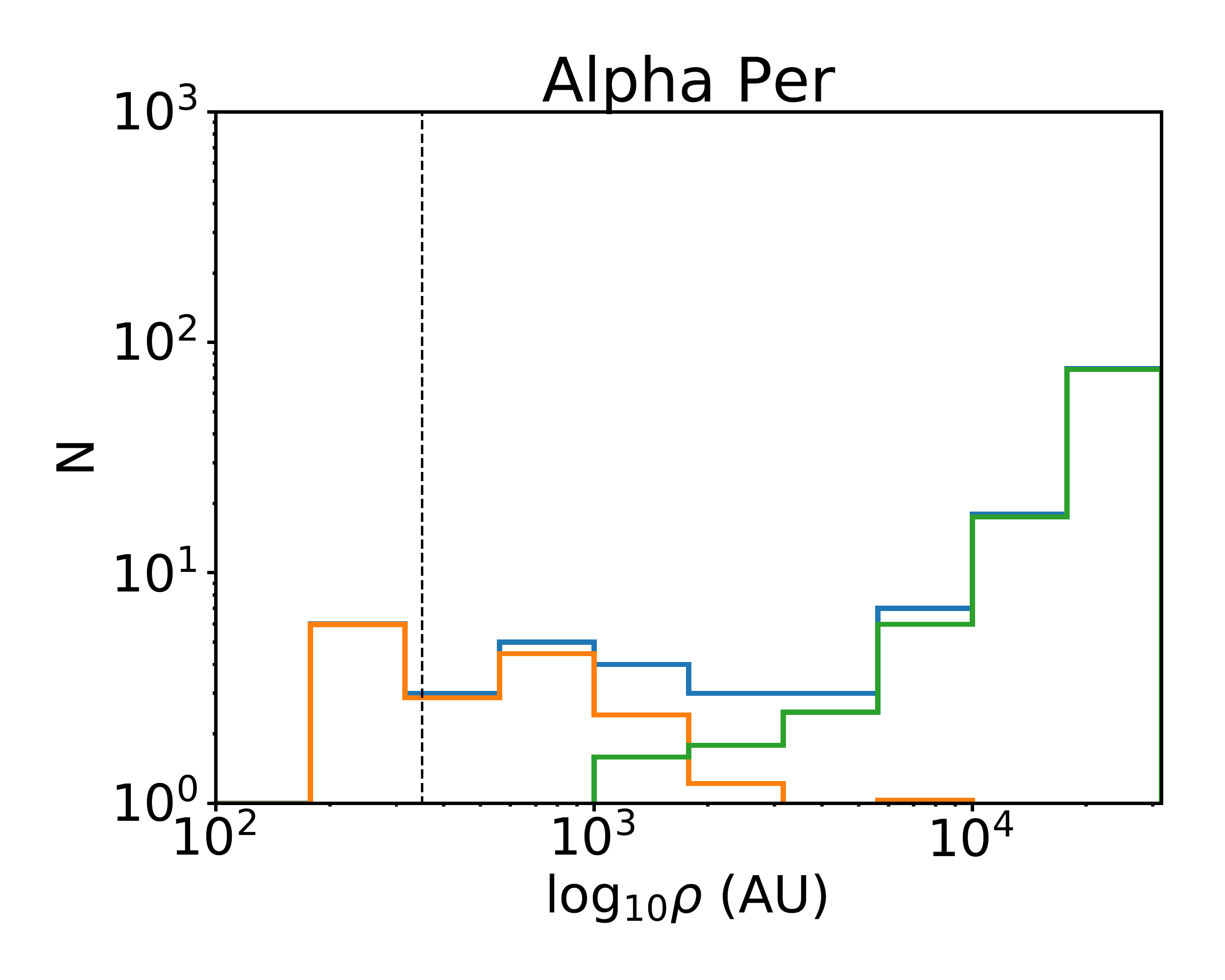}&
 \includegraphics[scale=0.21]{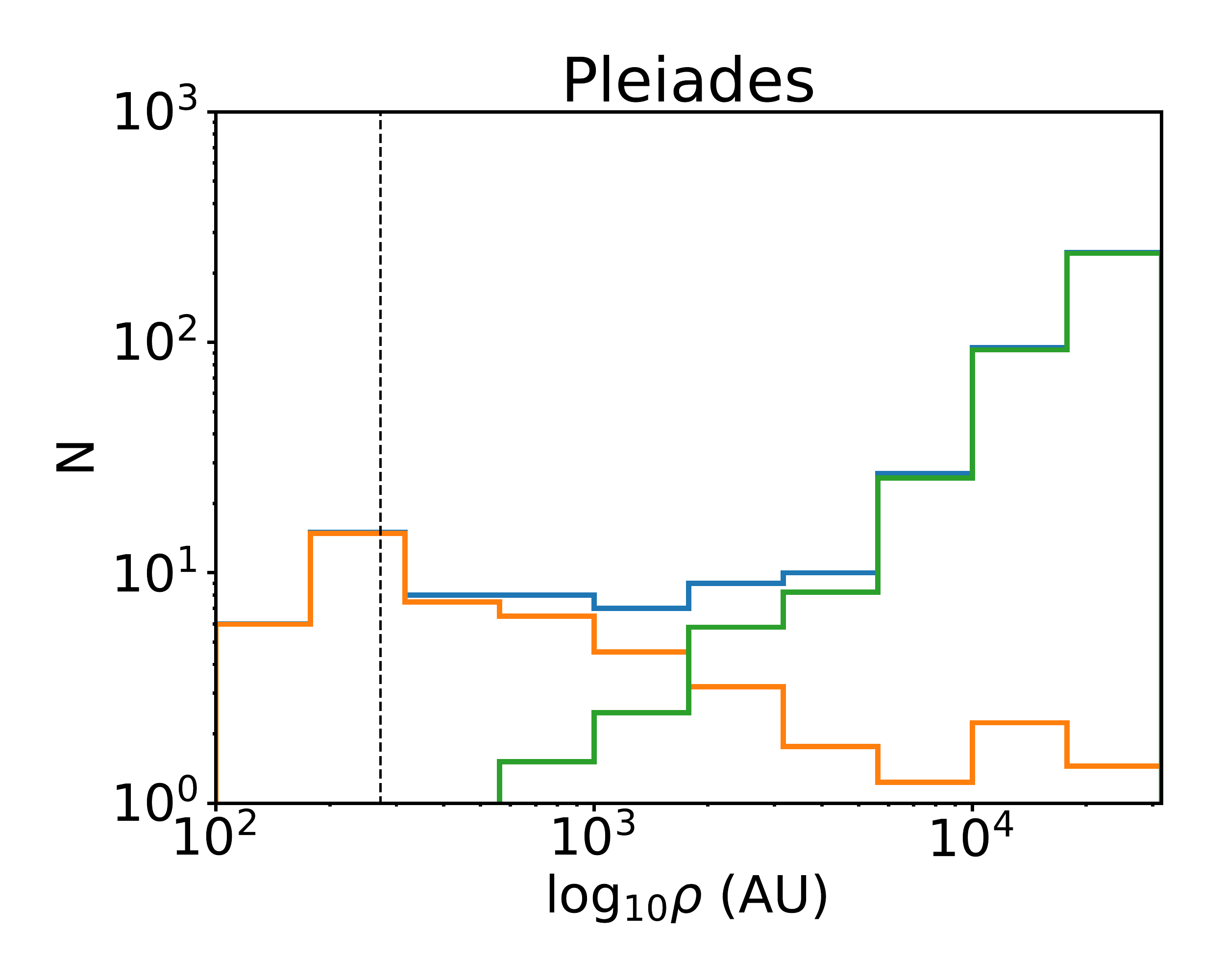}&\includegraphics[scale=0.21]{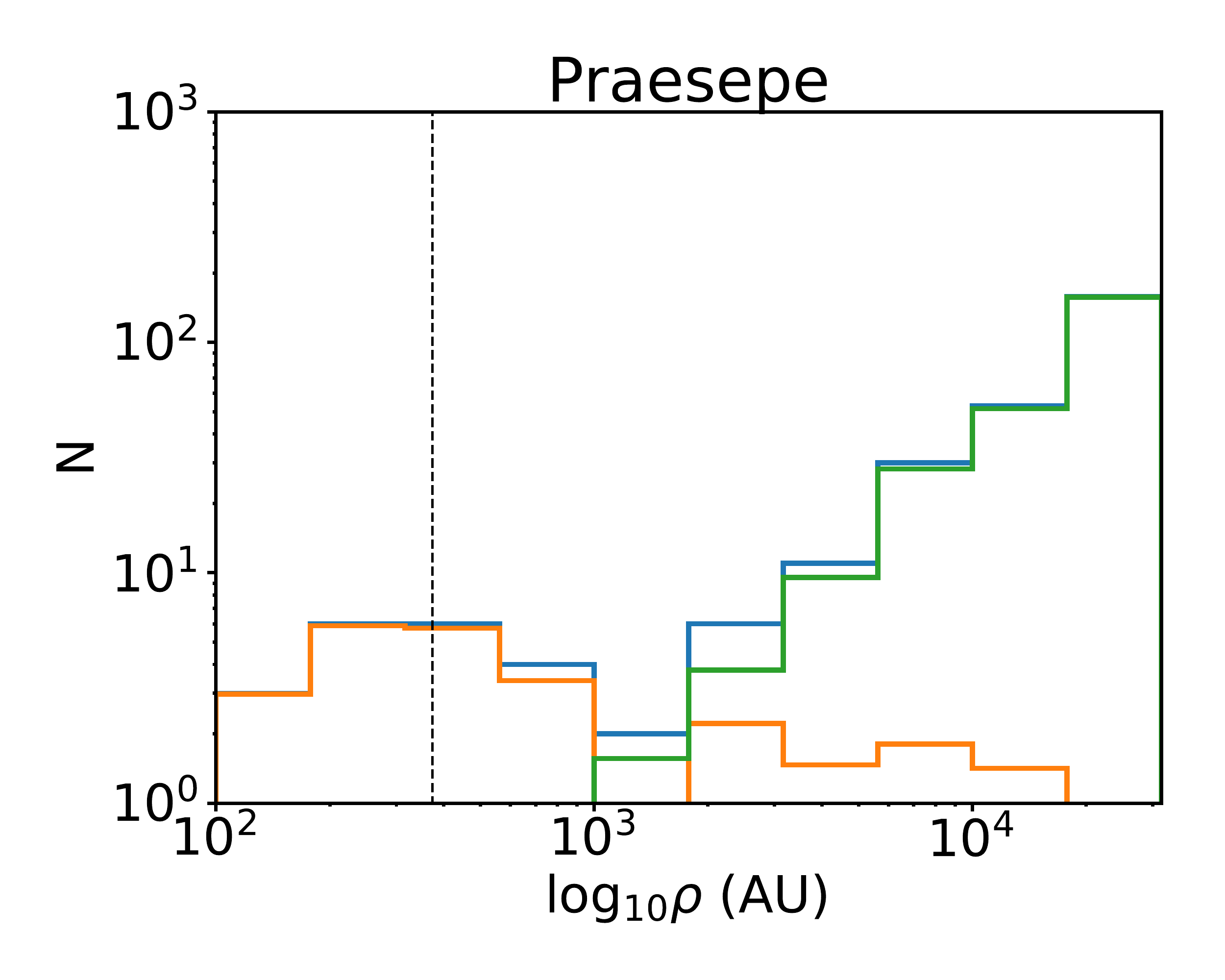}
 \end{tabular}
 \caption[]{Separation histograms for our three clusters. The blue line shows the total number of objects at each separation. The orange line shows the sum of the binary probabilities in each separation bin. The green line is the blue line minus the red line and shows the expected population of chance alignments. There is clearly an excess of pairs above the number of chance alignments below separations of around 2000AU} 
  \label{sep_hists}
 \end{figure*}

\begin{figure*}
 \setlength{\unitlength}{1mm}
 \begin{tabular}{cc}
 \includegraphics[scale=0.49]{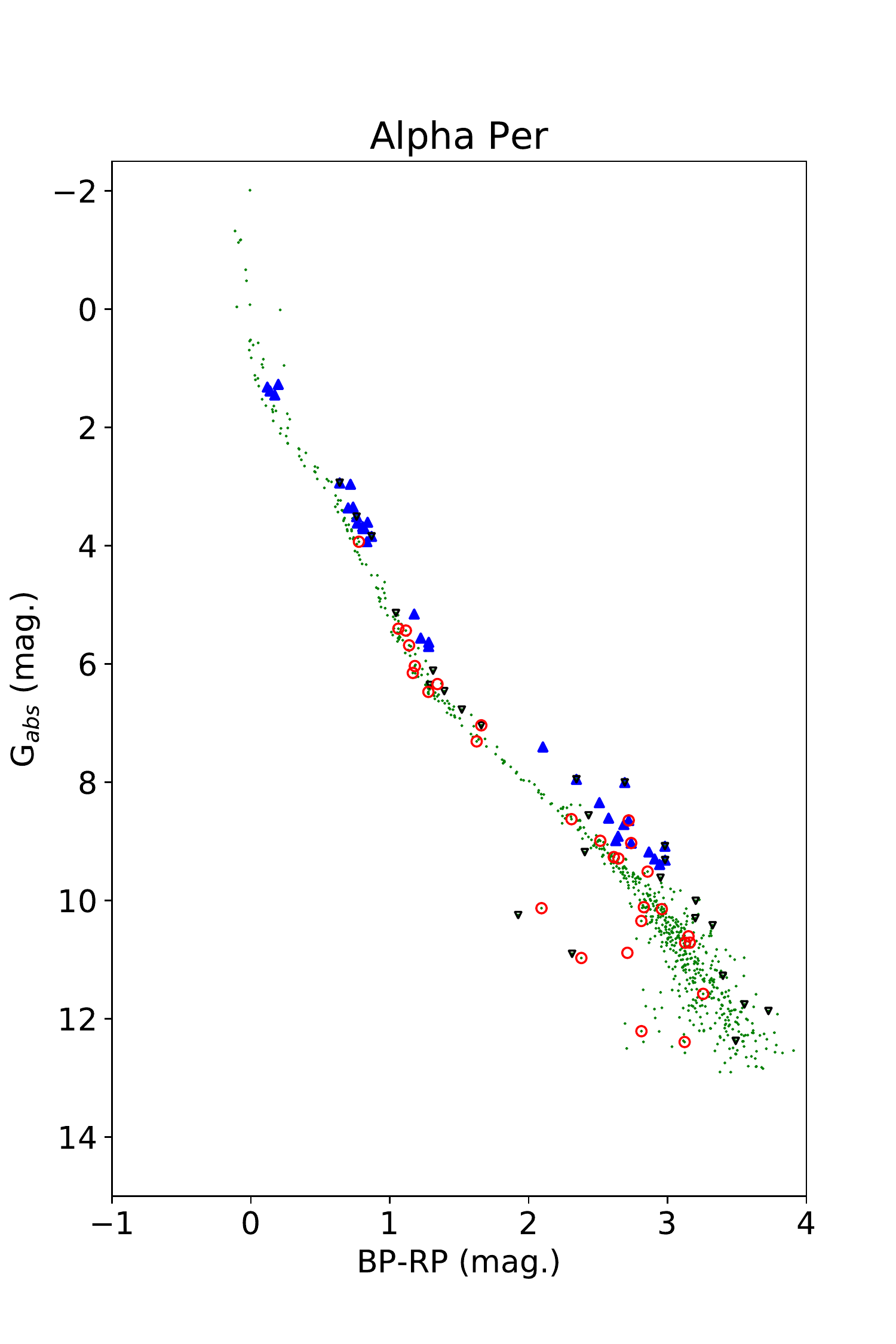}&
 \includegraphics[scale=0.49]{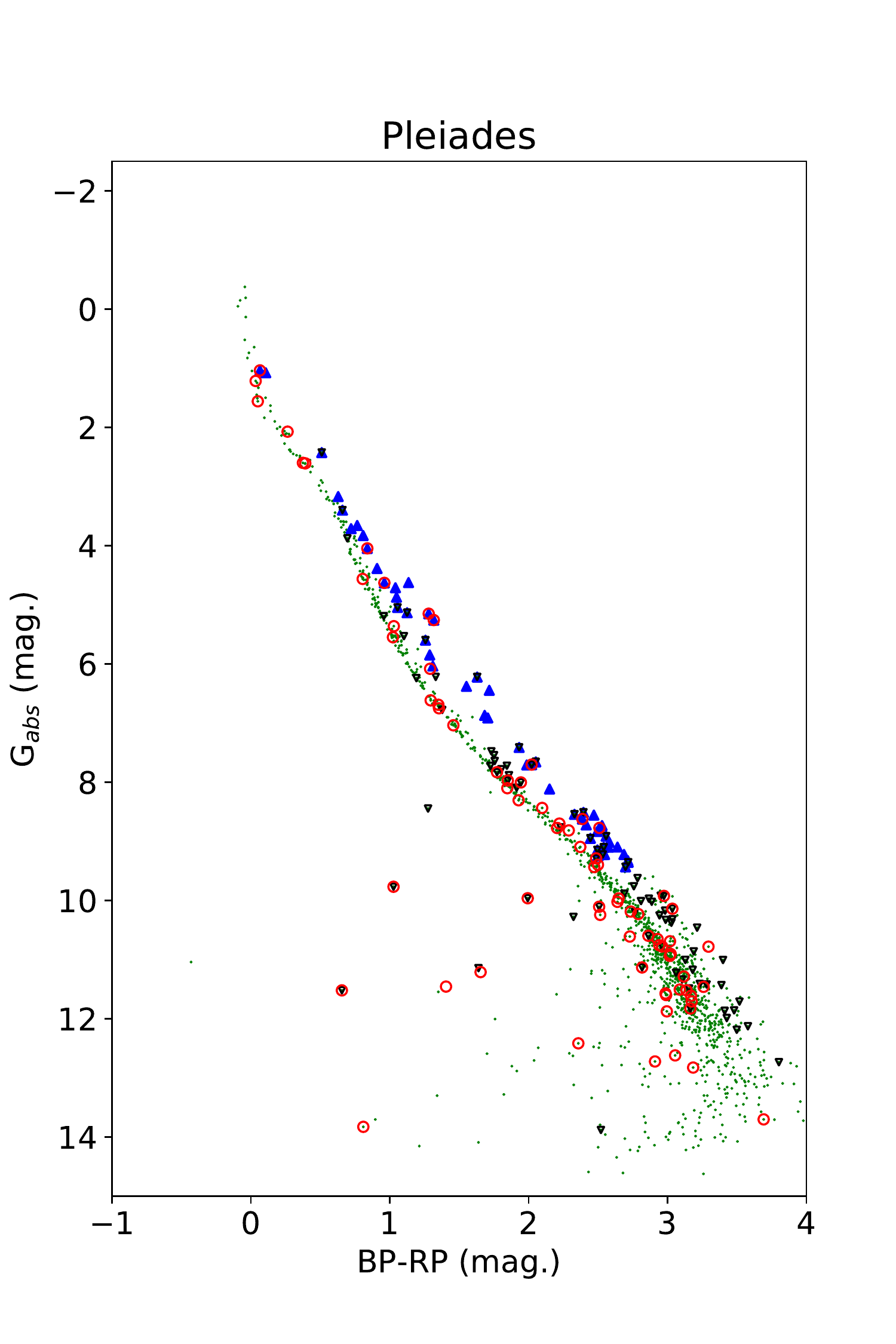}\\
 \includegraphics[scale=0.49]{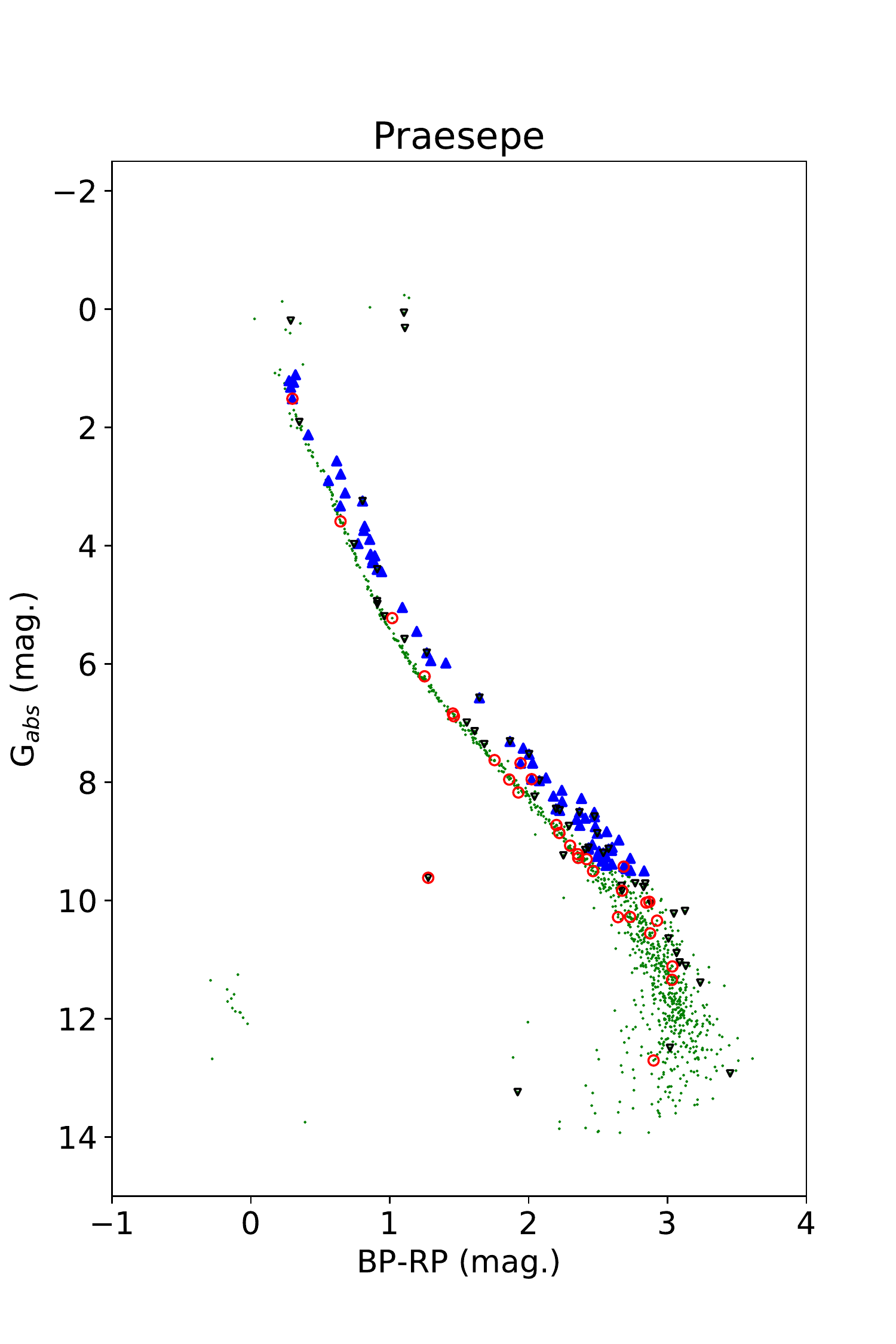}
 \end{tabular}
 \caption[]{Hertzprung-Russell diagrams for $p_{memb}>0.5$ members of our three clusters. Red circles show objects in $p_{bin}>0.5$ binaries, blue triangles are objects flagged as possible overluminous binaries and downward-pointing black triangles are objects flagged as having elevated astrometric noise. We clearly select stars above the cluster sequences as overluminous. Note also how stars with elevated astrometric noise lie preferentially above the main sequence.} 
  \label{unresolved_cmd}
 \end{figure*}

Our binary search will of course be constrained by the magnitude limit of our Gaia study. To quantify how complete we are, we estimated the binary mass ratios we could reach for each of our cluster members in each cluster (treating each as a possible primary star). We first estimated the mass of a star five magnitudes fainter than each possible primary using the PARSEC models for each cluster. This gives us a potential limiting mass for a potential companion in the 2--4 arcsecond range. If this mass was less than our lower mass limit for the cluster the star is a member of, then we used the cluster limiting mass as our limiting mass. We then divided this limiting mass by the mass of the potential primary to estimate the mass ratio above which we would be complete. We then assumed a completeness function that is zero below this mass ratio and 100\% above it. Then we repeated this process for all stars in the cluster, summing the completeness functions and dividing by the number of cluster members in our sample then gives us the completeness between two and four arcseconds. These functions are shown in Figure~\ref{mass_comp}. We also estimated the completeness for systems wider than four arcseconds by using the same process but only setting the limiting companion mass for each star to the lower limiting mass for the appropriate cluster. These completeness limits are also shown in Figure~\ref{mass_comp}. If we assume a flat mass ratio distribution we can estimate the completeness of our sample for each cluster in each companion separation range by taking the mean of our completeness functions. Hence we find completenesses of 64\% (2"--4") \& 71\% ($>4"$) for Alpha~Per, 64\% (2"--4") \& 70\% ($>4"$) for the Pleiades and 57\% (2"--4") \& 63\% ($>4"$) for Praesepe. It is likely that our completeness is higher than the quoted percentages. Our main source of incompleteness is undetected faint companions. Our faint limits for companions to these higher mass stars go to lower mass ratios than for companions to lower mass. These higher mass stars will be more likely to host binaries. Therefore we are likely to be more complete than our calculations around the stars most likely to host binaries. Additionally M dwarf primaries are more likely to host high mass ratio companions \citep{Duchene2013}. This means that the objects we are the least sensitive to, low mass ratio companions to low mass stars, are rare.
\subsection{Estimating the binary fraction}
We can estimate the binary fraction between 300\,AU and 3000\,AU for all three of our clusters. To do this we sum the binary probabilities for all pairs in this projected separation range. We then divide this by the total number of members of each cluster in our sample. Our calculation only includes binaries wider than two arcseconds in this  as we are likely substantially incomplete for binaries closer than this. We then correct our calculated binary fractions by dividing by factors of $\log_{10}3000 - \log_{10}2d_{cluster}$. This assumes a flat distribution in log-separation (i.e. Opik's law \citealt{Opik1924}) which broadly agrees with observations that show the separation distribution in this region is either flat \citep{Kraus2011} or gradually declining \citep{Raghavan2010,Tokovinin2012}. This takes into account the fact that cutting at separations of two arcseconds means we are incomplete for projected separations closer than $2d_{cluster}$. We also account for incompleteness for lower mass companions using the completeness estimates from Section~\ref{bin_comp}. As we are dealing with small number statistics for our binaries we determine one sigma confidence limits using the relations of \cite{Gehrels1986}. This gives a better representation the uncertainties based on a small number of detections and leads to different upper and lower uncertainty bounds.

The calculated binary fractions in the 300--3000\,AU projected separation range for stars in our three clusters are $f_{bin}=2.0\pm^{0.9}_{0.6}$\% for Alpha~Per, $f_{bin}=1.9\pm^{0.6}_{0.4}$\% for the Pleiades and $f_{bin}=2.8\pm^{0.7}_{0.6}$\% for Praesepe. The binary fractions are consistent within uncertainties. Table~\ref{bin_calc_table} shows the binary fractions for each cluster. We also calculate fractions for primary stars with mass estimates in the 0.5--1.5\,$M_{\odot}$ range (roughly corresponding to FGK stars) and in the $<0.5$\,$M_{\odot}$ range (roughly corresponding to M dwarfs). We note that the lower mass bin in each cluster has a lower binary fraction than the FGK star bin. Combining all the clusters together, the difference between the FGK and M dwarf fractions is 2.2\,$\sigma$. This is consistent with previous work in young cluster \citep{Kraus2009} and the field \citep{Dhital2010} that shows that M dwarfs are less likely to be the host stars from wide binary systems than FGK stars.

\begin{table*}
\caption{\label{bin_calc_table} The binary fractions for different types of stars in our clusters. We use uncertainty estimates calculated with the relations of \protect\cite{Gehrels1986} as these are more appropriate for small number statistics. In each case we estimate the binary fraction divided by the log separation range covered $\frac{df_{bin}}{d\log_{10}x}$. Our two subranges cover approximately FGK dwarfs ($0.5<M/M_{\odot}<1.5$) and M dwarfs ($M<0.5M_{\odot}$)}
\begin{center}
\begin{tabular}{lrrrrr}
\hline
&$\sum p_{bin}$&$N_{stars}$&Completeness&$d\log_{10}r$&$\frac{df_{bin}}{d\log_{10}x}$\\
\hline
\multicolumn{6}{c}{Alpha Per}\\
\hline
All dwarfs&11.0&815&64\%&0.932&2.2$\pm^{0.9}_{0.7}$\%\\
FGK dwarfs&8.5&250&72\%&0.932&5.1$\pm^{2.4}_{1.7}$\%\\
M dwarfs&2.4&506&53\%&0.932&1.0$\pm^{1.1}_{0.6}$\%\\
\hline
\multicolumn{6}{c}{Pleiades}\\
\hline
All dwarfs&21.7&1477&68\%&1.00&2.2$\pm^{0.6}_{0.5}$\%\\
FGK dwarfs&7.0&388&80\%&1.00&2.3$\pm^{1.2}_{0.8}$\%\\
M dwarfs&9.2&1010&59\%&1.00&1.5$\pm^{0.7}_{0.5}$\%\\
\hline
\multicolumn{6}{c}{Praesepe}\\
\hline
All dwarfs&11.8&1181&55\%&0.904&2.0$\pm^{0.8}_{0.6}$\%\\
FGK dwarfs&5.3&320&72\%&0.904&2.5$\pm^{1.2}_{1.1}$\%\\
M dwarfs&5.3&754&49\%&0.904&1.6$\pm^{1.0}_{0.7}$\%\\
\hline
\multicolumn{6}{c}{All clusters combined}\\
All dwarfs&&&&&2.1$\pm^{0.4}_{0.2}$\%\\
FGK dwarfs&&&&&3.0$\pm^{0.8}_{0.7}$\%\\
M dwarfs&&&&&1.4$\pm^{0.4}_{0.3}$\%\\
\end{tabular}
\end{center}
\end{table*}

   \clearpage
 \begin{deluxetable}{lcrrrrrrcl}
 \tablecolumns{10}
 \tablewidth{0pc}
\tablecaption{\label{pairs} A list of our probable wide binaries in Alpha~Per, the Pleiades and Praesepe. All astrometry and photometry is from Gaia DR2.}
\tabletypesize{\tiny}
\tablehead{
\colhead{Gaia ID}&\colhead{Position}&\colhead{$\mu_{\alpha}\cos{\delta}$}&\colhead{$\mu_{\delta}$}&\colhead{$\pi$}&\colhead{$G$}&\multicolumn{2}{c}{Projected}&\colhead{$p_{bin}$}&\colhead{Other name}\\
&&&&&&\multicolumn{2}{c}{Separation}\\
&\colhead{(Eq.=J2000 Ep.=2015.5)}&\colhead{(mas/yr)}&\colhead{(mas/yr)}&\colhead{(mas)}&\colhead{(mag)}&\colhead{('')}&\colhead{(AU)}
}
\startdata
\multicolumn{10}{c}{Alpha Per}\\
\hline
\hline
437596389381227648&02 58 22.98 +48 04 28.8&10.3$\pm$0.2&$-16.5\pm0.2$&4.5$\pm$0.1&17.3&0.9&186&1.00\\
437596389384924544&02 58 22.99 +48 04 27.9&20.6$\pm$1.1&$-18.3\pm$0.5&4.9$\pm$0.4&18.9\\
\hline
439718962223158016&03 07 58.13 +50 24 35.8&24.1$\pm$0.1&$-$12.1$\pm$0.1&5.5$\pm$0.1&15.3&4.8&869&0.94& 	
UGCS J030758.10+502436.0\\
439718962223158272&03 07 57.63 +50 24 36.1&24.0$\pm$0.1&$-$23.6$\pm$0.1&5.5$\pm$0.1&15.6&&&&UGCS J030757.60+502436.3 \\
\hline
436345969787424256\tablenotetext{1}& 13 48.54 +48 59 11.1&23.8$\pm$0.1&$-23.8\pm0.1$&5.4$\pm$0.1&12.0&2.6&489&0.98&TYC 3319-557-1\\
436345969787424896&13 48.73 +48 59 09.3&22.6$\pm$0.1&$-21.8\pm$0.1&5.4$\pm$0.1&13.7\\
\hline
436441176331772928&03 17 04.40 +49 20 14.6&23.6$\pm$0.1&$-$25.3$\pm$0.1&5.6$\pm$0.1&12.3&6.5&1165&0.93&\\ 
436441176324119680&03 17 03.80 +49 20 11.7&21.5$\pm$0.1&$-$25.5$\pm$0.1&5.6$\pm$0.1&12.6&&&&\\ 
\hline 
435680765247601536&03 17 42.13 +49 01 46.2&23.3$\pm$0.1&$-$24.3$\pm$0.1&5.6$\pm$0.1&11.7&1.5&274&0.98&AP121\\ 
435680765251589632\tablenotemark{\ddag}&03 17 42.21 +49 01 44.8&22.5$\pm$0.5&$-$28.1$\pm$0.5&5.5$\pm$0.2&16.1&&&&\\ 
\hline
435184610626097408\tablenotemark{\dag}&03 18 01.20 +47 00 07.6&23.5$\pm$0.1&$-$24.3$\pm$0.1&5.8$\pm$0.1&15.2&1.7&295&0.99&DH 83\\
435184610626097536\tablenotemark{\ddag}&03 18 01.29 +47 00 06.1&23.5$\pm$0.1&$-$25.1$\pm$0.1&5.8$\pm$0.1&15.4&\\
\hline
242294983663641088&03 20 15.30 +45 00 10.0&19.9$\pm$0.3&$-21.2\pm$0.2&4.3$\pm$0.13&17.7&1.1&236&1.00\\
242294983667977472&03 20 15.25 +45 00 11.0&19.9$\pm$0.4&$-21.1\pm0.3$&4.67$\pm$0.2&18.4\\
\hline
441640668031763584&03 22 32.80 +49 11 16.29&22.7$\pm$0.1&$-$26.0$\pm$0.1&5.7$\pm$0.1&16.4&2.9&484&0.96&\\ 
441640668027534720&03 22 32.52 +49 11 15.49&23.2$\pm$0.3&$-$26.1$\pm$0.2&6.0$\pm$0.1&17.7&&&&AP143\\ 
\hline 
441716774852013568&03 22 40.95 +49 40 40.94&23.9$\pm$0.1&$-$25.7$\pm$0.1&5.8$\pm$0.1&12.6&4.6&783&0.87&AP107\\ 
441716774852012032&03 22 41.08 +49 40 45.32&24.3$\pm$0.2&$-$25.7$\pm$0.2&5.8$\pm$0.1&16.5&&&&\\ 
\hline 
242736712463377280&03 22 55.54 +46 06 57.4&25.9$\pm$0.1&$-27.4\pm0.1$&6.2$\pm$0.1&12.2&3.9&623&0.87&UCAC4 681-020405\\
242736712461022976&03 22 55.73 +46 06 54.1&25.8$\pm$0.2&$-27.1\pm0.2$&6.2$\pm$0.1&17.0&&&&UGCS J032255.70+460654.2\\
\hline
442705648117153792\tablenotemark{\ddag}&03 23 20.34 +51 19 06.1&22.7$\pm$0.1&$-$23.3$\pm$0.1&5.4$\pm$0.1&13.4&1.1&205&1.00&\\
442705648121901696\tablenotemark{\ddag}&03 23 20.31 +51 19 07.2&21.5$\pm$0.1&$-$23.7$\pm$0.1&5.4$\pm$0.1&13.6\\
\hline
243108622271431424\tablenotemark{1,2}&03 26 22.67 +47 16 09.0&23.0$\pm$0.1&$-$25.3$\pm$0.1&5.7$\pm$0.1&11.7&7.1&1243&0.58&V627 Per AP 156 RSCVn\\ 
243108617973878272&03 26 21.98 +47 16 09.7&23.4$\pm$0.8&$-$26.1$\pm$0.4&5.9$\pm$0.3&18.5&&&&\\ 
\hline 
442671563260694016&03 26 35.97 +51 22 32.5&21.1$\pm$0.1&$-$24.0$\pm$0.1&5.4$\pm$0.1&15.0&2.9&546&0.93&UGCS J032635.95+512232.6\\
442671563256206336&03 26 35.96 +51 22 29.6&20.5$\pm$0.7&$-$25.0$\pm$0.5&5.6$\pm$0.4&18.5\\
\hline
441405922299247616&03 26 50.15 +48 47 31.6&22.1$\pm$0.1&$-$26.5$\pm$0.1&5.6$\pm$0.1&10.2&4.5&813&0.83&AP51\\ 
441405922299247488&03 26 49.91 +48 47 35.4&22.3$\pm$0.2&$-$26.4$\pm$0.1&5.8$\pm$0.1&16.3&&&&\\ 
\hline 
441596172170885888&03 27 50.11 +49 54 18.6&24.0$\pm$0.1&$-$26.2$\pm$0.1&5.9$\pm$0.1&15.4&14.9&2542&0.52&UGCS J032750.07+495418.9\\ 
441596172170886912&03 27 48.58 +49 54 17.9&24.2$\pm$0.2&$-$26.2$\pm$0.1&6.1$\pm$0.1&16.2&&&&APX 155D\\ 
\hline
249111577803259264\tablenotemark{\dag}&03 28 21.96 +47 36 05.6&23.3$\pm$0.2&$-$26.6$\pm$0.1&5.8$\pm$0.1&15.7&1.1&191&1.00&\\ 
249111577799138816\tablenotemark{\ddag}&03 28 21.91 +47 36 06.6&24.3$\pm$0.2&$-$24.5$\pm$0.2&6.0$\pm$0.1&16.2&&&&DH 184\\ 
\hline 
249265062754562048\tablenotemark{\dag}&03 31 16.13 +48 25 11.9&20.9$\pm$0.4&$-$24.6$\pm$0.3&5.3$\pm$0.2&15.0&0.9&176&1.00&DH 222\\ 
249265062749872768&03 31 16.20 +48 25 11.2&22.8$\pm$0.3&$-$26.9$\pm$0.2&5.8$\pm$0.1&16.3&&&&\\ 
\hline
249033787353638400&03 37 28.45 +48 27 24.9&21.4$\pm$0.2&$-$24.6$\pm$0.2&5.2$\pm$0.1&17.1&4.2&803&0.95&UGCS J033728.43+482725.2\\ 
249033787353638912&03 37 28.39 +48 27 20.7&21.5$\pm$0.3&$-$24.8$\pm$0.2&5.2$\pm$0.1&17.1&&&&\\  
\hline
\multicolumn{10}{c}{Pleiades}\\
\hline
\hline
69585140280997760&03 31 17.96 +26 01 43.0&21.2$\pm$0.1&$-42.6\pm$0.1&7.2$\pm$0.1&12.3&2.5&336&0.96&DANCe J03311794+2601434 \\
69585140280997888&03 31 17.81 +26 01 42.6&17.0$\pm$0.86&$-43.9\pm$0.5&6.0$\pm$0.5&17.3\\
\hline
64597927335800064\tablenotemark{\ddag}&03 36 24.36 +22 37 24.9&21.4$\pm$0.2&$-$44.5$\pm$0.1&7.3$\pm$0.1&13.5&2.5&372&0.96&DH 56\\ 
64597927334523904\tablenotemark{\ddag}&03 36 24.20 +22 37 23.9&21.2$\pm$0.4&$-$43.9$\pm$0.3&6.6$\pm$0.2&16.5&&&&\\  
\hline
68356367317952256\tablenotemark{2}&03 40 40.06 +24 44 08.4&21.7$\pm$0.2&$-$46.9$\pm$0.1&7.5$\pm$0.1&15.6&4.0&525&0.91&HZ Tau\\ 
68356367317952128&03 40 39.76 +24 44 08.1&21.6$\pm$0.3&$-$46.1$\pm$0.2&7.7$\pm$0.2&17.3&&&&\\ 
\hline
68364544935515392\tablenotemark{\ddag}&03 42 03.90 +24 42 44.3&21.1$\pm$0.5&$-$46.8$\pm$0.4&8.4$\pm$0.3&13.5&1.3&160&1.00&\\ 
68364544933829376\tablenotemark{\ddag,2}&03 42 03.82 +24 42 45.1&19.5$\pm$0.4&$-$47.8$\pm$0.3&7.6$\pm$0.1&13.6&&&&V610 Tau\\ 
\hline 
65063707949772544\tablenotemark{\dag}&03 43 24.56 +23 13 32.6&20.8$\pm$0.1&$-$41.5$\pm$0.1&7.2$\pm$0.1&10.3&3.6&504&0.90&HII 102\\ 
65063707949772672\tablenotemark{\ddag}&03 43 24.42 +23 13 29.6&21.3$\pm$0.2&$-$42.5$\pm$0.1&7.3$\pm$0.1&15.6&&&& 	UGCS J034324.40+231330.1  Known Bouvier97\\ 
\hline
65266494828710400&03 43 36.72 +24 13 55.6&21.1$\pm$0.1&$-$45.7$\pm$0.1&7.1$\pm$0.1&14.2&1.8&260&0.99&Bouvier97\\ 
65266499126062080\tablenotemark{\ddag,2}&03 43 36.58 +24 13 55.6&18.4$\pm$0.2&$-$41.2$\pm$0.1&6.9$\pm$0.1&14.5&&&&LZ Tau\\ 
\hline
65247704349267584&03 44 13.08 +24 01 50.24&21.4$\pm$0.1&$-$45.9$\pm$0.1&7.5$\pm$0.1&11.2&5.7&808&0.94&HII 299\\ 
65248460263511552\tablenotemark{\ddag}&03 44 12.76 +24 01 53.75&17.3$\pm$0.5&$-$49.0$\pm$0.4&7.0$\pm$0.3&11.8&&&&HII 298 Known Bouvier97\\ 
\hline
69930283851373184\tablenotemark{\dag\ddag,2}&03 44 13.95 +25 32 14.9&17.2$\pm$0.2&$-$40.4$\pm$0.3&6.7$\pm$0.2&16.0&1.2&173&0.99&V515 Tau\\
69930283853148544\tablenotemark{\ddag}&03 44 14.04 +25 32 14.9&19.5$\pm$0.7&$-$42.2$\pm$0.6&6.9$\pm$0.5&16.8\\
\hline
65249250535404928\tablenotemark{\dag}&03 44 14.65 +24 06 06.9&20.7$\pm$0.1&$-$43.5$\pm$0.1&7.2$\pm$0.1&10.9&1.8&265&1.00&HII 303 Known Bouvier97\\ 
65249250537488128\tablenotemark{\dag}&03 44 14.70 +24 06 05.1&17.6$\pm$0.1&$-$40.9$\pm$0.1&6.9$\pm$0.1&11.0&&&&\\ 
\hline
69823940463098752\tablenotemark{\dag,\ddag}&03 44 27.32 +24 50 37.5&21.0$\pm$0.1&$-$49.2$\pm$0.1&7.6$\pm$0.1&13.3&4.6&609&0.81&HII 347 Pair known in Bouvier97 lists secondary as field contaminant\\ 
69823940463098112&03 44 27.20 +24 50 41.8&20.4$\pm$0.5&$-$48.0$\pm$0.3&7.4$\pm$0.2&18.3&&&&\\ 
\hline 
65241313435901568&03 44 32.02 +23 52 29.4&19.7$\pm$0.1&$-$45.4$\pm$0.1&7.4$\pm$0.1&14.3&1.5&197&0.99&\\ 
65241313437941504\tablenotemark{\dag}&03 44 31.97 +23 52 30.7&19.6$\pm$0.1&$-$46.5$\pm$0.1&7.3$\pm$0.1&14.4&&&&HII 370\\ 
\hline
69864313155605120\tablenotemark{1}&03 45 15.38 +25 17 21.4&19.9$\pm$0.1&$-$45.8$\pm$0.1&7.4$\pm$0.1&11.0&3.9&556&0.83&HII 571 known Bouvier1997a also SB\\ 
69864313154046592\tablenotemark{\ddag}&03 45 15.64 +25 17 22.9&18.5$\pm$0.4&$-$48.0$\pm$0.3&7.0$\pm$0.2&17.2&&&&\\ 
\hline 
65282716922610944&03 45 37.81 +24 20 07.7&17.8$\pm$1.2&$-44.9\pm$1.1&9.7$\pm$0.7&7.1&5.3&665&0.73&HD 23387\\
65282716920396160&03 45 37.98 +24 20 02.9&19.7$\pm$0.4&$-47.3\pm$0.3&7.9$\pm$0.3&15.3&&&&UGCS J034537.95+242003.4\\
\hline
64956127609464320&03 45 48.84 +23 08 49.0&20.8$\pm$0.2&$-$45.7$\pm$0.1&7.4$\pm$0.1&6.9&3.6&492&0.98&HD 23410 Known binary WDS\\ 
64956123313498368&03 45 48.73 +23 08 52.2&19.0$\pm$0.1&$-$45.6$\pm$0.1&7.3$\pm$0.1&10.1&&&&TYC 1799-1443-1\\ 
\hline
70035351636365824\tablenotemark{2}&03 45 52.69 +25 51 41.5&20.5$\pm$0.1&$-$43.8$\pm$0.1&7.2$\pm$0.1&14.5&1.6&240&0.99&V444 Tau In bouvier97\\ 
70035355933666304\tablenotemark{\ddag}&03 45 52.58 +25 51 41.6&19.4$\pm$0.4&$-$46.7$\pm$0.3&6.6$\pm$0.2&15.7&&&&listed as a binary itself in Bouvier97\\ 
\hline
66800180408292992&03 46 05.66 +24 36 43.8&21.3$\pm$0.2&$-$44.5$\pm$0.1&7.4$\pm$0.1&15.1&5.5&78&0.82&SK 488\\ 
66800244832257280&03 46 05.71 +24 36 49.2&21.2$\pm$0.4&$-$45.3$\pm$0.2&7.3$\pm$0.2&17.3&&&&UGCS J034605.69+243649.7\\ 
\hline
66799763794633856\tablenotemark{1,2}&03 46 06.52 +24 34 02.0&20.0$\pm$0.1&$-$46.5$\pm$0.1&7.3$\pm$0.1&12.4&17.5&2377&0.62&V813 Tau\\ 
66799768088857856\tablenotemark{2}&03 46 06.93 +24 33 45.4&19.7$\pm$0.1&$-$47.0$\pm$0.1&7.4$\pm$0.1&12.7&&&&V789 Tau possible WDS eroneous entry\\ 
\hline 
66801623514684416\tablenotemark{\dag,2}&03 46 09.91 +24 40 24.2&20.3$\pm$0.2&$-$47.0$\pm$0.1&7.4$\pm$0.1&14.3&1.4&186&0.99&BD+22 548\\ 
66801623517294848&03 46 09.97 +24 40 25.3&17.7$\pm$0.8&$-$43.9$\pm$0.4&7.5$\pm$0.1&16.3&&&&\\ 
\hline 
65204342359530112&03 46 14.35 +23 51 02.2&18.5$\pm$0.4&$-$44.8$\pm$0.3&7.3$\pm$0.2&18.1&1.4&196&0.99&HHJ 56\\ 
65204342357685632&03 46 14.38 +23 51 00.8&20.1$\pm$0.7&$-$43.5$\pm$0.5&7.2$\pm$0.4&19.0&&&&\\ 
\hline 
65203758244032512&03 46 22.22 +23 52 40.6&19.7$\pm$0.2&$-$43.6$\pm$0.2&7.2$\pm$0.1&16.3&14.5&2020&0.51&DH 441\\
65203723882387200&03 46 22.27 +23 52 26.1&16.9$\pm$1.3&$-$44.6$\pm$0.9&7.3$\pm$0.6&19.5&&&&SHF 31\\
\hline 
65011481147922560&03 47 06.12 +23 45 19.6&18.5$\pm$0.3&$-$43.4$\pm$0.2&7.1$\pm$0.2&17.3&15.1&2125&0.54&DANCe J03470610+2345202 \\
65011476851589632&03 47 05.81 +23 45 34.2&19.2$\pm$0.4&$-$45.0$\pm$0.3&7.3$\pm$0.2&18.5&&&&UGCS J034705.79+234534.7\\
\hline
66728089380492160\tablenotemark{\ddag}&03 47 11.88 +24 13 53.2&19.2$\pm$0.6&$-$43.6$\pm$0.3&7.1$\pm$0.3&17.6&11.7&1651&0.69&HHJ 92\\ 
66728089380491648\tablenotemark{2}&03 47 11.04 +24 13 50.9&20.0$\pm$0.4&$-$46.0$\pm$0.2&7.2$\pm$0.2&17.6&&&&QS Tau\\ 
\hline
66728128037119616\tablenotemark{1}&03 47 18.15 +24 13 50.8&18.8$\pm$0.4&$-$47.0$\pm$0.2&7.9$\pm$0.2&17.0&0.8&101&1.00&V1279 Tau A\\
66728128034922624&03 47 18.16 +24 13 50.0&20.7$\pm$0.4&$-$47.0$\pm$0.2&7.7$\pm$0.2&17.6&&&&V1279 Tau B\\
\hline
65212691775922048\tablenotemark{\ddag}&03 47 18.19 +24 02 10.4&21.3$\pm$0.2&$-$45.4$\pm$0.2&7.6$\pm$0.1&13.6&1.3&172&0.99&Known Bouvier97\\ 
65212691773969280\tablenotemark{\ddag,2}&03 47 18.11 +24 02 11.1&21.9$\pm$0.3&$-$44.6$\pm$0.3&7.4$\pm$0.1&14.7&&&&V646 Tau\\ 
\hline
65207709611941376&03 47 24.44 +23 54 52.1&20.5$\pm$0.1&$-$44.3$\pm$0.1&7.2$\pm$0.1&7.3&6.3&871&0.95&HD 23631 known SB\\ 
65207709613871744\tablenotemark{\dag}&03 47 23.98 +23 54 51.4&20.2$\pm$0.1&$-$45.3$\pm$0.1&7.3$\pm$0.1&9.7&&&&HD 23631B binary system known Bouvier97\\ 
\hline
70085929172184704\tablenotemark{\ddag}&03 47 56.69 +26 31 50.8&21.9$\pm$0.2&$-$43.5$\pm$0.1&7.3$\pm$0.1&15.0&1.3&178&0.99&HHJ 423\\ 
70085933468406784\tablenotemark{\ddag}&03 47 56.62 +26 31 49.9&21.1$\pm$0.7&$-$44.7$\pm$0.4&7.0$\pm$0.1&16.0&&&&\\ 
\hline
 66517468482370304&03:48:17.14 +23:53:24.6&17.9$\pm$0.1&$-$45.6$\pm$0.1&7.3$\pm$0.1&10.2&12.9&1760&0.56&HD 282972\\
 66517468482370176&03:48:18.03 +23:53:28.6&19.1$\pm$0.1&$-$45.6$\pm$0.1&7.3$\pm$0.1&14.0&HII 1805\\
\hline
66828870789879168&03 48 20.32 +24 54 54.2&21.4$\pm$0.4&$-$43.3$\pm$0.3&8.4$\pm$0.2&16.1&0.5&62&0.99&\\ 
66828870787370624\tablenotemark{2}&03 48 20.30 +24 54 54.7&21.4$\pm$0.6&$-$45.1$\pm$0.5&7.1$\pm$0.3&16.4&&&&V344 Tau\\ 
\hline 
64933759417769984\tablenotemark{1}&03 48 43.92 +23 15 34.6&18.7$\pm$0.1&$-$45.2$\pm$0.1&7.2$\pm$0.1&8.3&5.0&696&0.62&HD 23791\\ 
64933759417767424\tablenotemark{\ddag}&03 48 44.20 +23 15 37.7&16.7$\pm$0.7&$-$42.3$\pm$0.4&7.2$\pm$0.3&17.6&&&&HD 23791 B\\
\hline 
66765855029760768\tablenotemark{2}&03 48 45.37 +24 37 25.6&18.4$\pm$0.3&$-$47.0$\pm$0.2&7.2$\pm$0.2&17.3&9.4&1311&0.76&V463 Tau\\ 
66765855029900288&03 48 44.71 +24 37 22.9&19.9$\pm$0.6&$-$45.1$\pm$0.5&6.8$\pm$0.3&18.6&&&&CFHT 5\\
\hline 
66525061984664832\tablenotemark{\ddag}&03 49 16.82 +24 00 57.9&22.3$\pm$0.4&$-$46.3$\pm$0.2&7.3$\pm$0.2&16.8&1.4&190&0.99&HHJ 308\\ 
66525096340629504&03 49 16.86 +24 00 59.2&17.6$\pm$0.7&$-$45.6$\pm$0.4&7.0$\pm$0.3&17.7&&&&\\ 
\hline 
66507469798631808\tablenotemark{\dag}&03 49 58.08 +23 50 54.5&18.7$\pm$0.2&$-$46.3$\pm$0.1&7.1$\pm$0.1&6.8&3.3&452&0.98&HD 23964 Known\\ 
66507469794885120&03 49 57.88 +23 50 52.6&20.7$\pm$0.2&$-$48.3$\pm$0.2&7.3$\pm$0.1&10.1&&&&HD 23964B\\ 
\hline
66555573432261376&03 50 12.88 +24 21 05.8&18.1$\pm$0.1&$-$44.4$\pm$0.1&7.2$\pm$0.1&14.5&1.9&268&0.99&UGCS J035012.86+242106.2\\ 
66555573432261120\tablenotemark{\ddag}&03 50 13.01 +24 21 06.5&19.1$\pm$0.2&$-$46.8$\pm$0.1&7.0$\pm$0.1&15.3&&&&UGCS J035012.99+242106.9\\ 
\hline  
70403589251055232&03 50 39.72 +26 34 19.7&20.2$\pm$0.2&$-$46.6$\pm$0.2&7.4$\pm$0.1&16.6&2.4&305&0.98& 	UGCS J035039.70+263420.0\\ 
70403589251055488&03 50 39.72 +26 34 17.4&19.4$\pm$0.2&$-$47.2$\pm$0.2&7.7$\pm$0.1&17.0&&&&UGCS J035039.69+263417.7\\ 
\hline 
67368829780861696&03 51 18.88 +26 03 08.2&19.4$\pm$0.2&$-$48.2$\pm$0.1&7.2$\pm$0.1&15.1&7.0&946&0.76&SK 237\\ 
67368834079044224&03 51 18.73 +26 03 14.9&18.9$\pm$0.3&$-$46.2$\pm$0.1&7.4$\pm$0.1&16.9&&&&UGCS J035118.71+260315.4\\ 
\hline 
63826448131945984&03 51 51.92 +21 49 21.2&20.5$\pm$0.2&$-$45.7$\pm$0.1&7.5$\pm$0.1&15.9&6.8&918&0.84&UGCS J035151.89+214921.6\\ 
63826070174824192&03 51 51.58 +21 49 16.3&20.4$\pm$0.2&$-$45.2$\pm$0.1&7.4$\pm$0.1&16.3&&&&UGCS J035151.56+214916.8\\ 
\hline 
66490977124308352\tablenotemark{\dag,\ddag,2}&03 51 53.92 +24 02 51.0&19.6$\pm$0.2&$-$44.8$\pm$0.2&7.1$\pm$0.1&15.7&1.8&249&0.98&V475 Tau\\ 
66490977124308480&03 51 54.04 +24 02 50.4&19.6$\pm$0.3&$-$43.3$\pm$0.2&7.0$\pm$0.2&17.2&&&&\\ 
\hline 
66665902550940672&03 51 58.83 +24 40 03.8&17.7$\pm$0.2&$-$45.0$\pm$0.1&7.3$\pm$0.1&15.6&8.8&1202&0.78&UGCS J035158.82+244004.3\\ 
66665902550940800\tablenotemark{2}&03 51 59.33 +24 39 58.2&17.3$\pm$0.2&$-$44.6$\pm$0.1&7.4$\pm$0.1&15.8&&&&V387 Tau\\ 
\hline 
66642641008348416&03 52 17.56 +24 27 19.2&19.2$\pm$0.2&$-$45.6$\pm$0.2&7.4$\pm$0.1&16.6&1.9&236&0.98&BPL 259\\ 
66642641008348544&03 52 17.66 +24 27 17.8&20.8$\pm$0.7&$-$44.7$\pm$0.5&8.0$\pm$0.4&18.9&&&&\\ 
\hline
65864427293364864&03 55 58.42 +24 32 59.0&18.7$\pm$0.1&$-$46.9$\pm$0.1&7.5$\pm$0.1&12.3&3.4&448&0.93&PELS 115\\ 
65864427293364992\tablenotemark{\ddag}&03 55 58.18 +24 32 59.1&19.0$\pm$0.3&$-$46.8$\pm$0.2&7.3$\pm$0.2&15.8&&&&\\ 
\hline 
67103405099117568&03 56 15.83 +25 29 15.0&18.1$\pm$0.4&$-$43.4$\pm$0.2&7.5$\pm$0.2&17.1&1.5&197&0.99&BPL 335\\ 
67103405096845056&03 56 15.77 +25 29 13.8&17.1$\pm$0.8&$-$45.6$\pm$0.3&7.4$\pm$0.4&18.6&&&&\\ 
\hline 
65437644983224320&03 56 28.14 +23 09 00.2&18.6$\pm$0.1&$-$45.6$\pm$0.1&7.3$\pm$0.1&8.3&7.4&979&0.56&3334\\ 
65437644983223936&03 56 28.65 +23 09 02.9&18.7$\pm$0.3&$-$46.7$\pm$0.1&7.6$\pm$0.1&15.8&&&&UGCS J035628.63+230903.4\\ 
\hline 
66167205308371328\tablenotemark{\ddag}&03 58 56.45 +24 18 31.6&20.2$\pm$0.2&$-$46.8$\pm$0.1&7.6$\pm$0.2&16.4&1.4&184&0.99&\\ 
66167411464494848&03 58 56.35 +24 18 31.7&20.1$\pm$0.3&$-$44.6$\pm$0.3&7.8$\pm$0.3&17.6&&&&HHJ 292\\ 
\hline
50649832062271360&04 01 04.15 +20 22 20.6&18.3$\pm$0.1&$-$44.6$\pm$0.1&7.7$\pm$0.1&14.7&0.9&131&1.00&DH 896 Known Hillenbrand\\
50649832065099008&04 01 04.15 +20 22 21.5&15.7$\pm$0.3&$-$46.7$\pm$0.1&7.2$\pm$0.1&16.1\\
\hline
66289044940711168\tablenotemark{\dag}&04 01 55.87 +24 44 02.0&18.6$\pm$0.3&$-$42.4$\pm$0.1&7.3$\pm$0.1&16.4&2.1&288&0.98&UGCS J040155.85+244402.3\\
66289044938594816&04 01 55.91 +24 43 59.9&17.7$\pm$1.5&$-$42.2$\pm$0.7&6.6$\pm$0.7&19.5&\\
\hline
53892669813156736&04 03 49.58 +23 43 13.3&19.0$\pm$0.2&$-$49.9$\pm$0.2&7.9$\pm$0.2&17.0&11.7&1468&0.70&DH 910\\
53892768594980096&04 03 49.74 +23 43 24.8&19.2$\pm$0.8&$-$49.3$\pm$0.5&7.8$\pm$0.6&19.2&&&&UGCS J040349.73+234325.1\\
\hline
66111886127026944\tablenotemark{\dag}&04 04 45.66 +24 41 19.0&17.6$\pm$0.2&$-$44.5$\pm$0.1&7.1$\pm$0.2&16.5&0.9&125&1.00&DH 912\\
66111886130402944&04 04 45.68 +24 41 18.1&15.9$\pm$0.3&$-$46.4$\pm$0.2&7.1$\pm$0.2&17.3\\
\hline
\multicolumn{10}{c}{Other notable pairs}\\
\hline
66798496781121792&03 45 54.50 +24 33 15.5&20.1$\pm$0.2&$-$46.4$\pm$0.1&7.6$\pm$0.1&5.7&8.8&1168&0.48&Asterope binary in Herschel but likely spurious\\ 
66798526845337344&03 45 55.14 +24 33 16.6&20.1$\pm$0.1&$-$45.5$\pm$0.1&7.4$\pm$0.1&13.8&&&&\\ 
\hline
\hline
\hline
\multicolumn{10}{c}{Praesepe}\\
\hline
\hline
676263049100512384&08 22 51.82 +21 40 10.9&$-$32.7$\pm$0.1&$-$15.0$\pm$0.1&5.0$\pm$0.1&15.6&2.3&459&0.96&2MASS J08225179+2140108\\
676263049100512256&08 22 51.66 +21 40 10.1&$-$31.9$\pm$0.1&$-$14.1$\pm$0.1&5.2$\pm$0.1&15.7\\
\hline
663299429047117952&08 23 27.28 +18 59 58.6&$-30.4\pm0.2$&$-11.4\pm$0.1&4.9$\pm$0.1&16.6&1.2&241&0.98&2MASS J08232733+1859585\\
663299429048377856&08 23 27.33 +18 59 57.6&$-30.64\pm$0.7&$-12.2\pm$0.2&4.7$\pm$0.2&17.7&&&&\\ 
\hline
662917726714872960&08 31 16.89 +19 21 18.8&$-33.8\pm$0.1&$-13.0\pm$0.1&5.6$\pm$0.2&15.6&0.8&135&0.99&HSHJ 14 \\
662917726713395968&08 31 16.90 +19 21 19.5&$-36.7\pm$0.6&$-12.6\pm$0.3&6.4$\pm$0.4&17.5\\
\hline
664625302630788608\tablenotemark{1}&08 35 56.91 +20 49 34.5&$-35.7\pm$0.1&$-14.3\pm$0.1&5.2$\pm$0.2&11.6&1.3&249&0.98&JS 102\\
664625302632292224\tablenotemark{\ddag}&08 35 56.96 +20 49 33.4&$-28.0\pm$0.9&$-13.2\pm$0.4&5.7$\pm$0.5&15.7\\
\hline
664406087501474048&08 36 39.43 +20 22 33.5&$-35.5\pm$0.2&$-15.5\pm$0.2&5.5$\pm$0.2&16.7&0.8&146&0.99&JS 141\\
664406087499336448&08 36 39.44 +20 22 34.3&$-37.0\pm$0.6&$-13.1\pm$0.4&3.9$\pm$0.5&18.3\\
\hline
659248858275152128&08 37 13.80 +17 30 48.5&$-$35.4$\pm$0.1&$-$11.7$\pm$0.1&5.3$\pm$0.1&13.3&2.7&503&0.96&2MASS J08371388+1730487\\ 
659248862571213184&08 37 13.99 +17 30 48.8&$-$34.4$\pm$0.1&$-$11.5$\pm$0.1&5.3$\pm$0.1&14.3&&&&UGCS J083714.00+173048.9\\ 
\hline
664293387559886464\tablenotemark{\dag}&08 38 14.19 +19 47 23.2&$-$35.2$\pm$0.1&$-$13.5$\pm$0.1&5.3$\pm$0.1&14.0&4.2&787&0.91&JC 121\\ 
664293387559886592&08 38 13.89 +19 47 22.8&$-$35.0$\pm$0.1&$-$14.3$\pm$0.1&5.2$\pm$0.1&15.2&&&&UGCS J083813.90+194722.9\\ 
\hline 
661260212936751616&08 38 50.99 +19 18 33.7&$-$36.1$\pm$0.3&$-$11.8$\pm$0.2&4.8$\pm$0.2&17.7&0.9&165&0.99&HSHJ 229\\ 
661260212933886464&08 38 50.97 +19 18 32.9&$-$35.8$\pm$0.4&$-$13.0$\pm$0.2&5.4$\pm$0.3&17.9&&&&\\ 
\hline 
659674614089017984&08 39 23.44 +18 39 59.1&$-$35.6$\pm$0.1&$-$11.7$\pm$0.1&5.3$\pm$0.1&15.2&3.5&660&0.87&JS 159\\ 
659674614089018112&08 39 23.34 +18 39 55.9&$-$35.1$\pm$0.3&$-$12.4$\pm$0.2&5.3$\pm$0.2&17.7&&&&UGCS J083923.35+183956.0\\ 
\hline 
661295461730107392&08 40 12.28 +19 38 22.1&$-$37.5$\pm$0.1&$-$13.2$\pm$0.1&5.8$\pm$0.1&9.8&4.4&756&0.74&BD+20 2160 SB\\ 
661295466028469120\tablenotemark{\ddag}&08 40 12.38 +19 38 17.9&$-$37.2$\pm$0.3&$-$12.1$\pm$0.1&5.5$\pm$0.1&15.9&&&&\\ 
\hline 
661211147230556160&08 40 21.30 +19 10 54.3&$-$36.2$\pm$0.1&$-$12.7$\pm$0.1&5.4$\pm$0.1&14.0&2.5&457&0.97&JC 201\\ 
661211142934329088&08 40 21.26 +19 10 51.9&$-$37.5$\pm$0.1&$-$14.0$\pm$0.1&5.4$\pm$0.1&14.5&&&&UGCS J084021.27+191052.1\\ 
\hline 
661303471844671104&08 40 39.38 +19 42 55.3&$-$35.1$\pm$0.2&$-$12.7$\pm$0.1&5.5$\pm$0.1&16.6&1.3&232&0.99&2MASS J08403942+1942553\\ 
661303471847393664&08 40 39.35 +19 42 54.1&$-$36.3$\pm$0.4&$-$12.1$\pm$0.2&5.4$\pm$0.2&18.3&&&&\\ 
\hline 
661413800965826816&08 40 53.75 +19 59 55.9&$-$36.3$\pm$0.2&$-$13.9$\pm$0.1&5.1$\pm$0.1&16.5&2.1&415&0.95&UGCS J084053.77+195956.1\\ 
661413800965826688&08 40 53.89 +19 59 55.0&$-$36.1$\pm$0.2&$-$12.8$\pm$0.1&5.4$\pm$0.1&16.6&&&&UGCS J084053.91+195955.2\\ 
\hline 
661401431460014976&08 41 09.76 +19 56 07.1&$-$36.5$\pm$0.1&$-$11.8$\pm$0.1&5.3$\pm$0.1&13.2&10.3&1930&0.51&JS 408\\
661401431460014720\tablenotemark{\dag}&08 41 10.48 +19 56 06.5&$-$37.3$\pm$0.1&$-$13.6$\pm$0.1&5.5$\pm$0.1&15.7&&&&UGCS J084110.50+195606.6\\
\hline 
661224581888230400&08 41 18.36 +19 15 39.4&$-36.7\pm$0.1&$-12.7\pm$0.1&5.1$\pm$0.1&8.0&2.2&423&0.99&V* HI Cnc A known\\
661224577591160448&08 41 18.40 +19 15 37.3&$-35.6\pm0.2$&$-12.0\pm$0.1&4.9$\pm$0.1&10.2&&&&V* HI Cnc AB\\
\hline 
660995230632845184&08 41 30.67 +18 52 18.5&$-$36.3$\pm$0.1&$-$12.3$\pm$0.1&5.4$\pm$0.1&12.5&1.7&310&0.97&JC 257 Known Hillenbrand\\ 
660995230632845440\tablenotemark{\ddag}&08 41 30.56 +18 52 17.6&$-$35.6$\pm$0.3&$-$12.2$\pm$0.2&5.5$\pm$0.2&16.2&&&&\\ 
\hline 
664923064123308416\tablenotemark{\ddag,1}&08 42 25.94 +21 13 50.8&$-$35.9$\pm$0.7&$-$14.1$\pm$0.4&5.6$\pm$0.4&16.1&1.2&215&0.98&\\ 
664923064123308672&08 42 26.03 +21 13 50.8&$-$36.9$\pm$0.2&$-$14.1$\pm$0.1&5.5$\pm$0.1&16.6&&&&2MASS J08422601+2113510\\ 
\hline 
661036840275967872&08 43 07.38 +19 14 14.9&$-$38.3$\pm$0.1&$-$13.7$\pm$0.1&5.7$\pm$0.1&15.7&4.2&732&0.88&JS 519\\ 
661036840275967744&08 43 07.41 +19 14 19.0&$-$37.9$\pm$0.2&$-$14.3$\pm$0.1&5.7$\pm$0.1&16.6&&&&UGCS J084307.43+191419.2\\ 
\hline
664909835625008256&08 44 53.83 +21 37 06.3&$-$39.0$\pm$0.1&$-$14.6$\pm$0.1&5.5$\pm$0.1&15.5&2.5&447&0.92&2MASS J08445387+2137065\\ 
664909831331050624&08 44 53.78 +21 37 08.8&$-$39.1$\pm$0.7&$-$16.6$\pm$0.4&5.7$\pm$0.5&18.9&&&&UGCS J084453.80+213708.9\\ 
\hline 
689285355577523584&08:48:56.22 +23:19:56.3&$-39.4\pm$0.1&$-14.9\pm$0.1&5.8$\pm$0.1&14.1&1.0&175&0.99&UCAC4 567-043746\\
689285355576359168&08:48:56.23 +23:19:57.3&$-43.2\pm$0.5&$-16.6\pm$0.4&5.6$\pm$0.3&17.2&\\
\hline 
\tablenotetext{\dag}{Object lies above the cluster main sequence}
\tablenotetext{\ddag}{Object has elevated Gaia astrometric noise}

\tablenotetext{1}{Known binary}
\tablenotetext{2}{Known flare or variable star}
\enddata
\end{deluxetable}%

\clearpage
\subsection{Higher order multiple frequency}
To probe the frequency of higher order multiples we estimated the fraction of binary components flagged as potential multiples using the methods outlined in Section~\ref{un_res}. This is not a perfect process and it is likely that some of the stars we flag as possible multiples are not true multiples but have elevated astrometric noise or lie above the cluster main sequence for some other reason. We also note that our flagging is incomplete as we exclude fainter cluster members from being flagged as overluminous. The brightness condition for this changes between clusters with stars fainter than $G_{abs}=12$\,mag. being excluded from our overluminous flagging for Alpha Per and the Pleiades, with a shallower limit of $G_{abs}=9.5$ for Praesepe. In Alpha~Per we find that for cluster members ($p_{memb}>0.5$) our multiple flagged fraction is $f_{flagged}=11\pm1$\% while for components of wide binaries $p_{bin}>0.5$ it is $f_{flagged}=22\pm^{11}_{7}$\%. For the Pleiades we find $f_{flagged}=12\pm1$\% for cluster members and $f_{flagged}=39\pm^{8}_6$\% for wide binary components. In Praesepe we find $f_{flagged}=9\pm1$\% for cluster members and $f_{flagged}=25\pm^{11}_8$\% for wide binary components. Note that Praesepe will have fewer objects flagged as overluminous due to our shallower flagging limit of $G<9.5$\,mag. In two of the three clusters the components of wide binaries are more likely to be flagged as potential multiples than cluster members in general. Combining the three clusters we have a member flagging rate of 11$\pm$1\% and a binary component flagging rate of 32$\pm^5_4$\%. This difference is more than 4$\sigma$.

\section{Comparison with other populations}
\subsection{Comparison with Young Moving Group members}
Young Moving Groups (YMGs) represent another population of young stars in the solar neighbourhood. They likely represent the result of lower-density, more dispersed formation events than bound clusters like Alpha~Per, the Pleiades or Praesepe. We use these YMGs as an alternative laboratory to study the wide binary fraction. 

We began by assembling a sample of bona fide moving group members defined by \cite{Gagne2018a}  and \cite{Gagne2018}. We restricted ourselves to the following moving groups: AB~Dor, beta~Pic, Carina, Columba, Tucana-Horologium, TW Hydra (TWA) and Carina Near. The first six of these groups were selected as they have ages measured by \cite{Bell2015} who judge them to have well-defined memberships. We also supplement this list with Carina~Near, a $\sim$200\,Myr-old moving group defined by \cite{Zuckerman2006}.  We selected GAIA DR2 counterparts for objects listed in \cite{Gagne2018a}, the additional bona fide members in \citealt{Gagne2018} have their Gaia IDs listed in that paper. We excluded a handful of moving group members that were more distant than 100\,pc. We then used PARSEC models for the appropriate-aged populations (taking \citealt{Bell2015} ages for our first six groups and the \citealt{Zuckerman2006} age for Carina~Near) to generate mass estimates based on absolute Gaia magnitude. This allowed us to identify stars in our sample that were similar to our previous mass-based division of open cluster stars into FGK stars and M stars. This left us with 190 stars which are either single stars or primary stars of multiple systems. Of these, 130 fall in to our $0.5<M/M_{\odot}<1.5$ FGK star mass range and 19 fall in to the $0.1<M/M_{\odot}<0.5$ M dwarf mass range. We exclude a handful of later-type objects from our sample of primaries.

We searched the Gaia archive for binary companions within 10\,arcminutes of each of our primaries. We then used the astrometric difference defined in Equation~\ref{dast_eq} to estimate the significance of the difference in proper motion and parallax and again selected only binaries with $n_{\sigma}<5$. To prevent us being overwhelmed by coincident pairings with unrelated stars with poor astrometric solutions, we excluded any star with a parallax error larger than 0.5\,milliarcseconds. We also excluded any pairing where the primary star met this condition. We list only our selected 300--3000\,AU pairs in the printed version of Table~\ref{ymg_tab}. The online version of this table contains all of our selected YMG binaries.

 \begin{figure*}
 \begin{center}
 \setlength{\unitlength}{1mm}
  \begin{tabular}{cc}
    \includegraphics[scale=0.5]{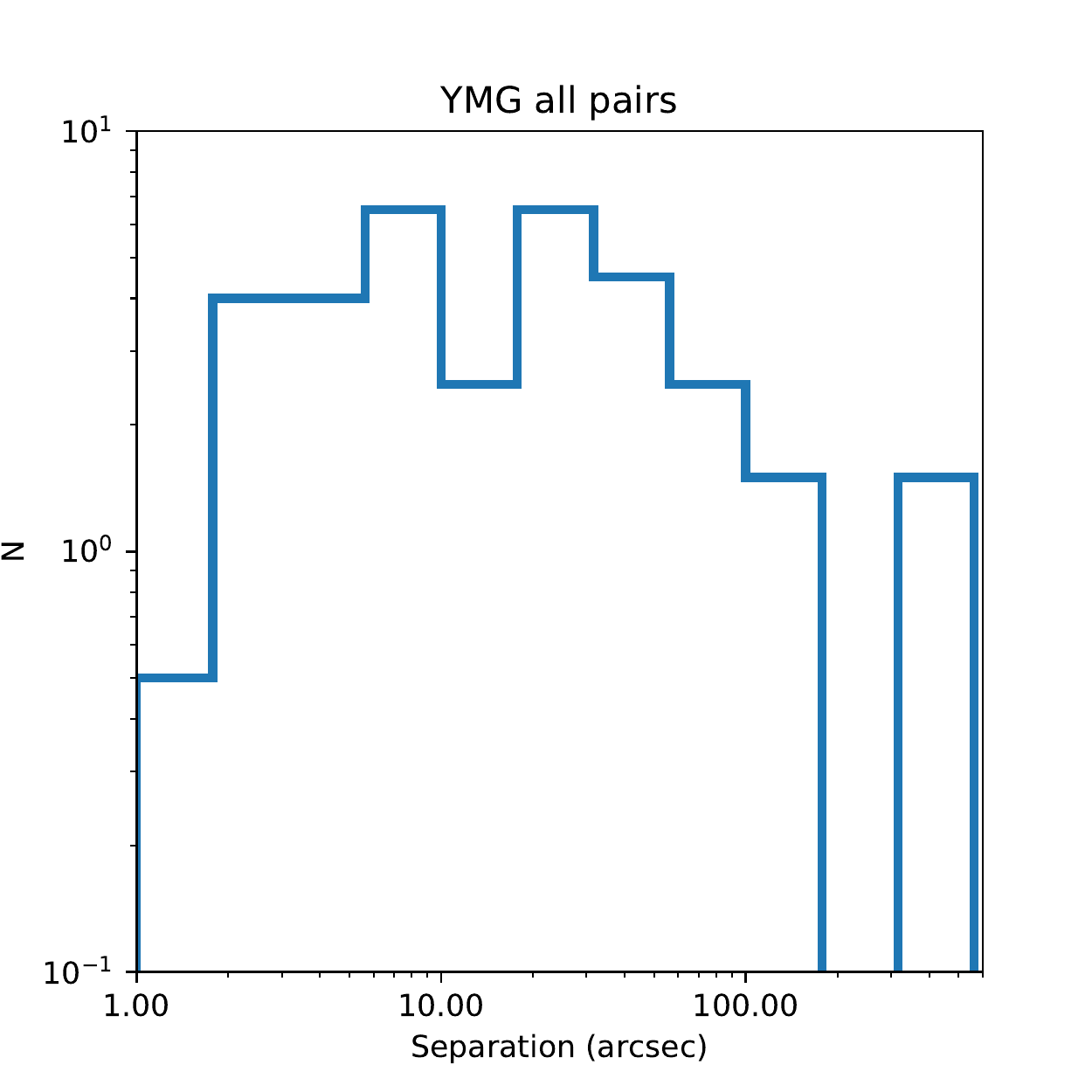}&
  \includegraphics[scale=0.5]{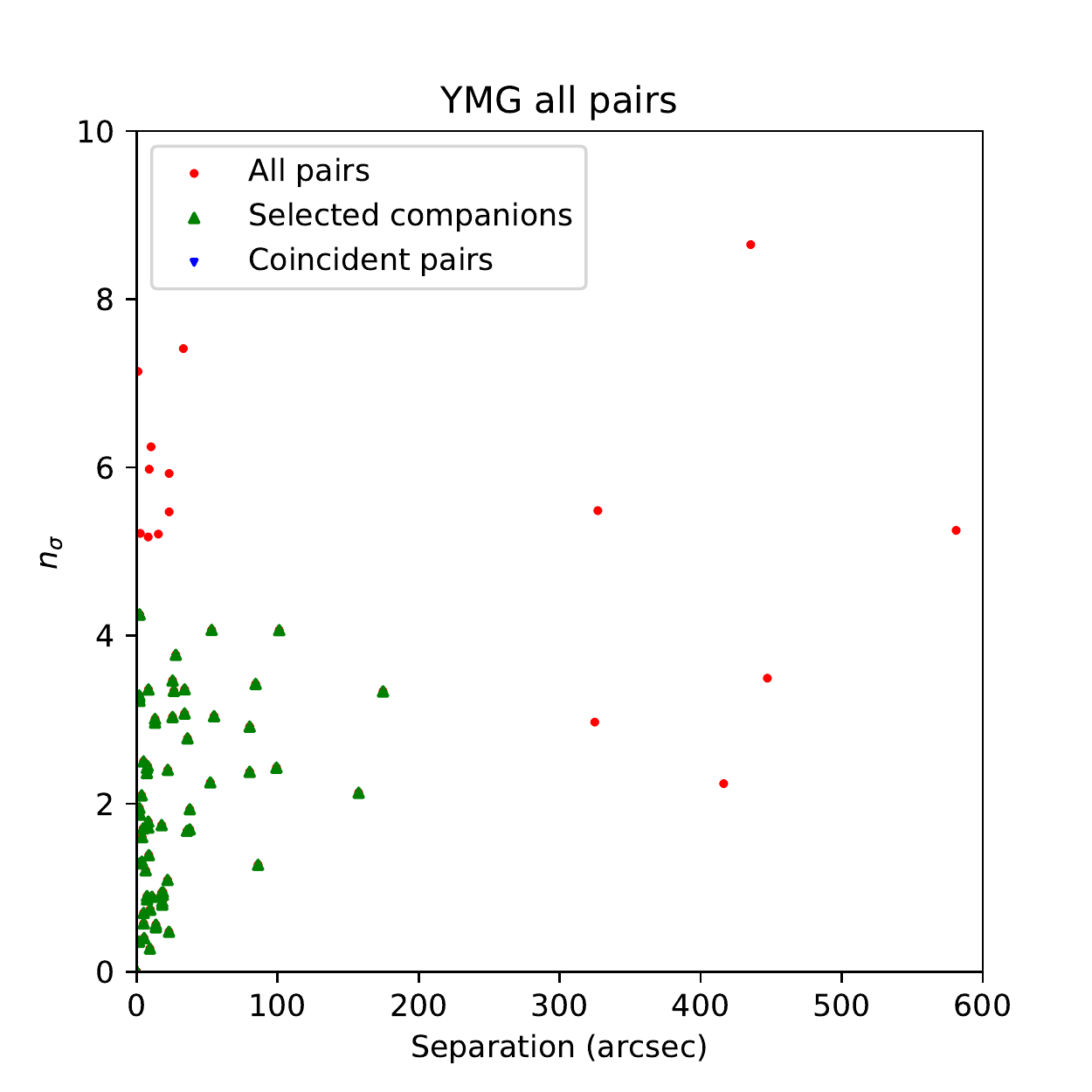}\\
 \includegraphics[scale=0.5]{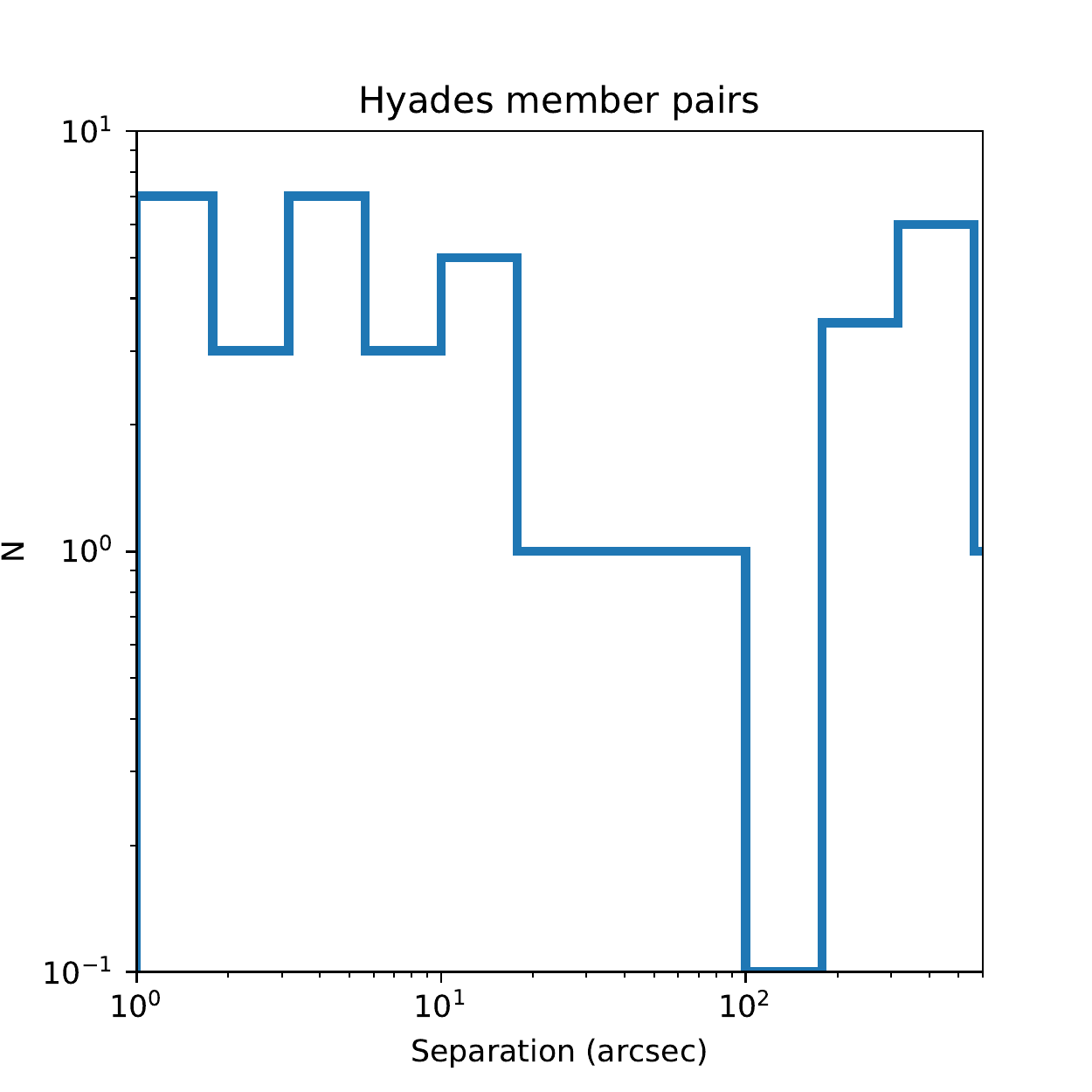}&
 \includegraphics[scale=0.5]{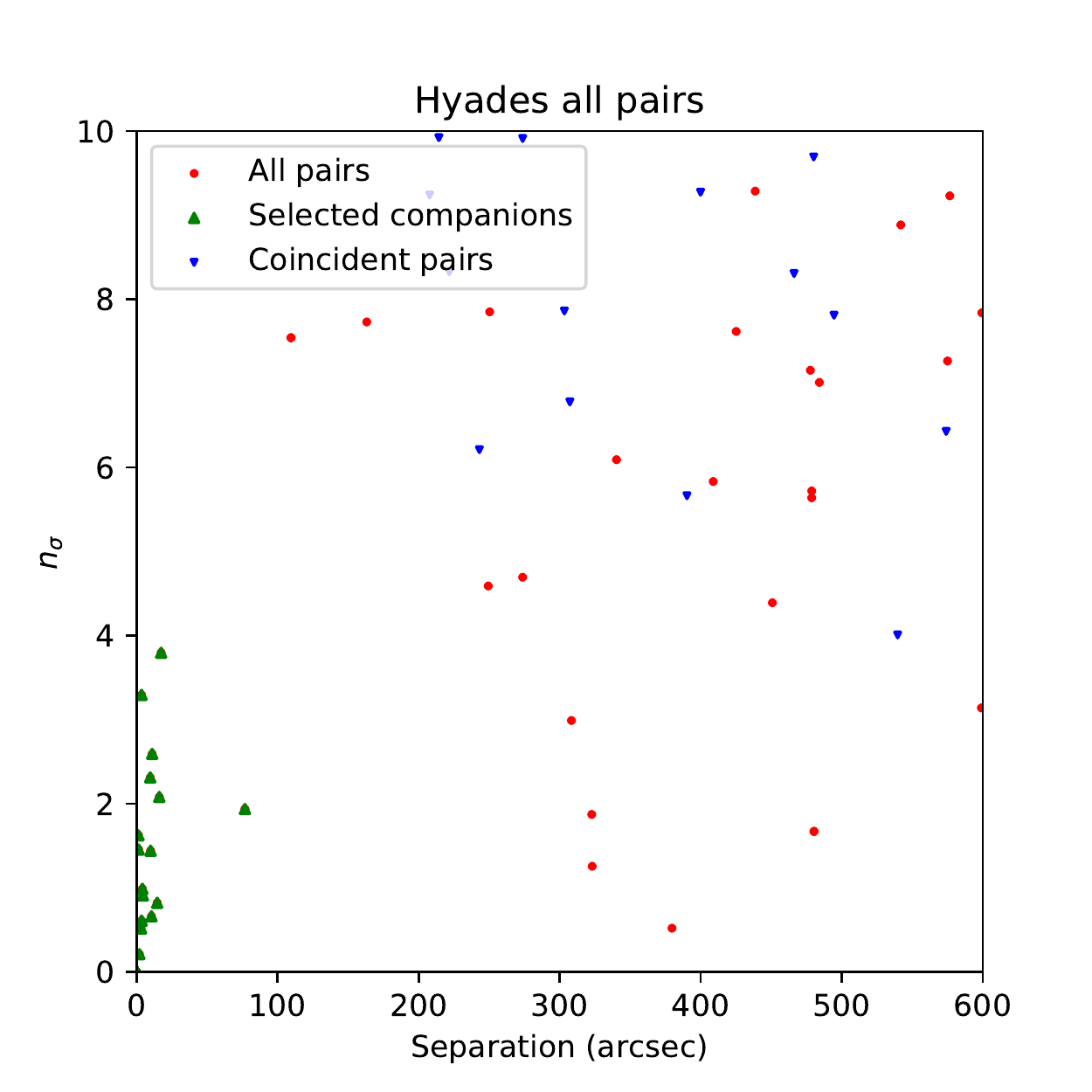}\\
  \includegraphics[scale=0.5]{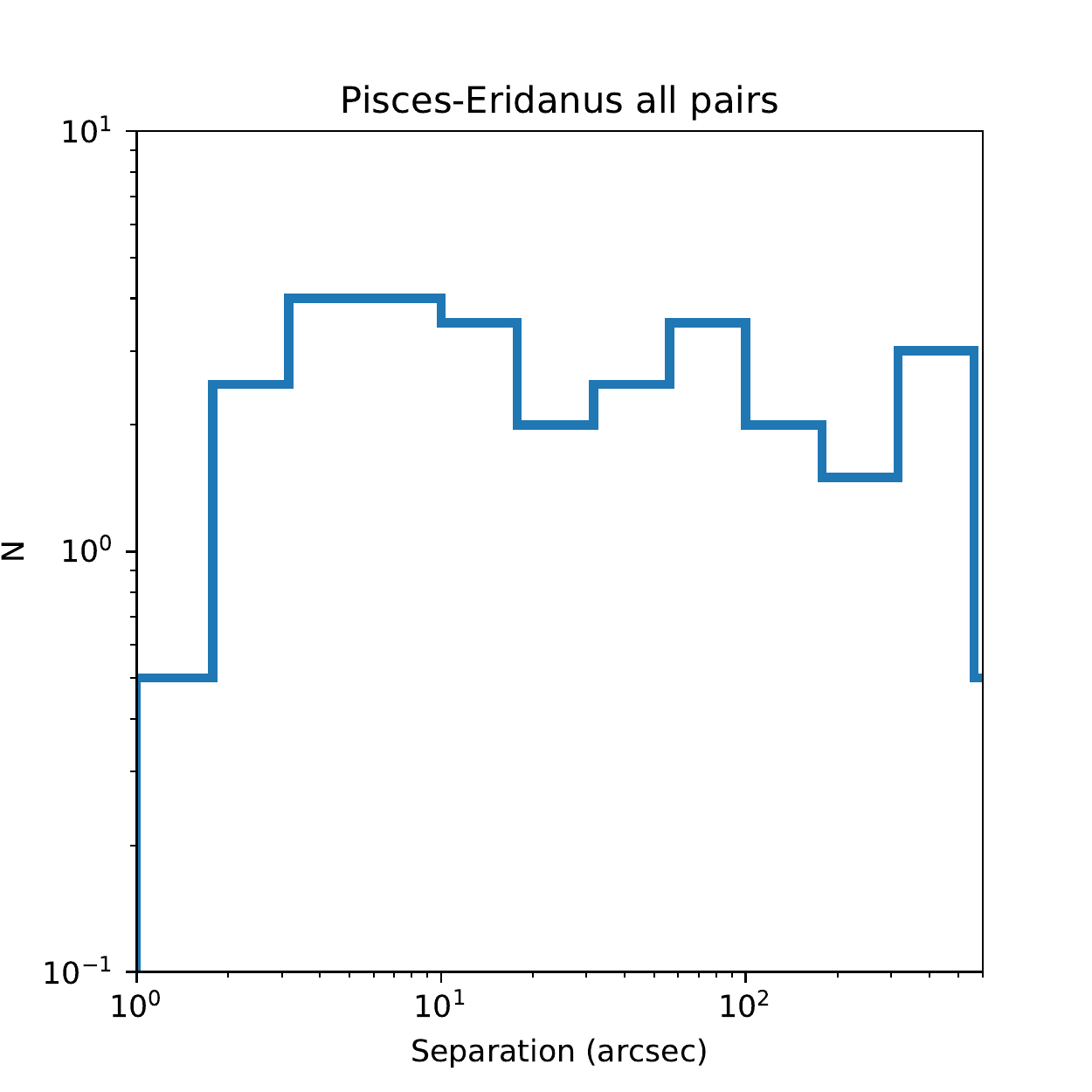}&
 \includegraphics[scale=0.5]{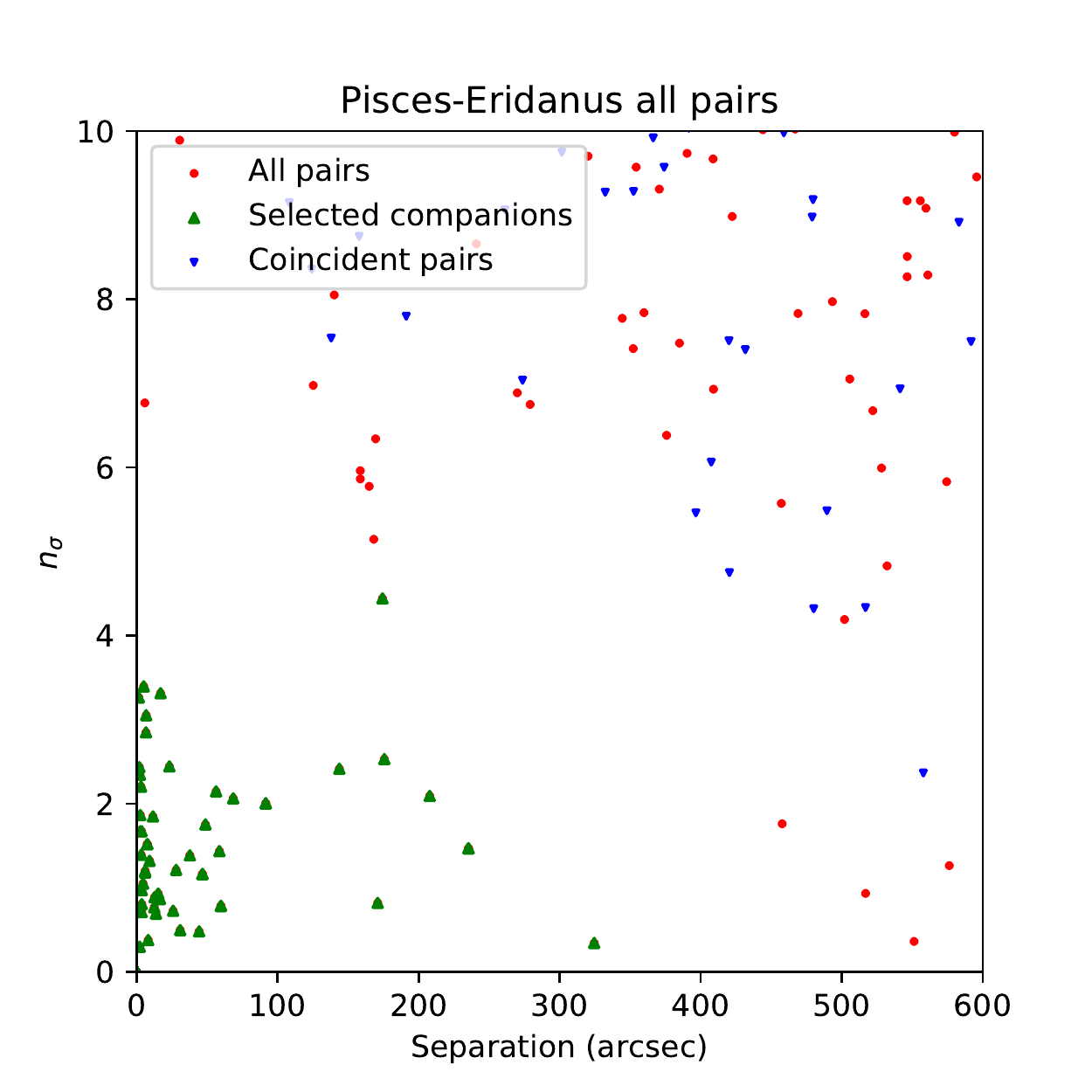}
 \end{tabular}
 \caption[]{The pair separation distribution for our three other clusters/populations. {\bf Top row:} pairs with bona-fide YMG members of \protect\cite{Gagne2018} and \protect\cite{Gagne2018a} {\bf middle row:} pairs with Hyades members from \protect\cite{Lodieu2019} and {\bf bottom row:} pairs with Pisces-Eridanus members from \protect\cite{Meingast2019}, {\bf bottom row:}.
 {\bf Left:} The number of pairs with $ n_{\sigma}<5$. {\bf Right:} All pairs with members from our other clusters/populations plotted alongside an offset pairing populations created using the method of \protect\cite{Lepine2007}. Note the lack of coincident pairs with $ n_{\sigma}<5$ below 200\,arcseconds for the Young Moving Groups, 120\,arcseconds for the Hyades and 400\,arcseconds for Pisces-Eridanus and. In all cases we exclude pairs where either component has a parallax error above half a milliarcsecond.} 
  \label{other_plot}
  \end{center}
 \end{figure*}

\cite{Deacon2016b} suggest that chance alignments between moving group members at similar separations to our search will be relatively rare due to their low density on the sky. To test for chance alignments with field stars we used the offset pairing method of \cite{Lepine2007}. In this we moved the YMG members by two degrees in Right Ascension and re-ran our pairing analysis. This should result in only chance alignments with unrelated stars. As shown in Figure~\ref{other_plot} there are no coincident pairs from the \cite{Lepine2007} offset method with $n_{\sigma}<5$ within 600" and there is no obvious increase in the number of pairs with YMG members with separation below 200". This suggests our binary sample is free of chance alignments for separations under 200". We note a number of close pairs with $5<n_{\sigma}<$10. These all appear to have at least one component with an elevated RUWE value and have higher values of $n_{\sigma}$ due to discrepant parallax values. These could well be binaries where one component is itself a close binary leading to an inaccurate astrometric solution. We include these binaries in our list of pairs but flag them as having higher $n_{\sigma}$ values. While these binaries are likely true, bound systems, we exclude them from our binary frequency analysis as to do otherwise would be inconsistent. As a result our Young Moving Group wide binary fractions are likely underestimates.

We estimated our completeness using the method outlined in Section~\ref{bin_comp}. We also restricted ourselves to pairs wider than two arcseconds as we are relatively complete at such separations. We divided our binary fraction analysis into three separation bins, 30-300\,AU, 300-3000\,AU and 3000-10000\,AU. Our lower and upper separation limits and the varying distances of our target stars meant that we sometimes did not cover all of each of our three projected separation ranges for all stars. In each projected separation range we then summed the product of the completeness and the log-separation range covered for each star. This was then the denominator for our binary fraction calculations. Our numerator was the number of companions found in each projected separation range. As before we calculated the binary fractions for all stars, FGK stars and M dwarfs. The results are shown in Table~\ref{other_calc_table}. We find that in the 300--3000\,AU projected separation range, FGK stars in young moving groups have a binary fraction of 14.6$\pm^{4.4}_{3.5}$\%, higher than we found in our open clusters. Our M dwarf primary sample is very small so while we find very high wide binary fractions for Young Moving Groups, these binary fractions are highly uncertain.

\subsection{Comparison with the Hyades}
In our work on open clusters we have selected members of three open clusters and have looked for wide binary companions in all three. One additional check we can make is by comparing with the sample in the Hyades, cluster similar in age to Praesepe (750\,Myr; \citealt{Brandt2015a}) using members identified by \cite{Lodieu2019}. While this work relies on the same Gaia dataset as our studies of our three open clusters, it uses a different statistical method, the method used in our YMG search above, for identifying companions. The Hyades is also a much closer cluster ($\sim$47\,pc) rather than the larger distances ($>$130\,pc) for our three other clusters. This means that any bias against close pairs which we have for some reason missed will likely manifest itself in binaries with smaller projected separations than it would in our three clusters.

We searched the Gaia archive for co-moving companions within 600\,arcseconds of \cite{Lodieu2019}'s Hyades members, following the authors' suggestions and restricting ourselves to objects within 30\,pc of the cluster centre. As with our YMG companion search, we excluded objects with parallax uncertainties greater than half a milliarcsecond. There will be two possible sources of contamination for Hyades binaries, chance alignments of cluster members and chance alignments with field stars. To estimate the contamination from chance alignments with other Hyades members we paired Hyades members from \cite{Lodieu2019} together and applied our $n_{\sigma}<5$ cut. A histogram of these pairings is shown in the left-hand panel of Figure~\ref{other_plot}. Chance alignments with other cluster members should increase with separation, just like in our three open clusters. We do not see any clear ramp-up in numbers below 120\,arcseconds. This indicates this region is free of chance alignments of cluster members. 

We repeated the chance alignment with field star analysis from the previous section. We found that out to pairing distances of eight arcminutes there were no chance alignments with $n_{\sigma}<5$. Hence we believe a sample of pairs that meet these criteria will be relatively free of contamination. The middle-right-hand panel of Figure~\ref{other_plot} shows our selected companions along with the population of chance alignments. There is a clear distinction in this diagram between our pairs and the chance alignments.

 We identified 37 wide binary systems in the Hyades that have separations less than 120\,arcseconds and $n_{\sigma}<5$. Of those, five have primaries in our FGK mass range and have projected separations in the 300-3000\,AU range. We list these pairs in Table~\ref{hyades_pairs} with our full list of pairs available in the electronic version of the table. 
 
 We estimated our completeness using the method outlined in Section~\ref{bin_comp} using the mass-absolute magnitude relation used by \cite{Lodieu2019} for Hyades members. We also restricted our analysis to pairs wider than two arcseconds as our census is relatively complete at such separations. However we note that there are 12 binary pairs with separations below our two arcsecond lower limit, the majority of which are M dwarf primaries. We divide our binary fraction analysis into two separation bins, 30-300\,AU and 300-3000\,AU. We do not detect any binaries with FGK or M primaries in the  3000-10000\,AU range (although we have one binary with an A star primary). Hence we measure binary fractions of zero in this range. We have no binaries wider than 3500\,AU, suggesting binaries wider than this are rare or lacking in the Hyades. We followed the same process outlined in the previous section and calculated the binary fractions for FGK stars and M dwarfs. The results are shown in Table~\ref{other_calc_table}. The binary fractions of both FGK and M stars in the 300-3000\,AU range agree well with the number found in our three open clusters.

 \clearpage
\begin{deluxetable}{lcrrrrrcl}
 \tablecolumns{10}
 \tablewidth{0pc}
\tablecaption{\label{ymg_tab} The multiple systems found for Young Moving Group members from \protect\cite{Gagne2018} and \protect\cite{Gagne2018a}. Here we list only the objects with primaries which fell into our FGK mass range, had $n_{\sigma}<5$ and projected separations in the 300-3000\,AU range with projected separations greater than two arcseconds.}
\tabletypesize{\tiny}
\tablehead{
\colhead{Gaia ID}&\colhead{Position}&\colhead{$\mu_{\alpha}\cos{\delta}$}&\colhead{$\mu_{\delta}$}&\colhead{$\pi$}&\colhead{$G$}&\multicolumn{2}{c}{Projected}&\colhead{Other name}\\
&&&&&&\multicolumn{2}{c}{Separation}\\
&\colhead{(Eq.=J2000 Ep.=2015.5)}&\colhead{(mas/yr)}&\colhead{(mas/yr)}&\colhead{(mas)}&\colhead{(mag)}&\colhead{('')}&\colhead{(AU)}
}
\startdata
\hline
\multicolumn{9}{c}{AB Dor}\\

\hline
374400957846408192\tablenotemark{\ddag,1} & 01:03:40.30 +40:51:26.7 & 126.9$\pm$0.1&-161.3$\pm$0.1 & 32.2$\pm$0.1 & 10.1 & 27.9 & 864.9&G 132-50 \\
374400893423204992&01:03:42.23 +40:51:13.5&132.9$\pm0.1$&$-$155.9$\pm$0.01&32.5$\pm$0.1&12.3&&&G 132-51 B\\
 374400893422315648\tablenotemark{\ddag} & 01:03:42.45 +40:51:13.2 & 130.4$\pm$0.2 & -162.5$\pm$0.2 & 32.7$\pm$0.1 & 13.2&&&G 132-51 A\\
\hline
248004472671281664\tablenotemark{1}& 03:33:13.59 +46:15:23.8 & 68.6$\pm$0.1&-175.3$\pm$0.1 & 27.5$\pm$0.1 & 8.0 & 9.5 & 346.1&HD 21845 \\
 248004472671505152 & 03:33:14.15 +46:15:16.3 & 69.0$\pm$0.1 & -172.6$\pm$0.1 & 27.5$\pm$0.1 & 10.5&&&HD 21845 B\\
\hline
4795598309045006208\tablenotemark{1}& 05:36:56.89 -47:57:52.9 & 23.3$\pm$0.1&-1.1$\pm$0.1 & 40.6$\pm$0.1 & 7.5 & 18.3 & 449.6&UY Pic\\
 4795596831576255488 & 05:36:55.14 -47:57:47.9 & 28.7$\pm$0.1 & 3.4$\pm$0.1 & 40.6$\pm$0.1 & 9.3&&&HIP 26369\\
\hline
4802947444765743104\tablenotemark{1}&05:37:12.93 -42:42:55.8 & 8.4$\pm$0.1&-10.3$\pm$0.1 & 12.5$\pm$0.1 & 9.5 & 4.0 & 320.8&HD 37551A\\
 4802947444765742848 & 05:37:13.26 -42:42:57.5 & 11.9$\pm$0.1 & -8.6$\pm$0.1 & 12.4$\pm$0.1 & 10.3&&&HD 37551B\\
\hline
\multicolumn{9}{c}{beta Pic}\\
\hline
107774202769886848\tablenotemark{1}& 02:17:25.39 +28:44:41.0 & 87.1$\pm$0.1&-74.1$\pm$0.1 & 25.1$\pm$0.1 & 6.9 & 13.8 & 547.2&HD 14082 \\
 107774198474602368 & 02:17:24.84 +28:44:29.3 & 86.0$\pm$0.1 & -71.1$\pm$0.1 & 25.2$\pm$0.1 & 7.6&&&HD 14082B\\
\hline
132362959259196032\tablenotemark{1}& 02:27:29.35 +30:58:23.5 & 79.5$\pm$0.1&-72.0$\pm$0.1 & 24.4$\pm$0.1 & 9.7 & 22.0 & 903.2&BD+30 397\\
 132363027978672000 & 02:27:28.15 +30:58:39.2 & 82.7$\pm$0.1 & -73.5$\pm$0.1 & 24.4$\pm$0.1 & 11.4&&&BD+30 397\\
\hline
5935776714456619008\tablenotemark{\ddag} & 16:57:20.24 -53:43:32.9 & -21.0$\pm$0.2&-84.1$\pm$0.1 & 19.8$\pm$0.1 & 11.3 & 11.0 & 555.4&TYC 8726-1327-1\\
 5935776710115544832 & 16:57:21.41 -53:43:29.2 & -16.2$\pm$0.2 & -85.7$\pm$0.1 & 19.8$\pm$0.1 & 15.9&&&2MASS J16572144-5343277\\
\hline 
5811866422581688320\tablenotemark{1}& 17:17:25.45 -66:57:05.9 & -21.5$\pm$0.1&-137.1$\pm$0.1 & 32.8$\pm$0.1 & 6.4 & 34.1 & 1040.7&HD 155555 A\\
 5811866358170877184 & 17:17:31.25 -66:57:07.7 & -14.8$\pm$0.1 & -145.1$\pm$0.1 & 33.0$\pm$0.1 & 11.4&&&HD 155555 C\\
\hline
6702775135228913280\tablenotemark{1}& 18:03:03.41 -51:38:57.8 & 2.3$\pm$0.1&-86.1$\pm$0.1 & 20.2$\pm$0.1 & 6.9 & 6.5 & 322.1&HD 164249 A\\
 6702775508886369408 & 18:03:04.11 -51:38:57.7 & 5.8$\pm$0.1 & -95.0$\pm$0.1 & 20.2$\pm$0.1 & 12.3&&&HD 164249 B\\
\hline
 6400161703868444800\tablenotemark{\ddag,1} & 21:21:24.75 -66:54:58.9 & 95.7$\pm$0.1 & -100.3$\pm$0.1 & 31.3$\pm$0.1 & 8.7& 26.5 & 836.1&V390 Pav\\
6400160947954197888\tablenotemark{\ddag} & 21:21:28.99 -66:55:07.8 & 116.0$\pm$0.7&-85.3$\pm$0.8 & 31.7$\pm$0.4 & 10.0&&&TYC 9114-1267-1 \\
\hline
\multicolumn{9}{c}{Carina Near}\\
\hline
  3466924200065405184\tablenotemark{1}& 11:56:42.10 -32:16:05.5 & -172.0$\pm$0.1&-8.2$\pm$0.1 & 28.2$\pm$0.1 & 7.5 & 18.7 & 664.2&HD 103743\\
 3466924200065405824 & 11:56:43.56 -32:16:02.8 & -178.9$\pm$0.1 & -6.7$\pm$0.1 & 28.1$\pm$0.1 & 7.6&&&HD 103742\\
\hline
 1541667932396172800\tablenotemark{1}& 12:28:04.18 +44:47:39.4 & -181.8$\pm$0.1&-4.7$\pm$0.1 & 22.0$\pm$0.1 & 7.3 & 9.7 & 442.3&HD 108574\\
 1541667932396172416 & 12:28:04.54 +44:47:30.5 & -180.4$\pm$0.1 & 0.4$\pm$0.1 & 21.9$\pm$0.1 & 7.9&&&HD 108575\\
\hline
\multicolumn{9}{c}{Tuc Hor}\\
\hline
4714764481913306496\tablenotemark{1}&02:07:26.31 $-$59:40:46.2&92.7$\pm$0.1&$-$18.3$\pm$0.01&21.9$\pm$0.1&7.4&52.3&2387&HD 13246\\
4714764447553568640&02:07:32.40 $-$59:40:21.4&93.6$\pm$0.1&$-$21.7$\pm$0.1&22.0$\pm$0.1&9.9&&&CD-60 416\\
\hline
4742040410461492096\tablenotemark{\ddag}&02:41:47.00 $-$52:59:52.6&97.8$\pm$0.1&$-$14.2$\pm$0.1&22.8$\pm$0.1&9.7&22.2&972&CD-53 544\\
4742040513540707072\tablenotemark{\ddag}&02:41:47.47 $-$52:59:30.8&93.5$\pm$0.1&$-$11.6$\pm$0.2&23.0$\pm$0.1&11.1&&&AF Hor\\
\hline
4842275841819363200\tablenotemark{1}&04:00:32.08 $-$41:44:54.4&68.2$\pm$0.1&$-$7.0$\pm$0.1&19.2$\pm$0.1&8.2&8.8&456&HD 25402 A\\
4842275837523665664&04:00:32.35 $-$41:45:02.6&71.7$\pm$0.1&$-$0.4$\pm$0.1&19.19$\pm$0.1&12.3&&&HD 25402 B\\
\hline
4891725758804030208\tablenotemark{1}&04:38:44.01 $-$27:02:02.0&56.3$\pm$0.1&$-$10.9$\pm$0.1&18.3$\pm$0.1&8.3&23.0&1255&HD 29615 A\\
4891725758804028672&04:38:45.73 $-$27:02:02.2&56.8$\pm$0.1&$-11.7\pm$0.1&18.4$\pm$0.1&14.4&&&HD 29615 B\\
\hline
\multicolumn{9}{c}{TWA}\\
\hline
5399220743767211776\tablenotemark{1}& 11:21:17.13 -34:46:45.8 & -66.0$\pm$0.1&-18.1$\pm$0.1 & 16.7$\pm$0.1 & 10.9 & 5.1 & 303.5&CD-34 7390 A\\
 5399220743767211264 & 11:21:17.36 -34:46:50.0 & -69.0$\pm$0.1 & -16.9$\pm$0.1 & 16.7$\pm$0.1 & 10.9&&&CD-34 7390 B\\
\hline
\enddata

\tablenotetext{\ddag}{Object has elevated Gaia astrometric noise}
\tablenotetext{1}{Known binary}
\end{deluxetable}%

\begin{table*}
\caption{\label{other_calc_table} The binary fractions for different types of stars in our six young moving  groups, the Hyades and the Pisces-Eridanus stream. We use uncertainty estimates calculated with the relations of \protect\cite{Gehrels1986} as these are more appropriate for small number statistics. In each case we estimate the binary fraction divided by the log separation range covered $\frac{df_{bin}}{d\log_{10}x}$. Our two subranges cover approximately FGK dwarfs ($0.5<M/M_{\odot}<1.5$) and M dwarfs ($M<0.5M_{\odot}$). For Pisces Eridanus we only cover binaries with primaries that are approximately FGK dwarfs ($0.5<M/M_{\odot}<1.5$) as there are no lower-mass objects in the membership list of \protect\cite{Meingast2019}. The x-axis error bars represent the projected separation range covered.}
\begin{center}
\begin{tabular}{lrrrr}
\hline
&$N_{companions}$&$N_{stars}$&$\sum Completeness \times d\log r$&$\frac{df_{bin}}{d\log_{10}x}$\\
\hline
\hline
\multicolumn{5}{c}{Young Moving Groups}\\
\hline
\hline
\multicolumn{5}{c}{FGK stars}\\
\hline
30-300\,AU&6&130&54.6&11.0$\pm^{6.7}_{4.8}$\%\\
300-3000\,AU&17&130&116.5&14.6$\pm^{4.4}_{3.5}$\%\\
3000-10000\,AU&1&130&49.5&2.0$\pm^{4.9}_{1.7}$\%\\
\hline
\multicolumn{5}{c}{M stars}\\
\hline
30-300\,AU&4&19&10.1&39.6$\pm^{32.0}_{19.2}$\%\\
300-3000\,AU&3&19&12.4&24.2$\pm^{23.7}_{13.4}$\%\\
3000-10000\,AU&1&19&2.9&34.5$\pm^{65.5}_{30.0}$\%\\
\hline
\hline
\multicolumn{5}{c}{Hyades}\\
\hline
\hline
\multicolumn{5}{c}{FGK stars}\\
\hline
30-300\,AU&7&196&66.6&10.5$\pm^{5.7}_{3.9}$\%\\
300-3000\,AU&5&196&176.6&2.5$\pm^{2.0}_{1.2}$\%\\
3000-20000\,AU&0&196&98.4&0.0$\pm^{2.0}_{0.0}$\%\\
\hline
\multicolumn{5}{c}{M stars}\\
\hline
30-300\,AU&8&390&94.0&8.5$\pm^{4.2}_{3.0}$\%\\
300-3000\,AU&3&390&232.4&1.3$\pm^{1.3}_{0.7}$\%\\
3000-20000\,AU&0&390&134.2&0.0$\pm^{1.5}_{0.0}$\%\\
\hline
\hline
\multicolumn{5}{c}{Pisces-Eridanus}\\
\hline
\hline
\multicolumn{5}{c}{FGK stars}\\
\hline
300-3000\,AU&22&249&204.9&10.7$\pm^{2.8}_{2.3}$\%\\
3000-20000\,AU&11&249&183.6&6.0$\pm^{2.4}_{1.8}$\%\\
\hline
\end{tabular}
\end{center}
\end{table*}

  \clearpage
  \begin{deluxetable}{lcrrrrrc}
 \tablecolumns{10}
 \tablewidth{0pc}
\tablecaption{\label{hyades_pairs} Wide companions to Hyades members listed in \protect\cite{Lodieu2019}. Here we list only the companions where the primaries have mass estimates in the $0.5<M/M_{\odot}<1.5$ range and with projected separations in the 300-3000\,AU range with projected separations greater than two arcseconds. A full version of this table with all pairs will be available electronically.}
\tabletypesize{\tiny}
\tablehead{
\colhead{Gaia ID}&\colhead{Position}&\colhead{$\mu_{\alpha}\cos{\delta}$}&\colhead{$\mu_{\delta}$}&\colhead{$\pi$}&\colhead{$G$}&\multicolumn{2}{c}{Projected}\\
&&&&&&\multicolumn{2}{c}{Separation}\\
&\colhead{(Eq.=J2000 Ep.=2015.5)}&\colhead{(mas/yr)}&\colhead{(mas/yr)}&\colhead{(mas)}&\colhead{(mag)}&\colhead{('')}&\colhead{(AU)}
}
\startdata
\hline
108421608959951488\tablenotemark{1}&02:58:05.48 +20:40:07.0&234.0$\pm$0.1&$-$31.3$\pm$0.1&30.6$\pm$0.1&5.7&14.7&479.0\\
108421402801188864&02:58:06.45 +20:40:01.3&238.1$\pm$0.2&$-$24.5$\pm$0.2&30.6$\pm$0.1&13.5\\
\hline
125343573948444800\tablenotemark{1}&03:13:03.02 +32:53:46.3&184.5$\pm$0.1&$-$61.9$\pm$0.1&25.3$\pm$0.1&8.0&10.6&419.2\\
125343608307015296&03:13:03.83 +32:53:49.3&186.2$\pm$0.1&$-$65.6$\pm$0.1&25.4$\pm$0.1&12.9\\
\hline
56447041481918720\tablenotemark{1}&03:37:33.53 +17:51:14.1&170.6$\pm$0.1&$-$30.2$\pm$0.1&25.8$\pm$0.1&11.7&16.2&626.3\\
56447037184877824&03:37:34.09 +17:51:00.0&170.5$\pm$0.1&$-$27.9$\pm$0.1&25.6$\pm$0.1&12.1\\
\hline
50756411675761920&03:49:06.16 +18:50:11.6&104.5$\pm$0.1&$-$20.9$\pm$0.1&16.4$\pm$0.1&12.2&9.7&595.4\\
50756415973201152&03:49:05.54 +18:50:15.8&106.2$\pm$0.1&$-$23.1$\pm$0.1&16.6$\pm$0.1&14.3\\
\hline
145293181643038336\tablenotemark{\ddag}&04:26:18.61 +21:28:13.0&103.2$\pm$0.1&$-$38.4$\pm$0.1&20.7$\pm$0.1&7.2&11.1&536.9\\
145293181643038208&04:26:19.31 +21:28:07.5&105.8$\pm$0.2&$-$40.3$\pm$0.1&20.9$\pm$0.1&15.1\\
\hline
\enddata 
\tablenotetext{1}{Known binary}
\tablenotetext{\ddag}{Object has elevated Gaia astrometric noise}
\end{deluxetable}%
  \begin{deluxetable}{lcrrrrrc}
 \tablecolumns{10}
 \tablewidth{0pc}
\tablecaption{\label{pe_pairs} Wide companions to Pisces-Eridanus stream members listed in \protect\cite{Meingast2019}. Here we only list objects in the 300-3000\,AU projected separation range with projected separations greater than two arcseconds. A full version of this table with all pairs will be available electronically.}
\tabletypesize{\tiny}
\tablehead{
\colhead{Gaia ID}&\colhead{Position}&\colhead{$\mu_{\alpha}\cos{\delta}$}&\colhead{$\mu_{\delta}$}&\colhead{$\pi$}&\colhead{$G$}&\multicolumn{2}{c}{Projected}\\
&&&&&&\multicolumn{2}{c}{Separation}\\
&\colhead{(Eq.=J2000 Ep.=2015.5)}&\colhead{(mas/yr)}&\colhead{(mas/yr)}&\colhead{(mas)}&\colhead{(mag)}&\colhead{('')}&\colhead{(AU)}
}
\startdata
\hline
2741273390752119680&00 18 42.99 +05 06 35.6&20.0$\pm$0.1&$-$15.8$\pm$0.0&8.3$\pm$0.0&13.3&2.5&301\\
2741273395049220992&00 18 43.13 +05 06 34.3&19.5$\pm$0.1&$-$16.1$\pm$0.1&8.4$\pm$0.0&14.2\\
\hline
2799470060173799680&00 23 04.65 +21 27 50.9&13.1$\pm$0.1&$-$18.1$\pm$0.1&5.9$\pm$0.1&12.4&3.6&615\\
2799470060175375872&00 23 04.44 +21 27 52.9&13.3$\pm$0.2&$-$18.0$\pm$0.2&6.0$\pm$0.1&16.7\\
\hline
2528797033588184960&00 44 19.62 $-$04 23 56.9&20.2$\pm$0.1&$-$11.1$\pm$0.1&10.1$\pm$0.1&8.7&8.3&820\\
2528796277673941376&00 44 20.05 $-$04 24 02.1&20.9$\pm$0.2&$-$10.8$\pm$0.1&10.1$\pm$0.1&15.1\\
\hline
4975223840046231424&00 47 38.48 $-$47 41 45.8&24.6$\pm$0.0&22.4$\pm$0.0&12.4$\pm$0.0&11.5&9.3&746\\
4975223840048745472&00 47 39.36 $-$47 41 48.6&24.0$\pm$0.1&22.4$\pm$0.1&12.3$\pm$0.1&15.2\\
\hline
2577307864562003456&01 02 29.70 +06 59 54.5&14.1$\pm$0.1&$-$17.3$\pm$0.1&9.8$\pm$0.1&9.8&12.6&1280\\
2577307860266860800&01 02 30.41 +07 00 01.3&15.0$\pm$0.5&$-$16.4$\pm$0.2&10.0$\pm$0.2&17.9\\
\hline
2590688680553786368&01 24 29.20 +15 30 36.2&13.3$\pm$0.1&$-$27.1$\pm$0.1&9.8$\pm$0.0&10.2&3.6&366\\
2590688680553786240\tablenotemark{\ddag}&01 24 29.27 +15 30 39.7&19.4$\pm$0.4&$-$25.6$\pm$0.1&9.8$\pm$0.1&15.1\\
\hline
2508964107269556864&01 42 37.20 $-$00 47 50.9&11.0$\pm$0.1&$-$12.2$\pm$0.1&8.9$\pm$0.1&12.2&4.7&522\\
2508964111565146752\tablenotemark{\ddag}&01 42 37.02 $-$00 47 54.7&11.8$\pm$0.5&$-$15.0$\pm$0.3&9.1$\pm$0.3&17.1\\
\hline
5141461678818746880&01 59 17.21 $-$17 46 38.0&10.0$\pm$0.1&$-$0.4$\pm$0.1&11.5$\pm$0.0&12.8&7.7&667\\
5141461683110761728&01 59 17.45 $-$17 46 44.8&9.7$\pm$0.0&$-$0.6$\pm$0.0&11.4$\pm$0.0&13.4\\
\hline
2462582789799726336&01 59 54.07 $-$09 30 16.6&7.9$\pm$0.1&$-$5.7$\pm$0.1&8.0$\pm$0.1&10.8&6.0&748\\
2462582794095085056&01 59 53.73 $-$09 30 19.6&1.4$\pm$3.3&$-$5.3$\pm$2.7&1.9$\pm$1.8&20.4\\ 
\hline
24667616384335872&02 27 47.44 +11 14 27.3&3.6$\pm$0.1&$-$19.5$\pm$0.1&7.6$\pm$0.0&10.7&15.4&2036\\
24667616384335744&02 27 48.49 +11 14 26.5&4.1$\pm$0.1&$-$20.0$\pm$0.1&7.6$\pm$0.1&15.1\\
\hline
5155187986271622912&03 20 33.29 $-$14 16 58.4&$-$0.7$\pm$0.1&$-$3.0$\pm$0.1&7.5$\pm$0.1&12.3&2.4&323\\
5155187986271622784\tablenotemark{\ddag}&03 20 33.17 $-$14 16 56.7&$-$1.7$\pm$0.1&$-$2.1$\pm$0.1&7.3$\pm$0.1&13.4\\
\hline
5085954625291316224&03 30 22.88 $-$24 22 26.1&$-$0.0$\pm$0.0&2.6$\pm$0.0&8.5$\pm$0.0&12.1&3.7&441\\
5085954625291316096&03 30 22.74 $-$24 22 22.8&$-$1.0$\pm$0.1&2.4$\pm$0.1&8.5$\pm$0.1&15.1\\
\hline
3193559873957065472&03 55 08.93 $-$10 10 11.8&$-$3.5$\pm$0.1&$-$1.3$\pm$0.1&5.6$\pm$0.0&9.4&12.8&2300\\
3193559839597327360&03 55 08.08 $-$10 10 14.7&$-$3.5$\pm$0.1&$-$1.4$\pm$0.1&5.7$\pm$0.1&16.6\\
\hline
5092558979319804672&04 13 00.24 $-$19 51 17.2&$-$6.1$\pm$0.1&2.5$\pm$0.1&5.9$\pm$0.0&9.1&17.1&2874\\
5092558979319804928&04 12 59.03 $-$19 51 18.6&$-$6.2$\pm$0.2&3.0$\pm$0.2&6.5$\pm$0.2&17.0\\
\hline
3205573756476323328&04 23 54.59 $-$02 33 43.4&$-$7.1$\pm$0.1&$-$5.9$\pm$0.0&6.4$\pm$0.0&11.6&16.7&2604\\
3205573756476257792&04 23 54.71 $-$02 33 59.9&$-$6.7$\pm$0.2&$-$5.6$\pm$0.1&6.5$\pm$0.1&17.1\\
\hline
3198734278756825856&04 26 27.10 $-$07 39 39.7&$-$9.6$\pm$0.0&$-$4.9$\pm$0.0&6.2$\pm$0.0&11.5&3.3&535\\
3198734278756825984&04 26 26.91 $-$07 39 41.6&$-$8.2$\pm$0.1&$-$4.8$\pm$0.0&6.2$\pm$0.0&14.5\\
\hline
3185678437170300800&04 34 42.76 $-$08 57 18.5&$-$9.2$\pm$0.0&$-$4.8$\pm$0.0&6.0$\pm$0.0&11.9&6.9&1143\\
3185678437170301056&04 34 42.38 $-$08 57 22.4&$-$8.8$\pm$0.0&$-$5.8$\pm$0.0&6.1$\pm$0.0&13.7\\
\hline
2709158614610269312&22 30 54.93 +06 09 13.8&18.3$\pm$0.1&$-$12.1$\pm$0.1&6.0$\pm$0.0&11.7&11.7&1941\\
2709158687624627584&22 30 54.15 +06 09 15.1&18.0$\pm$0.2&$-$11.2$\pm$0.2&6.2$\pm$0.1&15.8\\
\hline
2623655475128272384&22 38 32.87 $-$06 12 49.4&24.8$\pm$0.1&$-$9.8$\pm$0.1&7.4$\pm$0.1&10.3&6.7&904\\
2623655470833544704\tablenotemark{\ddag}&22 38 32.95 $-$06 12 42.7&29.3$\pm$0.1&$-$9.2$\pm$0.1&7.6$\pm$0.1&12.9\\
\hline
2596395760081700608&22 39 53.53 $-$16 36 23.3&24.5$\pm$0.1&$-$4.4$\pm$0.1&7.9$\pm$0.0&11.5&3.2&408\\
2596395764377286272\tablenotemark{\ddag}&22 39 53.74 $-$16 36 24.2&24.0$\pm$0.2&$-$1.8$\pm$0.3&7.7$\pm$0.1&15.6\\
\hline
2625503165763607808&22 43 16.91 $-$03 03 06.6&25.9$\pm$0.1&$-$12.2$\pm$0.1&6.9$\pm$0.1&10.8&5.1&746\\
2625503165766158848&22 43 17.18 $-$03 03 09.8&25.4$\pm$0.4&$-$11.4$\pm$0.4&5.5$\pm$0.4&18.2\\
\hline
2612073116562162304&22 52 26.88 $-$05 05 12.5&25.8$\pm$0.1&$-$10.3$\pm$0.1&7.8$\pm$0.0&12.1&2.7&344\\
2612073082202424320&22 52 26.71 $-$05 05 13.4&22.5$\pm$0.1&$-$11.0$\pm$0.1&7.7$\pm$0.0&14.3\\
\hline
2659702787750762624&23 25 32.83 +03 32 42.4&18.1$\pm$0.1&$-$12.6$\pm$0.1&7.0$\pm$0.1&12.8&13.8&1966\\
2659702787750762752&23 25 32.06 +03 32 34.7&18.1$\pm$0.1&$-$12.1$\pm$0.1&7.0$\pm$0.0&12.9\\
\hline
\enddata

\tablenotetext{\ddag}{Object has elevated Gaia astrometric noise}
\end{deluxetable}%

\clearpage
\subsection{Comparison with the Pisces-Eridanus stream}
The Pisces-Eridanus stream is a kinematic structure identified by \cite{Meingast2019} in Gaia data. They estimated that this stream is a $\sim$1\,Gyr-old dispersed remnant of a single star formation event that extends to 400\,pc in length. More recently \cite{Curtis2019} have used gyrochronology to estimate an age of $\sim$125\,Myr. Moreover \cite{Curtis2019} suggest that the Pisces-Eridanus stream formed in a dispersed manner with multiple cores, similar to nearby young associations. Hence this stream could represent a more dispersed star formation event which lies at a similar distance to the Pleiades and has a similar age. We set out to measure the wide binary fraction in this stream as any deviation from the Pleiades wide binary fraction would hint at an environmental dependence in the formation and/or survival of wide binaries. 

We repeated the analysis we did for the Hyades and the Young Moving Groups using the membership list from \cite{Meingast2019}. Again we examined coincident pairs between members of the stream and also searched for coincident pairs using the offset pairing method of \cite{Lepine2007}. Figure~\ref{other_plot} shows both of these analyses. It is clear that there is no significant population of chance alignments out to separations of 400". We list our 300-3000\,AU pairs in the printed version of Table~\ref{pe_pairs} with the full list in the electronic version of this table. We estimated the masses of both components of each pair using the same PARSEC models we used for the Pleiades (as the \citealt{Curtis2019} age suggest this cluster is the same age as Pisces-Eridanus). The members listed in \cite{Meingast2019} almost all fall into the $0.5<M/M_{\odot}<1.5$ range.  Hence we are only able to calculate the binary fraction for FGK stars. We do this using the same incompleteness analysis as was used for the Hyades and the Young Moving Groups. Our binary frequency calculations are shown in Table~\ref{other_calc_table}. As with the Young Moving Groups we find that the there is an excess of binaries in the 300-3000\,AU range with 10.7$\pm^{2.8}_{2.3}$\% of FGK stars having companions in the 300-3000\,AU projected separation range compared to our value of 3.6$\pm^{0.8}_{0.7}$\% in our three open clusters.

\section{Discussion}
\subsection{The environmental dependence of the wide binary fraction}
\begin{figure*}
 \setlength{\unitlength}{1mm}
 \begin{tabular}{cc}
 \includegraphics[scale=0.4]{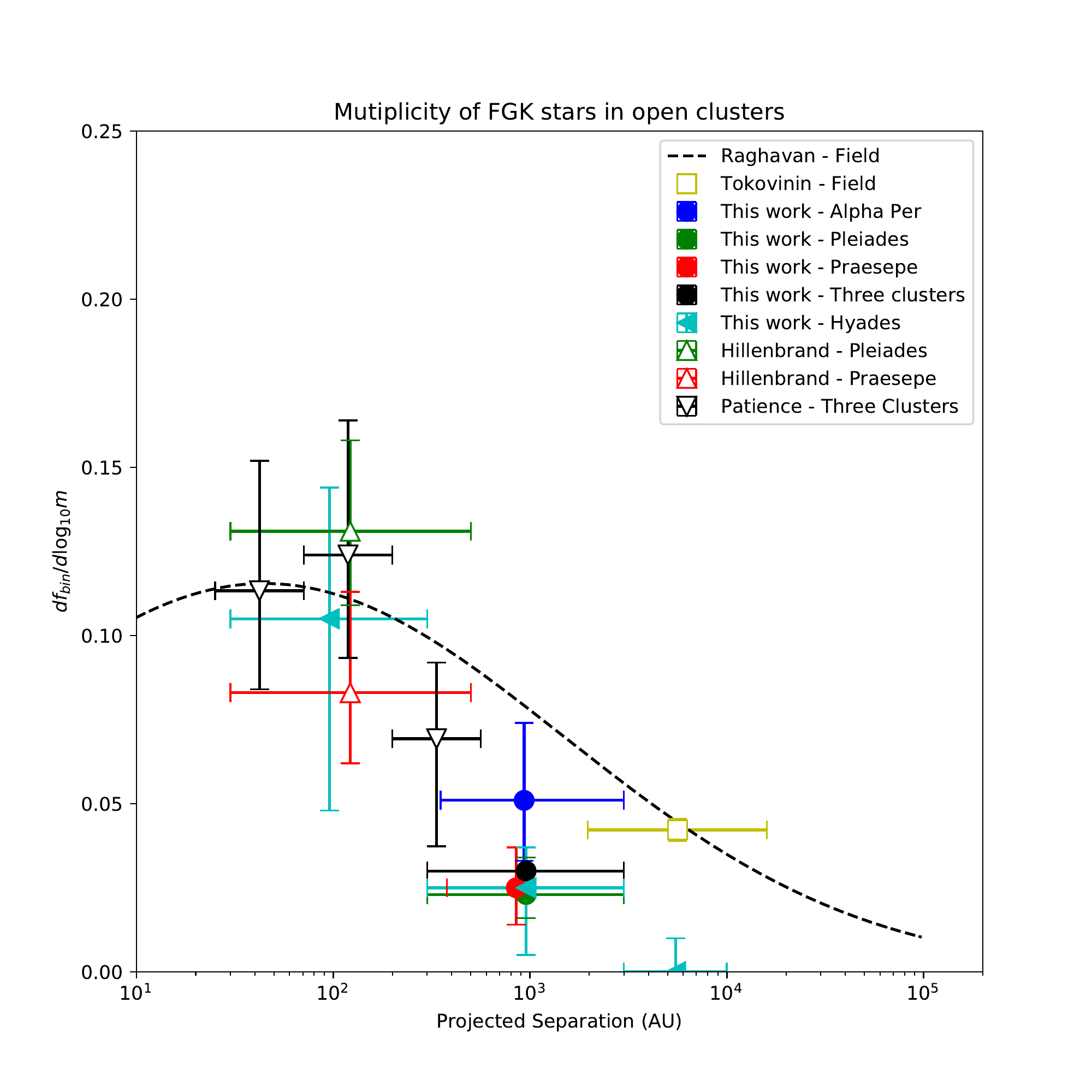}&
 \includegraphics[scale=0.4]{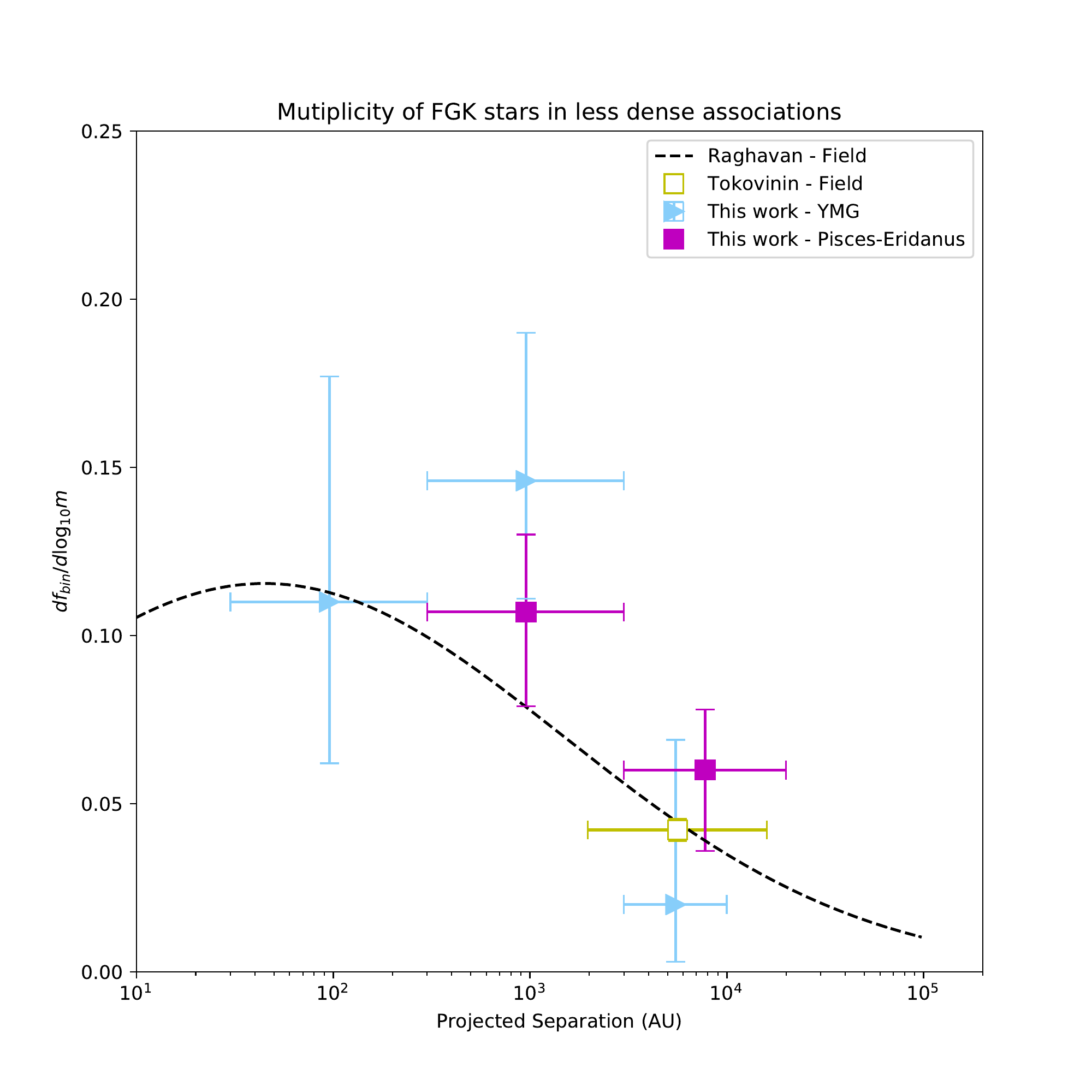}
 \end{tabular}
 \caption[]{The binary fractions from this work, \protect\cite{Raghavan2010}, \protect\cite{Patience2002} (incorporating the results of \protect\citealt{Bouvier1997a}), \protect\cite{Hillenbrand2018} and \cite{Tokovinin2012}. {\bf Left:} It appears that while closer binaries in Alpha~Per, the Pleiades, Praesepe and the Hyades are in agreement with the field population, there is a deficit of wider binaries. {\bf Right:} Conversely the less dense Young Moving Groups and Pisces-Eridanus stream show wide binary fractions at or above the field values.} 
  \label{f_bin_plot}
 \end{figure*}

Our results are plotted, along with the literature values and our Young Moving Group FGK binary fractions in Figure~\ref{f_bin_plot}. As discussed in the introduction, studies of close unresolved binarity in our three clusters indicate that the binary fraction matches the field distribution.  As these studies mostly concentrate on FGK stars we restrict ourselves to pairs where the our primaries have a mass range between 0.5 and 1.5 solar masses. \cite{Hillenbrand2018} refer to their primaries as KM stars but the vast majority have mass estimates in the 0.5--0.9\,$M_{\odot}$ range. We recalculate the uncertainties in the literature results using \cite{Gehrels1986} so that they are comparable with our uncertainty estimates. We also calculated the combined cluster binary fraction from \cite{Patience2002} using the figures in their Table~A1 and multiplying by a factor of $\frac{4}{3}$. This latter factor takes into account possible incompleteness as \cite{Patience2002} state they are likely complete for $q>0.25$ so assuming a flat mass ratio distribution they should have detected 75\% of binaries. We divide by the log separation range each study covers so that we have comparable values of $\frac{df_{bin}}{d\log_{10}x}$ (where $x$ is the angular separation). 
 
It is clear that for the Pleiades, Praesepe and the combined result for all three clusters, the wide binary fraction for FGK dwarfs falls below that expected in the field. The field binary fraction falls away at higher separations and our open cluster 300--3000\,AU wide binary fraction is more comparable with the field value found at separations wider than 2000\,AU by \cite{Tokovinin2012}. Alpha~Per has a higher, but more uncertain binary fraction indicating that this higher value could be due to noise. The deficit in wide companions in two of our three clusters and in the combined cluster result appear to be significant at the $\sim3\sigma$ level. By contrast the Pisces-Eridanus and YMG binary fractions seems to lie slightly above or on the field distribution.  It is possible that we have somehow biased ourselves against close pairs. However we have shown in Section~\ref{bin_comp} that cluster members can be reliably detected at separations wider than two arcseconds. Hence we suggest that this deficit in the number of wide binaries in our cluster could be due to some level of dynamical processing of wide binaries in open clusters. If this is correct then we would expect that the field population of wide binaries would have formed (and survived) preferentially in lower-density formation environments, albeit with some contribution from denser clusters as well. 
 
We can use the approach taken by \cite{Brandner1998} and \cite{Patience2002} to estimate the origins of field wide binaries and of stars in general. These works proposed a model where there were two possible origins for wide binaries, Pleiades-like clusters and lower density star forming regions similar to Pisces-Eridanus. This is an extreme model as there is actually a continuum of stellar formation environments with only 7\% of embedded clusters resulting in Pleiades-like open clusters \citep{Lada2003}. There are also more extreme dense, high-mass star forming regions than the progenitor of the Pleiades with multiple clusters in the Milky Way having masses over 10$^4$\,M$_{\odot}$ \citep{PortegiesZwart2010} which will be so dense that the average separation between members in the centre will be of the order of a few thousand AU. Nonetheless, if we assume that the binary fraction of field stars follows the distribution of \cite{Raghavan2010} this would mean that in our 300-3000\,AU separation range we would expect an occurrence rate of $\frac{df_{bin}}{d\log_{10}x}=7.8\%$. In the Pleiades we find $\frac{df_{bin}}{d\log_{10}x}=$2.3$\pm^{1.1}_{0.7}$\% in the same separation while in Pisces-Eridanus we have $\frac{df_{bin}}{d\log_{10}x}=$10.7$\pm^{2.8}_{2.3}$\%. We then assume the aforementioned toy model of star formation with only two environments, the progenitor environment of the Pleiades and the progenitor environment of Pisces-Eridanus. In this model $X\%$ of stars would form in the former environment and $100-X\%$ would form in the latter. We used a numerical simulation where we drew a hundred thousand random binomial distributions with our calculated 300-3000\,AU binary fractions for both the Pleiades and Pisces-Eridanus and with the correct number of stars in each around which we could detect binaries. For each of these hundred thousand simulated populations we calculated $X$ and from the distribution of $X$ calculated the median and standard deviation around that median. These simulations imply that $X=38\pm^{13}_{16}$\% of stars form in Pleiades-progenitor-like environments and the remainder in less dense, more distributed environments. This is below the 70$\pm^{10}_{15}\%$ or 80\% figure calculated by \cite{Patience2002} who assumed that the binary separation distributions in star forming regions would equate to the final binary separation distributions that would make their way from those star forming regions into the field. We again note that this method for estimating the origin of wide binaries forces star formation environments into two distinct types when there is a spectrum of star formation environments.

Our work suggests that the products of low density star formation regions have higher wide binary fractions than the products of higher-mass star-forming clusters. In the lower-density Taurus-Auriga \citep{Kraus2011} and Ophiuchus \citep{Cheetham2015} regions the wide binary fraction appears to be at or above the field level. The best nearby comparator we have for more massive star forming regions is the Trapezium Cluster within the Orion Nebula Cluster (ONC). \cite{Lada2003} used the low number of wide binaries found by \cite{Scally1999} as evidence for the rapid destruction of wide binaries in the progenitors of open clusters. However recently \cite{Jerabkova2019} found a wide (1000--3000\,AU) binary fraction of 5\% with primaries that are mostly M dwarfs. This work, like that of \cite{Scally1999} covered a much larger area than just the central Trapezium Cluster itself. In a survey of the central 20\,arcmins of the ONC \cite{Reipurth2007} found a binary fraction of 8.8$\pm$1.1\% between 67.5 and 675 AU, again with a sample dominated by M dwarfs (83\%). Even accounting for the ONC's younger age (1--3\,Myr; \citealt{Beccari2017}) the vast majority of these objects would have comparable masses to our M dwarf sample in our clusters. We find that across our three clusters that 1.4$\pm^{0.4}_{0.3}$\% of M dwarfs are the primary stars in binary systems with separations in the 300--3000\,AU range. We recalculate the \cite{Reipurth2007} binary fraction in their two widest bins 250--675\,AU and find a binary fraction of  $\frac{df_{bin}}{d\log_{10}x}=5.4\pm^{1.6}_{1.3}$\%. \cite{Reipurth2007} found the ratio of binaries with projected separations wider than $\sim$250\,AU to those wider than this limit varied with the distance from the centre of the Trapezium Cluster with relatively few wide binaries in the inner 400\,arcseconds of the cluster. \cite{Reipurth2007} also note that their survey of the central region of the ONC has roughly 2.2 times fewer wide binaries than studies of looser T-Tauri associations.

We note that using a total mass of 735\,M$_{\odot}$ and a half-mass radius of 3.66\,pc \citep{Pinfield1998} the hard-soft boundary in the Pleiades today would be roughly 1500\,AU. However much of the dynamical processing of a cluster like the Pleiades would likely have been done when it was younger and possibly more compact. See \cite{Parker2009}'s work on an ONC-like cluster.
\subsection{Other possible explanations}
We have found that open clusters such as the Alpha Per, the Hyades, the Pleiades and Praesepe host fewer wide binaries than associations which likely formed in less dense star forming regions such as Young Moving Groups and the Pisces-Eridanus stream. There are a number of arguments against the results of our survey being evidence for an environmental dependence in the wide binary frequency. It could be that we are missing a large number of binaries in the open clusters due to small-scale incompleteness in the Gaia data. However we have shown that the quality of  Gaia data are not affected by a close companion wider than three arcseconds and only minimally affected in the 2"--3" range. Furthermore if small-scale incompleteness were to affect the Pleiades and suppress its binary fraction, one would also expect it to affect the Pisces-Eridanus stream (which lies at a similar distance) in a similar way. Finally, we have shown that the Hyades also has a suppressed wide binary fraction at $>$300\,AU. For this to be caused by small-scale incompleteness the Gaia data would need to be incomplete at separations of 6 arcseconds. We have found no evidence to support such incompleteness.

It could be argued that the deficit in wide binaries in open clusters would be replenished by the cluster members capturing other cluster members as the cluster dissolves. Indeed it has been suggested by \citep{Kouwenhoven2010} this could be origin of 1--30\% of $>$1000\,AU wide binaries. However we note that the Hyades, a cluster which has dissolved to the point where capture between members seems incredibly unlikely, also has a deficit of wide binaries similar to the other open clusters. 

It could be possible that systems which currently have separations below our 300\,AU lower bound will evolve into wider binaries by the time they reach field age. \cite{Elliott2016} suggest a model based on work by \cite{Reipurth2012} where wide binaries are hidden higher-order multiples and slowly increase their separation over time, albeit at younger ages than our open clusters.  We think this process will be unlikely to have affected our results as the deficit in wide binaries appears similar between the Pleiades (125\,Myr) and the Hyades \& Praesepe ($\sim$750\,Myr). Further we also note that the Pleiades and the Pisces-Eridanus stream are approximately the same age and show significantly different wide binary fractions, suggesting an environment-dependent rather than age-dependent effect. It is however true that further evolution of wide binary systems before they reach field age would alter the fraction of wide binaries each star formation environment would donate to the field.

\section{Conclusions}
We have shown that there appears to be a deficit of wide (300--3000\,AU) binaries in open clusters. This deficit persists over a range of cluster distances, suggesting it is not the product of small-scale incompleteness in the Gaia data release. We also compared to young moving groups and the Pisces-Eridanus stream (both likely products of lower-density, less-clustered star formation events) and found these had wide binary fractions at or above the field level. These results suggest that wide binaries are more likely form and/or survive in low density formation environments than more dense star clusters.
\section*{Acknowledgments}
This work makes use of the TopCAT software developed by \citep{Taylor2005}. We thank our anonymous referee for their helpful and constructive comments. The first author would like to thank the organisers of the 20th Cambridge Workshop on cool stars, stellar systems and the Sun for putting him in a dorm flat with noted Pleiades expert John Stauffer. This reminded the author that he hadn't done anything on the open cluster wide binary project he'd been planning for a while.
This work has made use of data from the European Space Agency (ESA) mission
{\it Gaia} (\url{https //www.cosmos.esa.int/gaia}), processed by the {\it Gaia}
Data Processing and Analysis Consortium (DPAC,
\url{https //www.cosmos.esa.int/web/gaia/dpac/consortium}). Funding for the DPAC
has been provided by national institutions, in particular the institutions
participating in the {\it Gaia} Multilateral Agreement. N.R.D acknowledges acknowledges the support of the DFG priority program SPP 1992 "Exploring the Diversity of Extrasolar Planets (Characterising the population of wide orbit exoplanets)". This research made use of Astropy,\footnote{http://www.astropy.org} a community-developed core Python package for Astronomy \citep{Robitaille2013,Price-Whelan2018}
\bibliographystyle{mn2e} 
 \bibliography{./ndeacon} 
 \appendix
 \section{Deriving the cluster membership likelihood}
 \label{memb_like}
 We used a similar likelihood method to \cite{Deacon2004} (based on previous work such as \citealt{Sanders1971} and \cite{Hambly1995}) which fitted likelihood distributions in proper motion space for the cluster and field populations. We  used an altered version of this method in proper motion space. First we calculate the expected proper motion of a cluster member at the space position of each star using the $UVW$ velocities quoted in \cite{Lodieu2019a} and coordinate system transformations from the {\sc astropy} coordinates package \citep{Robitaille2013}. We then project the proper motion of each object to a basis where one direction is parallel to the expected cluster proper motion at the space position of each object and one direction perpendicular ($\mu_{\parallel obs}$ and $\mu_{\perp obs}$). In this basis a cluster member will be expected to have its proper motion entirely in the direction of cluster motion. Thus the total expected proper motion for each cluster member is $\mu_{\parallel exp}$ while $\mu_{\perp exp}=0$. We then subtract the expected proper motion in the direction of cluster motion such that $\Delta\mu_{\parallel}=\mu_{\parallel obs}-\mu_{\parallel exp}$. We then use $\Delta\mu_{\parallel}$ and $\mu_{\perp obs}$ as the basis for our proper motion analysis. The cluster forms a Gaussian distribution around roughly (0,0) while we found that the field population was also a Gaussian centred away from the origin. The width of the cluster distribution will be a combination of the true $UVW$ distribution of the cluster, the range of distances along the line of sight, uncertainties in the expected proper motions and the measurement uncertainties of the proper motions. We also
diverge from the method of \cite{Deacon2004} by adding a parallax terms to our likelihoods. Here the width of the parallax distribution will be a combination of parallax uncertainty and the line of sight depth of the cluster. As the typical parallax and proper motion uncertainties increase with magnitude we split our analysis into magnitude bins. This allows fainter stars to have higher parallax and proper motion dispersions than brighter stars. We neglect the covariances between proper motion and parallax and assume that the intrinsic proper motion dispersion does not change significantly across the cluster. Assuming the Pleiades is about 10\,pc deep we would expect the proper motion dispersion to vary by around 7\% with smaller dispersions for other clusters. Our selection only uses proper motions and parallaxes to determine membership probabilities while other methods use the spatial distribution of the stars to calculate membership probabilities. We choose not to do this as we do not want to assume a possibly incorrect functional form for the spatial distribution that could bias our selection.We start by defining a likelihood $\phi$ that is a sum of field and cluster components.
 \begin{equation}
\label{memb_like1}
\phi=f\phi_f+(1-f)\phi_c
\end{equation}
Where the field component is defined as,
 \begin{equation}
\label{memb_like2}
\phi_f=\frac{c_{\varpi}}{2\pi\Sigma_{\parallel}\Sigma_{\perp}}e^{\big(-\frac{(\mu_{\perp}-\mu_{\perp,f})^2}{2\Sigma_{\perp}^2}-\frac{(\Delta\mu_{\parallel}-\Delta\mu_{\parallel, f})^2}{2\Sigma_{\parallel}^2}-\frac{\varpi}{\tau_{\varpi}}\big)}
\end{equation}
Where the proper motion terms parallel and perpendicular to the direction of cluster motion ($\Delta\mu_{\parallel}$ and $\mu_{\perp}$) are Gaussian and the parallax terms ($\varpi$) is a declining exponential. \cite{Deacon2004} used a declining exponential for the $\mu_y$ term. We chose to use a declining exponential in $\varpi$ after examining a historgram of the parallax values for each the area around each cluster. As we use a declining exponential, we require a normalisation term for $\varpi$
\begin{equation}
\label{tau2}
c_{\varpi}=\frac{1}{\int_{\varpi_{min}}^{\varpi{max}} e^{-\frac{\varpi}{\tau_{\varpi}}}d\varpi}=\frac{1}{\tau_{\varpi}(e^{-\frac{\varpi_{min}}{\tau_\varpi}}-e^{-\frac{\varpi_{max}}{\tau_{\varpi}}})}
\end{equation}
For all our clusters we choose limits of $\varpi_{min}=3$\,mas and $\varpi_{max}=15$\,mas.

We define the cluster component of our likelihood as,
 \begin{equation}
\label{memb_like3}
\phi_c=\frac{1}{(2\pi)^{3/2}\sigma_{\varpi}\sigma_{\perp}\sigma_{\parallel}}e^{\big(-\frac{(\mu_{\perp}-\mu_{\perp c})^2}{2\sigma_{\perp}^2}-\frac{(\Delta\mu_{\parallel}-\Delta\mu_{\parallel c})^2}{2\sigma_{\parallel}^2}-\frac{(\varpi-\varpi_{c})^2}{2\sigma_{\varpi}^2}\big)}
\end{equation}
Where we use Gaussians for both the parallax and both components of the proper motion. Unlike in \cite{Deacon2004} we use separate standard deviations for the proper motion components of the Gausssian.

We then find the best fitting parameters by maximising the derivative of the logarithm of the likelihood with respect to each of our likelihood parameters $\Theta$ such that,

\begin{equation}
\label{par_diff}
\sum_i \frac{\partial \ln \phi_i}{\partial \Theta}=0
\end{equation}
This then leads to individual equations
\begin{equation}
\label{eq_f}
f :\sum_i \frac{\phi_{f,i}-\phi_{c,i}}{\phi_i}=0
\end{equation}
\begin{equation}
\label{eq_sigma_mux}
\sigma_{\perp}: \sum_i \frac{\phi_{c,i}}{\phi_i}\left( \frac{(\mu_{\perp,i}-\mu_{\perp c})^2}{\sigma_{\perp}^2} -1\right)
\end{equation}
\begin{equation}
\label{eq_sigma_muy}
\sigma_{\parallel}: \sum_i \frac{\phi_{c,i}}{\phi_i}\left( \frac{(\Delta\mu_{\parallel,i}-\Delta\mu_{\parallel c})^2}{\sigma_{\parallel}^2} -1\right)
\end{equation}
\begin{equation}
\label{eq_sigma_plx}
\sigma_{\varpi}: \sum_i \frac{\phi_{c,i}}{\phi_i}\left( \frac{(\varpi_i-\varpi_c)^2}{\sigma_{\varpi}^2} -1\right)
\end{equation}
\begin{equation}
\label{eq_muxc}
\mu_{\perp c}: \sum_i \frac{\phi_{c,i}}{\phi_i}\left( \mu_{\perp,i}-\mu_{\perp c} \right)
\end{equation}
\begin{equation}
\label{eq_muyc}
\Delta\mu_{\parallel c}: \sum_i \frac{\phi_{c,i}}{\phi_i}\left(\Delta\mu_{\parallel,i}-\Delta\mu_{\parallel c} \right)
\end{equation}
\begin{equation}
\label{eq_l-plxc}
\varpi_{c}: \sum_i \frac{\phi_{c,i}}{\phi_i}\left( \varpi_{i}-\varpi_{c} \right)
\end{equation}
\begin{equation}
\label{eq_sigma_muxf}
\Sigma_{\perp}: \sum_i \frac{\phi_{f,i}}{\phi_i}\left( \frac{(\mu_{\perp,i}-\mu_{\perp f})^2}{\Sigma_{\perp}^2} -1\right)
\end{equation}
\begin{equation}
\label{eq_muxf}
\mu_{\perp f}: \sum_i \frac{\phi_{f,i}}{\phi_i}\left( \mu_{\perp,i}-\mu_{\perp f} \right)
\end{equation}
\begin{equation}
\label{eq_sigma_muyf}
\Sigma_{\parallel}: \sum_i \frac{\phi_{f,i}}{\phi_i}\left( \frac{(\mu_{\parallel,i}-\mu_{\parallel f})^2}{\Sigma_{\parallel}^2} -1\right)
\end{equation}
\begin{equation}
\label{eq_muyf}
\Delta\mu_{\parallel f}: \sum_i \frac{\phi_{f,i}}{\phi_i}\left( \mu_{\parallel,i}-\mu_{\parallel f} \right)
\end{equation}
\begin{equation}
\label{eq_tauplx}
\tau_{\varpi}:\sum_i \frac{\phi_{f,i}}{\phi_i}\left( \frac{\varpi_{i}}{\tau_{\varpi}} -1-c_{\varpi}(\varpi_{min}e^{-\frac{\varpi_{min}}{\tau{\varpi}}}-\varpi_{max}e^{-\frac{\varpi_{max}}{\tau{\varpi}}}) \right)
\end{equation}
We solve these equations iteratively using a bisection algorithm. We list the fitted parameters for each magnitude bin in each cluster in Table~\ref{fit_params}. Once the cluster parameters have converged we use these to calculate membership probabilities using,
 \begin{equation}
\label{memb_prob}
p_{memb,i}=\frac{(1-f)\phi_{c,i}}{f\phi_{f,i}+(1-f)\phi_{c,i}}
\end{equation}
\begin{table*}
\caption{\label{fit_params} The fitted parameters for each magnitude bin in each cluster.}
\tiny
\begin{center}
\begin{tabular}{|rrrrrrrrrrrrr}
\hline
Range&$f$&$\sigma_{\perp}$&$\sigma_{\parallel}$&$\sigma_{\varpi}$&$\mu_{\perp c}$&$\Delta\mu_{\parallel c}$&$\varpi_{c}$&$\Sigma_{\perp}$&$\Sigma_{\parallel}$&$\tau_{\varpi}$&$\mu_{\perp f}$&$\mu_{\parallel f}$\\
&&(mas/yr)&(mas/yr)&(mas)&(mas/yr)&(mas/yr)&(mas)&(mas/yr)&(mas/yr)&(mas)&(mas/yr)&(mas/yr)\\
\hline
\multicolumn{12}{c}{Alpha Per}\\
\hline
$G<$10&0.792&0.623&0.784&0.231&0.019&-0.239&5.678&14.622&19.411&2.195&3.065&-8.216\\
10$<G<$12&0.941&0.434&0.565&0.197&0.065&-0.209&5.660&14.757&19.758&1.541&3.056&-7.429\\
12$<G<$14&0.957&0.387&0.460&0.219&-0.150&-0.295&5.678&15.240&19.450&1.332&2.452&-6.634\\
14$<G<$16&0.942&0.464&0.586&0.206&0.167&-0.417&5.695&15.147&20.390&1.606&2.444&-7.066\\
16$<G<$18&0.933&0.480&0.948&0.280&0.226&-0.580&5.749&15.280&20.569&1.349&2.817&-7.070\\
18$<G<$19&0.967&0.764&1.627&0.415&0.265&-0.436&5.713&13.557&19.341&1.039&2.347&-8.317\\
\hline
\multicolumn{12}{c}{Pleiades}\\
\hline
$G<$10&0.737&0.876&1.172&0.267&0.043&-0.409&7.358&17.523&17.231&2.804&-0.179&-12.161\\
10$<G<$12&0.882&0.831&0.904&0.261&-0.052&-0.326&7.303&17.621&18.369&1.637&-1.510&-9.447\\
12$<G<$14&0.902&0.848&0.855&0.222&-0.117&-0.320&7.336&18.039&18.887&1.588&-2.265&-8.892\\
14$<G<$16&0.885&0.845&1.207&0.289&0.020&-0.486&7.352&18.654&19.079&1.747&-2.203&-8.847\\
16$<G<$18&0.870&0.822&1.437&0.338&0.089&-0.417&7.349&18.073&18.681&1.451&-1.818&-8.873\\
18$<G<$19&0.948&0.991&2.254&0.483&0.363&-0.399&7.315&16.795&18.271&1.124&-2.239&-8.096\\
19$<G<$20&0.979&1.267&4.067&0.673&0.334&-0.581&7.378&14.611&16.511&0.989&-1.847&-10.132\\
\hline
\multicolumn{12}{c}{Praesepe}\\
\hline
$G<$10&0.740&0.581&1.207&0.182&-0.036&-0.317&5.310&17.665&18.512&2.504&-6.653&-13.481\\
10$<G<$12&0.885&0.756&0.768&0.107&-0.099&-0.790&5.364&17.165&17.425&1.523&-8.847&-12.274\\
12$<G<$14&0.891&0.624&0.735&0.160&-0.020&-0.947&5.369&17.780&18.600&1.342&-8.781&-11.434\\
14$<G<$16&0.881&0.637&1.186&0.157&0.005&-0.669&5.371&18.076&19.170&1.554&-9.584&-12.476\\
16$<G<$18&0.897&0.531&1.640&0.249&0.017&-0.351&5.336&18.073&18.772&1.352&-9.826&-11.923\\
18$<G<$19&0.900&0.669&2.471&0.386&0.008&-0.416&5.338&16.986&17.969&1.085&-9.969&-12.680\\
19$<G<$20&0.969&0.876&4.261&0.534&-0.152&-1.289&5.461&14.938&16.860&0.948&-9.320&-14.098\\
\hline
\end{tabular}
\end{center}
\normalsize
\end{table*}
 \section{Deriving the binary companion likelihood}
 \label{likemath}
 We begin by thinking of the pairs as the function of four variables, the separation on the sky between the two components of the pair $x$, the masses of two components $m_1$ and $m_2$, and the distance of the pair from the cluster centre $r$. We use the distance of the primary (brighter) component of each pair from the cluster centre. As our pairs are a few tens of arcseconds in separation and the clusters are several degrees across this should not affect our results significantly.  We define the number of pairs as a function of cluster radius, separation and primary and secondary masses as,

\begin{equation}
\label{like1}
\begin{aligned}
n(r,x,m_1,m_2)=&f_{comp}n_{comp}(r,x,m_1,m_2)\\
&+(1-f_{comp})n_{coincident}(r,x,m_1,m_2) 
\end{aligned}
\end{equation}
Where $f_{comp}$ is the fraction of companions in our sample. We also define the number of true binary pairs as,
\begin{equation}
\label{n_comp}
n_{comp}(r,x,m_1,m_2)=c_{comp,m}c_{comp,\rho}\rho_{obj}(r) c_{comp,x}(xd)^{-1}
\end{equation}
where $d$ is the distance to the primary derived from inverting the Gaia parallax, $\rho(r)=e^{(-r/r_0)}$ is the number density of cluster members at a distance $r$ from the cluster centre and $c_{comp,rho}$ normalises the number of comparies such that if $f_{comp}=1$ then the total number of pairs from $r=0$ to $r=r_{max}$ is $N_{tot}$ the total number of pairs in our sample. The variable $c_{comp,x}$ normalises the separation distribution and $c_{comp,m}$ normalises the mass distribution.
Note that the number depends on the number density of stars. Each true binary system will consist of two stars, 
a primary and a secondary. The distribution of primary stars will follow the number density of stars in the cluster. However we assume a constant binary fraction across the cluster so the probability of any particular primary star having a companion will not depend on the distribution of stars in the cluster. We note that in a dynamically relaxed cluster such as Praesepe (or the Pleiades; \citealt{Moraux2004}), binary systems will sink to the centre of the cluster. However modelling such a radially-dependent binary fraction is likely beyond our small sample of binaries. We also assume a log-flat distribution in projected separation in AU \citep{Opik1924}. We note that assuming a separation distribution that declines with separation would increase the (already close to unity) binary probabilities of closer binaries while decreasing the binary probabilities of already marginal wider binaries. This would have the overall effect of reducing the sum of the binary probabilities in the 300-3000\,AU range in our calculations. Note also there is no dependence on the secondary mass. This is because we assume a flat mass ratio distribution for binaries.

Now we turn to the normalisation factors. Firstly $c_{comp,\rho}$,
\begin{equation}
c_{comp,\rho}=\frac{N_{tot}}{2\pi\int_0^{r_{max}} r \rho(r)dr}=\frac{N_{tot}}{2\pi r_0^2(1-e^{(-r/r_0)}(1+r_{max}/r_0))}
\end{equation}
And $c_{comp,x}$ normalises the part of Equation~\ref{n_comp} that is dependent on separation such that
\begin{equation}
c_{comp,x}=\frac{1}{\int_{x_{min}}^{x_{max}} (xd)^{-1}dx}=\frac{d}{\log(\frac{x_{max}}{x_{min}})}
\end{equation}
Where $\log$ is the natural logarithm. Finally we look at the normalisation of a flat mass ratio distribution,
\begin{equation}
c_{comp,m}=\frac{1}{\int_{0}^{m_1}dm_2}=\frac{1}{m_1}
\end{equation}
Here $m_1$ is the primary and by definition the secondary cannot have a larger mass than this .

Hence our true physical binary population becomes 
\begin{equation}
n_{comp}(r,x)=\frac{1}{m_1}\frac{N_{tot} e^{-r/r_0}}{2\pi r_0^2(1-e^{(-r_{max}/r_0)}(1+r_{max}/r_0))}\frac{1}{x\log(\frac{x_{max}}{x_{min}})}
\end{equation}

The coincident population will take the form
\begin{equation}
\begin{aligned}
n_{coincidient}(r,x)=&c_{coincident,m_2}\xi(m_2)c_{coincident,\rho}(\rho_{obj}(r))^2\\
& c_{coincident,x}\times2\pi x
\end{aligned}
\end{equation}
note that this depends on density squared. Coincident pairs are one star in the cluster randomly having another star in the cluster at angular distance $x$ away from it. The first factor of density comes from the fact that each pair starts with a cluster member (let's call it "star one") and the density of those cluster members follows $\rho(r)$. The cluster member "star one" could then paired with a random nearby cluster member. The number of cluster members "star one" could be paired within an annulus at separation $x$ with a given bin width $dx$ is the sky area of that annulus $2\pi x$ times the number density of cluster members near to "star one". This number density introduces the second factor of $\rho (r)$. This second factor of $\rho (r)$ means that the population of coincident pairs in the cluster has half the scale length in $r$ than the population of true binary companions. The mass term is the value of the system mass function of the cluster at the mass of the secondary star in the pair. This is because a coincident secondary is simply a random draw from the population of the cluster and the system mass function describes the distribution of masses in the cluster. We do not account for mass segregation in the cluster in our likelihood. We use the system mass functions we calculated for each cluster, converting each from $\frac{dn}{d\log_{10}m}$ to $\frac{dn}{dm}$. We numerically solve the integral of each system mass function between the minimum and maximum masses in our isochrone for each object to derive our normalisation factor $c_{coincident,m_2}$.

The other normalisation factors are similar to the companion distribution, firstly assume that if all the pairs are background pairs then the integral over $(\rho_{obj}(r))^2$ will give $N_{tot}$.
\begin{equation}
c_{coincident,\rho}=\frac{N_{tot}}{2\pi\int_0^{r_{max}} r (\rho(r))^2 dr}=\frac{N_{tot}}{2\pi (2r_0)^2(1-e^{(-2r_{max}/r_0)}(1+2r_{max}/r_0))}
\end{equation}

and the normalisation in separation space is given by
\begin{equation}
c_{coincident,x}=\frac{1}{\int_{x_{min}}^{x_{max}}2\pi x dx}=\frac{1}{\pi(x_{max}^2-x_{min}^2)}
\end{equation}
hence we find that the coincident population takes the form
\begin{equation}
n_{coincidient}(r,x)=\frac{N_{tot}c_{coincident,m_2}\xi(m_2)e^{-2r/r_0}}{2\pi (2r_0)^2(1-e^{(-2r/r_0)}(1+2r_{max}/r_0))}\frac{2x}{(x_{max}^2-x_{min}^2)}
\end{equation}
Dividing both $n_{comp}$ and $n_{coinicident}$ by $N_{tot}$ yields the normalised likelihood,
\begin{equation}
\begin{aligned}
\phi&=f_{comp}\phi_{comp}+(1-f_{comp})\phi_{coincident}\\
\phi_{comp}&=\frac{e^{-r/r_0}}{2\pi r_0^2(1-e^{(-r/r_0)}(1+r_{max}/r_0))}\frac{1}{x\log(\frac{x_{max}}{x_{min}})}\\
\phi_{coincident}&=\frac{e^{-2r/r_0}c_{coincident,m_2}\xi(m_2)}{2\pi (2r_0)^2(1-e^{(-2r/r_0)}(1+2r_{max}/r_0))}\frac{2x}{(x_{max}^2-x_{min}^2)}
\end{aligned}
\end{equation}

\end{document}